\documentclass[11pt]{article}

\usepackage{amsmath}
\usepackage{amssymb} 
\usepackage{epstopdf}
\usepackage{graphicx}

\newcommand{\bra}[1]{\langle #1 |} 
\newcommand{\ket}[1]{| #1 \rangle } 
\newcommand{\vev}[1]{\langle #1 \rangle } 
\newcommand{\braket}[2] {\langle #1 | #2\rangle }

\newcommand{\half}{\frac{1}{2}}
\newcommand{\cqfd}{$\blacksquare$ }

\title{The Open Quantum Brownian Motion} 
\author{}
\date{}

\begin{document}
\maketitle
\pagestyle{empty}

\vskip -0.5 truecm 

\centerline{Michel Bauer ${}^{\spadesuit~}$, Denis Bernard ${}^{\clubsuit~}$ and Antoine Tilloy ${}^{\clubsuit~}$}

%\vskip 0.5 truecm

\begin{center}
{\footnotesize 
\noindent
${}^\spadesuit$ Institut de Physique Th\'eorique de Saclay, CEA-Saclay $\&$ CNRS, 91191 Gif-sur-Yvette, France.\\
$^\clubsuit$ Laboratoire de Physique Th\'eorique de l'ENS, CNRS $\&$ Ecole Normale Sup\'erieure de Paris, France.}
\end{center}
\vskip 0.5 truecm

\pagestyle{plain}

\begin{abstract}
Using quantum parallelism on random walks as original seed, we introduce new quantum stochastic processes, the open quantum Brownian motions. They describe the behaviors of quantum walkers -- with internal degrees of freedom which serve as random gyroscopes -- interacting with series of probes. These processes may also be viewed as the scaling limit of open quantum random walks and we develop this approach along three different lines: quantum trajectory, quantum dynamical map, and quantum stochastic differential equation. We also present a study of the simplest case, with a two level system as internal gyroscope, illustrating the interplay between ballistic and diffusive behaviors at work in these processes.
%Started April 2013, now \today.
\end{abstract}

\vskip 1.0 truecm 

\baselineskip = 6 pt
%\newpage
\tableofcontents
%\newpage
\baselineskip = 14 pt

\vskip 0.5 truecm

\noindent {\bf Notations:}\\
-- ${\cal H}_z$: orbital (walker) Hilbert space, $\mathbb{C}^{\mathbb{Z}}$ in the discrete, $L^2(\mathbb{R})$ in the continuum.\\
-- ${\cal H}_c$: internal (gyroscope) Hilbert space.\\
-- ${\cal H}_{\rm sys}={\cal H}_z\otimes{\cal H}_c$: system Hilbert space.\\
-- ${\cal H}_p$: probe Hilbert space, ${\cal H}_p=\mathbb{C}^2$.\\
-- $\rho^{\rm tot}_t$: density matrix for the total system (walker+gyroscope+probes).\\
-- $\bar \rho_t$: reduced density matrix on ${\cal H}_{\rm sys}$: $\bar\rho_t=\int dxdy\, \bar\rho_t(x,y)\otimes \ket{x}_z\bra{y}$.\\
-- $\hat \rho_t$: system density matrix in a quantum trajectory: $\hat\rho_t=\int dxdy\, \hat\rho_t(x,y)\otimes \ket{x}_z\bra{y}$.\;\\
.\hskip .8 truecm    If diagonal and localized in position: $\hat \rho_t=\rho_t\otimes\ket{X_t}_z\bra{X_t}$.\\
-- $\rho_t$: internal density matrix in a simple quantum trajectory.\\
-- $X_t$: walker position in a simple quantum trajectory.\\
-- $B_t$: normalized Brownian motion.\\
-- $\xi_t$, $\xi_t^\dag$: quantum noises.\\

\noindent
{\bf Acknowledgements}: \\
 This work was in part supported by ANR contract ANR-2010-BLANC-0414. \\

\section{Introduction}

The importance of random walks and Brownian motion \cite{feller}, both in
physics and in mathematics, needs not to be emphasized. A first version of
quantum random walks \cite{quantRW} was introduced two decades ago, with many potential applications, in particular to quantum information or quantum computing \cite{review,Qcomp}. Their behaviors are drastically different from those of their classical ancestors in two respects at least: first their diffusive properties are affected by multiple quantum interferences, and second the quantum coin they use is only reshuffled but not reinitialized at each step which reinforces the effect of interferences. A new version of quantum random walks \cite{Attal_etal} has recently been introduced with the aim of incorporating decoherence effects (hence the name ``open'' quantum random walks). Although subject to quantum effects, this later version keeps structural properties inherited from their classical analogues. In particular, a natural extension of it -- a dilation --  involves independent quantum coins at each steps, and as a consequence, it also naturally incorporates a quantum analogue of the notion of filtration, familiar in the theory of stochastic processes \cite{probas}, together with a time arrow associated to the increase of available information which goes with it. This, in some cases, allows to talk about quantum trajectories of those walks, a notion which was partially lost in the original definition. 

The aim of this paper is to construct ``open'' quantum Brownian motions either as a time and space continuous scaling limit of open quantum random walks or as an extension of quantum parallelism on random walks. This yields quantum stochastic processes with (we believe) many potential, and hopefully interesting, mathematical properties -- by construction they yield quantum processes with spatial structures, as other quantum stochastic processes they promote most of basic structures of classical stochastic processes to a non-commutative setting, etc. -- and physical applications -- they provide realizations of progressive non-demolition measurements, their quantum trajectories seem to possess a rich variety of distinct regimes, they can be used to model physical situations for open or out of equilibrium quantum systems, etc. These processes are different, but clearly related, to the quantum Brownian motions modeled by quantum particles in contact with a thermal bath as in the Caldeira-Leggett model \cite{CalLeg,Froehlich}.

As recently pointed out \cite{BBT13}, a competition between ballistic and
diffusive behaviors is at play on open quantum Brownian motions. We shall elaborate on
it in the last Section. There is a change of regime, from ballistic to
diffusion, which can be understood as follows. In some range of the parameter
space, the external quantum coins, which we call probes, induce mesoscopic
measurements of the internal walker gyroscope. As the characteristic time of
these mesoscopic measurements is shortened these measurements become sharper and
sharper, and this induces projections of the internal gyroscope state intertwined with Bohr quantum jumps \cite{Bohr}. The ballistic regime is a consequence of this phenomena.

One of the simplest way to grasp what open quantum Brownian motions could be is probably to start from a description of what quantum parallelized random walks could lead to, and this is the purpose of the two following (and hopefully pedagogical) sub-Sections. The manuscript aims are more precisely presented in the third sub-Section of this Introduction, and depending on her/his taste, the reader may choose to read these sub-Sections in the order she/he prefers.
 
\subsection{Quantum parallelism on random walks}

Let us first have a look at standard random walks but with a quantum mechanically biased view.
We consider a random walker that we view as a quantum system\footnote{The archetypal and intuitive example would be that of a quantum particle hoping on a 1D lattice but one could consider other systems, e.g. number of photons in cavity QED experiment as in \cite{Haroche}).} and suppose that the evolution of this particle is given by the following iterated scheme. At each time step, the walker is coupled with another system which we call a probe. These probes are going to play the role of (quantum) coins and they are renewed at each time step. We assume the space of probe states is $\mathbb{C}^2$ and we pick an orthonormal basis which we denote by $\ket{\pm}$. For a random walk on the line we introduce the walker state space $\mathbb{C}^\mathbb{Z}$ with specified basis $\ket{x}$, $x\in\mathbb{Z}$, with $\langle y\ket{x}=\delta_{y;x}$. The elementary evolution is then\footnote{Physically, this elementary evolution is that of the coupled system (walker+probe) and should have been written $ \ket{x} \otimes \ket{
\phi} \rightarrow \frac{1}{\sqrt{2}} \ket{x+1} \otimes \ket{+} + \frac{1}{\sqrt{2}} \ket{x-1} \otimes \ket{-}$ with $\ket{\phi}$ the initial state of the probe. We shall however imagine all probes initially prepared in the same state, and we do not write it explicitly here.},
\begin{equation}\label{eq:basic}
 \ket{x} \rightarrow \frac{1}{\sqrt{2}} \ket{x+1} \otimes \ket{+} + \frac{1}{\sqrt{2}} \ket{x-1} \otimes \ket{-}
\end{equation}
We view it as a map from $\mathbb{C}^\mathbb{Z}$ to $\mathbb{C}^\mathbb{Z}\otimes \mathbb{C}^2$, preserving the normalization of the state. We then imagine iterating this process (which physically may be obtained as a result of iterated interactions between the walker and a series of probes). After $n$ iterations, we get a state in $\mathbb{C}^\mathbb{Z}\otimes \left( \mathbb{C}^2\right)^{\otimes n}$. To make this clearer, consider a walker starting at position $x_0$ and iterating trice the elementary evolution. The total system (walker + probes) state $\ket{\psi_3}$ is then :
\begin{equation*}
\begin{split}
 \ket{\psi_3} =& \hskip .45 truecm \frac{1}{\sqrt{2^3}}\, \ket{x_0+3} \otimes \ket{+++} \\
 &+\frac{1}{\sqrt{2^3}}\, \ket{x_0+1} \otimes \left(\ket{++-}+\ket{+-+}+\ket{-++}\right) \\
 &+\frac{1}{\sqrt{2^3}}\, \ket{x_0-1} \otimes \left(\ket{-+-}+\ket{--+}+\ket{+--}\right) \\
 &+\frac{1}{\sqrt{2^3}}\, \ket{x_0-3} \otimes \ket{---} .
\end{split}
\end{equation*}
On this example we see that the resulting entangled state is the sum of states which are tensor products of a walker state at a localized position with a probe state in $\left( \mathbb{C}^2\right)^{\otimes 3}$ in one-to-one correspondence with a walk of length $3$ ending at this position. This sum saturates all possible walks of length three. This structure will clearly persist after $n$ iterations, so that the states after $n$ iterations can written as follows: 
\begin{equation}\label{eq:parallel}
 \ket{\psi_n}=\frac{1}{2^{n/2}}\, \sum_{[\omega_n]} \ket{X_n([\omega_n])}\otimes\ket{[\omega_n]}
\end{equation}
where the sum is over all walks $[\omega_n]$ of length $n$, starting at $x_0$ and with position $X_j([\omega_n])$ at their $j^{\rm th}$ step, and $\ket{[\omega_n]}$ are orthogonal vectors associated to each walk. This linear rewriting is what is called ``quantum parallelism".

This can be done in a more abstract or axiomatic way forgetting about repeated interactions. Let us once again consider a walker on a discrete line described by its position $X_n$. We can then decide to add a Hilbert space where we store the history of the process, i.e. its full trajectory. If we call $\ket{[\omega_n]}$ a basis of states corresponding to a given trajectory, we can simply decompose it into a product of fictitious probes encoding for every single increment of the trajectory,
\[
\ket{[\omega_n]}:= \ket{\pm}\otimes...\otimes\ket{\pm},
\]
which then formally gives our previous picture.

The state (\ref{eq:parallel}) enables us to book-keep all possible walks in an algebraic way.  At this point there is no more information in this sum than in the classical description of random walks. It also codes the probability of realization of any given sample: the probability of occurrence of the walk $[\omega_n]$ is the modulus square of the coefficient in front of $\ket{X_n([\omega_n])}\otimes\ket{[\omega_n]}$. These probabilities sum to one thanks the normalization of $\ket{\psi_n}$, $\langle\psi_n\ket{\psi_n}=1$.
This can be viewed from a quantum measurement perspective. Imagine measuring the observable $\sigma^z$ on each of the probes\footnote{Pauli matrices are defined by $\sigma^z=\bigl(\begin{smallmatrix} 1& 0 \\ 0& -1\end{smallmatrix}\bigr)$, $\sigma^x=\bigl(\begin{smallmatrix} 0& 1 \\ 1& 0\end{smallmatrix}\bigr)$, and $\sigma^y=\bigl(\begin{smallmatrix} 0& -i \\ i& 0\end{smallmatrix}\bigr)$.}. The output of the measurement of one probe is one of the $\sigma^z$ eigen-values $\pm$. The output of the series of measurements of all probes is a sequence of pluses and minuses, which are in bijection with the classical trajectories $[\omega_n]$. By von Neumann rules for quantum measurement, the probability of getting a given sequence $[\omega_n]$ as results of these measurements on state $\ket{\psi_n}$ is the modulus square of the coefficient in front of $\ket{X_n([\omega_n])}\otimes\ket{[\omega_n]}$. That is, von Neumann measurement on probes induces the appropriate measure on random walks.

Up to now, all this seems very formal and it may seem that the quantum rewriting did not add anything. In this paper we will observe genuine quantum effects by considering only slight extensions of this model. We will first allow for probe measurements that are not aligned with the basis vectors. The measurement results will not be in one-to-one correspondence with walker trajectories anymore and the position will not be well defined either. We will call this simple generalization \emph{tilted random walks} and study them first without any further complication in the next sub-Section. 

This quantum rewriting is also useful to introduce some non Markovian effects. Adding new internal degrees of freedom, we will give a different weight to the trajectories and thus make the process path dependent. This is made possible thanks to linearity axiom of quantum mechanics. This will be the route to open quantum random walks and their continuous limit, open quantum Brownian motions. In this extension -- which may pompously be named ``non commutative'' --, we will still be able to measure the probes which will give quantum trajectories, or average over them which will give quantum dynamical maps, or keep them all to allow for any further measurement or manipulation which will give quantum stochastic processes.

\subsection{Tilted random walks and non-demolition momentum measurements}

Let us consider the simplest example of what we call tilted random walks. We consider the exact same model as before but allow for different probe measurements. We are going to show that the notion of trajectory is lost, as usual in quantum mechanics, but that surprisingly tilted random walks carry out progressive von Neumann measurements \cite{vonNeu} of the walker momentum.

We write $\ket{\psi_t}=\sum_x \psi_{x,t}\, \ket{x}$, with $\sum_x |\psi_{x,t}|^2=1$, the walker state after $t$ steps assuming that we have already measured the $t$ first probes\footnote{The state of the total system (walker + $t$-first probes) is actually of the form  $\ket{\psi}_t\otimes \ket{s_1\cdots s_t}$ where $s_j$ are the results of the measurement on the $j$-th probe. We do not write explicitly the state $\ket{s_1\cdots s_t}$ as it is frozen at later time and does not play a role anymore.}. We replace the infinite line by a circle so that $x\in \mathbb{Z}_N$ and impose periodic boundary conditions on $\psi_{x,t}$. By linearity of quantum mechanics, the evolution is:
\[
\begin{split}
\ket{\psi_t} \longrightarrow &\quad \frac{1}{\sqrt{2}}\sum_x \psi_{x,t}\, \big(\ket{x+1} \otimes \ket{+}+\ket{x-1} \otimes \ket{-}\big)\\
&=\frac{1}{\sqrt{2}} \sum_x \ket{x}\otimes \big(\psi_{x-1,t}\,\ket{+}+\psi_{x+1,t}\,\ket{-}\big) .
\end{split}
\]

We now assume that we measure the probe observable with normalized eigen-vectors:
\begin{eqnarray*}
\ket{+^u} &=& e^{+i\varphi/2} \cos\vartheta/2\, \ket{+} + e^{-i\varphi/2} \sin\vartheta/2\, \ket{-} ,\\
\ket{-^u} & =& e^{-i\varphi/2} \cos\vartheta/2\, \ket{-} - e^{+i\varphi/2} \sin\vartheta/2\, \ket{+} .
\end{eqnarray*}
This corresponds to measuring the observable $\sigma_u={\bf u}\cdot\sigma$, i.e. the probe effective spin along a tilted direction ${\bf u}$ oriented with angles $\vartheta$ and $\varphi$ on the sphere. The outputs of these measurements are still $\pm$. After the tilted measurement of the $t$-th probe, the walker state is up-dated by von Neumann rules to the normalized state proportional to $\sum_x \ket{x}\, \bra{\pm^u}\left(\psi_{x-1,t}\ket{+}+\psi_{x+1,t}\ket{-}\right)$, because the probe state has to be projected $\ket{\pm^u}$, so that
\[
 \ket{\psi_{t+1}} = \frac{1}{\sqrt{2 p_\pm(t)}}\, \sum_x  \big(\psi_{x-1,t}\,\braket{\pm^u}{+}+\psi_{x+1,t}\,\braket{\pm^u}{-}\big)\, \ket{x},
 \]
with probability (by von Neumann rules),
\[ p_\pm(t):=\frac{1}{2}\sum_x \Big| \psi_{x-1,t}\, \braket{\pm^u}{+}+\psi_{x+1,t}\, \braket{\pm^u}{-}\Big|^2.\] 
These probabilities are that of getting $\pm$ as result of the tilted measurement of the $t$-th probe. 
For $\vartheta=0$ or $\pi$, this reproduces standard random walks because then $\braket{\pm^u}{+}=1$ or $0$ and symmetrically $\braket{\pm^u}{-}=0$ or $1$.
  
Notice that, even if we initially started with a state localized at a given lattice position, after only one iteration the walker state is not localized anymore and will never be localized again (for $\vartheta\not= 0,\pi$). Hence we cannot talk about walker trajectories (although we simply started from random walks but quantum parallelized).

The system is translation invariant and this suggests to use discrete Fourier
transforms. Writing $N$ for the number of lattice sites, we use the convention that $\psi_{x,t}=\frac{1}{\sqrt{N}}\sum_k {\phi_{k,t}}\,e^{2i\pi kx/N}$, with $\sum_k |\phi_{k,t}|^2=1$, and $|\phi_{k,t}|^2$ is the momentum probability distribution at time $t$. We then get a simple evolution for the discrete wave function in momentum space,
\begin{equation} \label{eq:recurrw}
  \phi_{k,t}\to \phi_{k,t+1}= \frac{1}{\sqrt{2 p_\pm(t)}}\left[\braket{\pm^u}{+}\, e^{-2i \pi kx/N}+\braket{\pm^u}{-}\, e^{2i \pi kx/N}\right] \phi_{k,t}, 
\end{equation}
with probability $p_\pm(t)$, as defined above, which can written as,
\begin{equation*}
  p_\pm(t)= \half \Big(1 \pm \sin \vartheta \; \big[ \sum_l \cos (4\pi l/N+\varphi) \: |\phi_{l,t}|^2\,\big]\Big).
\end{equation*}

We now compute the expectation of $|\phi_{k,t+1}|^2$ conditional on the past up to time $t$, which we denote as $\mathbb{E}\left[ |\phi_{k,t+1}|^2 \;|\; {\cal F}_t \right] $. From eq.(\ref{eq:recurrw}
) we get:
\begin{equation} \label{eq:phi2}
 |\phi_{k,t+1}|^2 =\frac{1}{2p_\pm(t)} \left[1 \pm \sin \vartheta\, \cos\big({4\pi k}/{N}+\varphi\big) \right] \big|\phi_{k,t}\big|^2
\end{equation}
When computing the conditional expectation we have to sum over all possible values of $|\phi_{k,t+1}|^2$, conditioned on the values of $|\phi_{k,t}|^2$ weighted by their probabilities of occurrence (which are $p_\pm(t)$). Hence, these probabilities in the denominator and in the numerator cancel out, and we get:
\begin{eqnarray}\label{eq:phymartin}
\mathbb{E}\left[ |\phi_{k,t+1}|^2 \;|\; {\cal F}_t \right] =|\phi_{k,t}|^2.
\end{eqnarray}
The process (\ref{eq:recurrw}) is defined on the probability space consisting of all possible measurement outputs, and ${\cal F}_t$ is the natural filtration associated to the probe measurement results up to time $t$. Eq.(\ref{eq:phymartin}) then tells us that $t\to  |\phi_{k,t}|^2$ is a bounded martingale (it is bounded because $\sum_k  |\phi_{k,t}|^2=1$) \cite{probas}.

\begin{figure}
\centering
\includegraphics[scale=0.38]{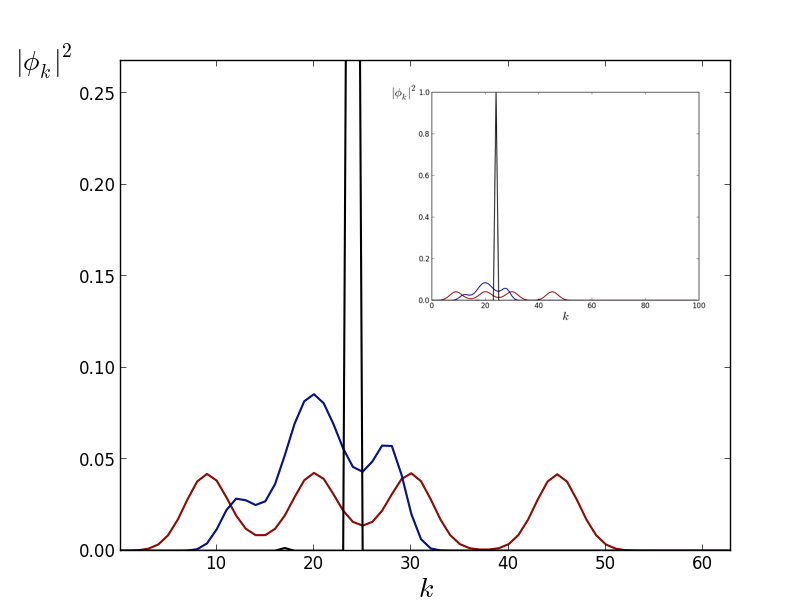}
\includegraphics[scale=0.38]{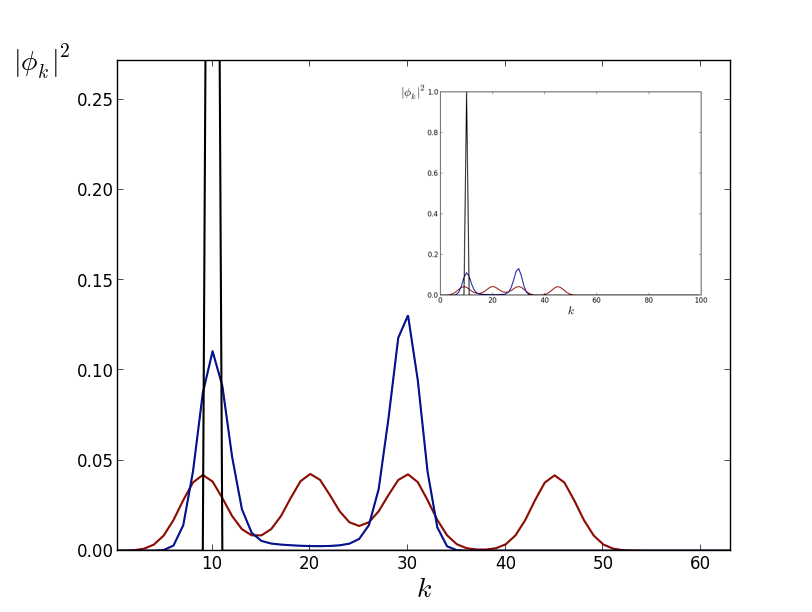}
\caption{\small Evolution of the momentum distribution $|\phi_{k,t}|^2$ in two samples of tilted random walk ($\varphi=\vartheta=0.6$ and $N=100$) starting with an identical but generic initial condition consisting in the sum of four Gaussian distributions (red curves). After $50$ iterations the intermediate distributions (blue curves) are more concentrated but still show a few peaks. The final distributions (black curves, computed after $5.10^4$ steps) are single Dirac peaks but randomly located.}
\label{fig:col}
\end{figure}

This result is remarkable because we then learn by the martingale convergence theorem \cite{probas} that $|\phi_{k,t}|^2$ converges almost surely at large time to a random value which we denote by $|\phi_{k,\infty}|^2$. See Fig.\ref{fig:col}. We now argue that\footnote{If initially $|\phi_{k,t=0}|^2$ is supported on an interval of length smaller than $N/2$.}, this limiting momentum distribution is peaked at a random target $k_\infty$:
\begin{eqnarray} \label{eq:moment}
|\phi_{k,\infty}|^2:= \lim_{t\to \infty} |\phi_{k,t}|^2 = \delta_{k;k_\infty} .
\end{eqnarray}
Indeed, since $p_\pm(\infty):= \sum_l q_\pm(l)\, |\phi_{l,\infty}|^2$, with
$q_\pm(l):=\half[1 \pm \sin \vartheta \cos (4\pi l/N+\varphi)] $, are non zero
(for $\vartheta\not= \pm\pi/2$), there exists a subsequence of times $t^+_j$
(resp. $t^-_j$) going to infinity such that the measurement outputs at times
$t^+_j$ (resp. $t^-_j$) are all $+$ (resp. $-$). Considering the large time
limit of eq.(\ref{eq:phi2}) for these subsequences yields the fixed point
conditions
\[ \Big(p_\pm(\infty) - q_\pm(k)\Big)\, |\phi_{k,\infty}|^2 =0.\] Since
$p_\pm(\infty)= \sum_l q_\pm(l)\, |\phi_{l,\infty}|^2$ with $q_\pm(l)$ all
different except at pairs of momenta differing by $\pm N/2$ (for
$\vartheta\not=0,\pi$ and $\varphi$ irrational), solutions\footnote{Since
  $q_\pm(k\pm N/2)=q_\pm(k)$ solutions of the fixed point condition can have
  support on pairs of momenta differing by $\pm N/2$. However, if the length of
  the support of the initial momentum distribution $|\phi_{k,t=0}|^2$ is smaller
  than $N/2$, then only one peak remains (because if $|\phi_{k,t=0}|^2$ vanishes
  initially it is null at any later time).} to this equation, with $\sum_k
|\phi_{k,\infty}|^2=1$, are peaked\footnote{In fact, our argument, which relies
  on the infinite number of apparitions of $+$ and $-$, is only of heuristic
  value.  The tricky point is that the outcomes are not independent random
  variables.  But they are exchangeable, and De Finetti's theorems (see e.g. the
  second reference in \cite{probas}) give a rigorous justification of the
  result.} at a random momentum $k_\infty$, as in eq.(\ref{eq:moment}).

Hence, for any tilted quantum trajectory the momentum distribution $|\phi_{k,t}|^2$ collapses almost surely to a Dirac mass at a random momentum $k_\infty$ which we may view as the result of a momentum measurement. The convergence is also in $L^1$, and because $|\phi_{k,t}|^2$ is a martingale, the probability of convergence towards a given momentum is the initial spectral momentum density at this point (because $\mathbb{E}[|\phi_{k,\infty}|^2]=|\phi_{k,t=0}|^2$ by the martingale property). That is, 
\begin{eqnarray} \label{eq:vNmeasure}
 \mathbb{P}[k_\infty=k]= |\phi_{k,t=0}|^2,
 \end{eqnarray}
 and this is in accordance with von Neumann rules for quantum measurement of the momentum. Thus tilted random walks provide realizations of progressive measurements of the walker momentum observable. This is similar to progressive non-demolition measurements analyzed in \cite{BB11,BBB12}.

\subsection{Outline}

The rest of the paper is devoted to the construction of open quantum Brownian motions (open QBM) starting from open quantum random walks (open QRW) by generalizing the quantum parallelism on random walks explained above. The main difference is that the walkers, still interacting with series of probes, now carry extra internal degrees of freedom which are going to behave as effective gyroscopes determining the direction of walker displacements. 

We are going to start from a discrete setting adapted to open QRW and then take the scaling limit which yields time and space continuous quantum processes, the open QBM. Next Section is devoted to open QRW. The four following ones deal with open QBM along different facets: quantum trajectories (when measuring probes), quantum dynamical maps (when keeping no track of the result of the probe measurements), and quantum stochastic differential equations (when neither tracing out the probe degrees of freedom nor measuring them). We refer to \cite{Qtrajec} for an introduction to quantum trajectories, to \cite{Qmap} for quantum dynamical maps, to \cite{Qnoise} for quantum stochastic differential equations, and to \cite{Qprocess} for general introductions to the theory of open quantum system and quantum stochastic processes. From some perspectives, the quantum stochastic differential equations (quantum SDE), or quantum stochastic noises, encompass the other facets because the latter are reduction of the former. But 
this is not the complete picture, as we are going to illustrate, because the quantum SDE can be reconstructed from the quantum trajectories or the quantum dynamical maps with tiny input hypothesis. Since it is more natural from both quantum mechanical and probability points of view, we choose to order the presentation by starting with quantum trajectory and quantum dynamical map and then quantum SDE. The last Section is devoted to the particular case in which the internal gyroscope is a two level system. Even in this most simple case, open QBM possesses a rich structure with an interplay between ballistic and diffusive behaviors.

Besides constructing the open QBM, one of the other aims of this manuscript is to present from a biased theoretical physics point of view (with a mixture of quantum mechanics, probability theory and field theory) different aspects of quantum stochastic processes and their classical reductions. Some of the results we present in a context adapted to open QBM may be transposed to other quantum stochastic processes. Some of the relations we develop are known in different scientific communities -- and some are new -- but we adopt (or we believe that we adopt) a different point of view closer to field theory, and as a consequence our proofs are different (and, we hope, shed extra light on the results).

We choose to present the results or the claims at the beginning of each sub-Section and then a sketch of a proof, alias ``esquisse d'une preuve" \footnote{Of course inspired by a well known mathematical essay\cite{Grothen}. We choose this name for the way it nicely sounds and not because we claim to approach its historical strength in any way.}, in each sub-Section. We hope that this presentation clarifies the concepts or the tools involved and makes more transparent the relations between the various approaches. The proofs we present may not always contain all the details necessary for claiming that they fully belong to the rigorous mathematical world, but they have some, and they are certainly proofs in the common sense of theoretical physics (and also maybe from some perspectives of mathematical physics).

\section{Open quantum random walks}

Although the main purpose of the present manuscript is to describe what the continuous limit of open quantum random walks (open QRW) is, that is what open quantum Brownian motions (open QBM) are, we start by recalling what discrete open QRW are, at least in order to fix the framework and the notations. Open QRW are generalizations of the quantum parallelism on random walks explained above but with walkers carrying extra internal degrees of freedom and interacting with series of probes. The quantum dynamical map  for open QRW was first introduced in \cite{Attal_etal}. A slightly different version (implementing a specific dilation of these maps) was introduced in \cite{BBT13}, in which some of the results presented in the following were announced.

\subsection{Discrete model}

Open quantum random walks on the line are defined on the total Hilbert space ${\cal H}_c\otimes{\cal H}_z\otimes {\cal H}_p^{\otimes \infty}$ with ${\cal H}_c$ the finite dimensional internal space (the ``gyroscope" Hilbert space), ${\cal H}_z$ the orbital Hilbert space (the ``walker" Hilbert space), and ${\cal H}_p= \mathbb{C}^2$ (the ``probe" Hilbert space). We take ${\cal H}_z=\mathbb{C}^{\mathbb{Z}}$ in the discrete setting and ${\cal H}_z=L^2(\mathbb{R})$ in the continuous limit. We denote by ${\cal H}_{\rm sys}:={\cal H}_c\otimes{\cal H}_z$ the Hilbert space of the {\it system} (walker + gyroscope) formed by the orbital and internal degrees of freedom (d.o.f.'s). We call the {\it total system} that formed by the (walker + gyroscope + probes).
For a while we assume that all probes are prepared in the same identical pure state $\ket{\phi}_p$. But this has to be -- and will be in a following Section -- generalized to density matrix $\rho_p$. We specify a basis of probe states $\ket{\pm}_p$ in ${\cal H}_p$ made of eigen-states of $\sigma^z$, with eigen-values $\pm$\footnote{We denote by $\sigma^{x,y,z}$ the standard Pauli matrices acting on ${\cal H}_p$.}. We also introduce the basis of ortho-normalized states $\ket{x}_z$ in the orbital space ${\cal H}_z$ localized at point $x$ on the line.

Let $U$ be the unitary operator acting on ${\cal H}_{\rm sys}\otimes{\cal H}_p$ coding for the system-probe interaction. We demand that its action on states $ \ket{\chi}_c\otimes\ket{x}_z\otimes\ket{\phi}_p$ gives the entangled normalized states
\begin{eqnarray} \label{eq:defU}
\boxed{U\ket{\chi}_c\otimes\ket{x}_z\otimes\ket{\phi}_p:=(B_+\ket{\chi}_c)\otimes\ket{x+\delta}_z\otimes\ket{+}_p +(B_-\ket{\chi}_c)\otimes\ket{x-\delta}_z\otimes\ket{-}_p,}
\end{eqnarray}
for any $\ket{\chi}_c\in{\cal H}_c$, with $\delta$ the lattice spacing. Of course $B_\pm$ depends linearly on $\ket{\phi}_p$.
Unitarity imposes  $B_+^\dag B_++ B_-^\dag B_-=\mathbb{I}$. 

Open quantum random walks are obtained by iterating system-probe interaction but with a new, identically prepared, probe at each step. If no measurement is done on the probes we get the quantum noise formulation, if measurements are done on the probes we get quantum trajectories. Quantum dynamical maps are obtained when tracing over the probes d.o.f.'s.

\underline{\it Quantum noise.}
By linearity, if the walker is in state $\ket{\psi}_z:=\sum_x \psi(x)\, \ket{x}_z$, then 
\[ U\ket{\chi}_c\otimes\ket{\psi}_z\otimes\ket{\phi}_p= (B_+\ket{\chi}_c)\otimes\ket{\psi_+}_z\otimes\ket{+}_p +(B_-\ket{\chi}_c)\otimes\ket{\psi_-}_z\otimes\ket{-}_p,\]
with $\ket{\psi_\pm}_z$ the state obtained from $\ket{\psi}_z$ by translation by one step either to the left or to the right, i.e. $\ket{\psi_\pm}_z=\sum_x \psi(x\mp\delta)\, \ket{x}_z$.
Alternatively, 
\begin{eqnarray}\label{eq:defUbis}
\boxed{ U\ket{\phi}_p= (B_+\otimes e^{-i\delta P})\otimes\ket{+}_p + (B_-\otimes e^{+i\delta P})\otimes\ket{-}_p,}
 \end{eqnarray} 
with $P:=-i\partial_x$ the translation operator, such that $e^{-i\delta P}\ket{x}_z=\ket{x+\delta}_z$, or wording differently, 
\begin{eqnarray}\label{eq:defU+-}
 U_\pm:= {}_p\bra{\pm}U\ket{\phi}_p=B_\pm\otimes e^{\mp i\delta P},
 \end{eqnarray}
as an operator equation in ${\cal H}_{\rm sys}={\cal H}_c\otimes{\cal H}_z$.
Quantum  random walks are obtained by iterating this interaction with a new probe at each step. Keeping track of all probes, the total Hilbert space is ${\cal H}_{\rm sys}\otimes {\cal H}_p^\infty$, and at the $n$-th step the unitary transformation acts on ${\cal H}_{\rm sys}$ and on the $n$-th copy of ${\cal H}_p$, and we denote it by $U_{0n}$. If the initial state of the system is $\ket{\chi}_c\otimes\ket{x_0}_z$, after $n$ steps the state of the total system is
\[ \Big(U_{0n}\cdots U_{02}\, U_{01}\, \ket{\chi}_c\otimes\ket{x_0}_z\otimes\ket{\phi}_p^{\otimes n}\Big)\,\otimes\ket{\phi}_p\otimes\ket{\phi}_p\otimes\cdots,\]
and only the $n$-th first probes have been affected by the interaction. Expanding the result of the action of the unitary operators on the probe states $\ket{\pm}_p$ yields a sum over states which may be indexed by random walks of $n$-step  $[\omega_n]: =(\varepsilon_1,\cdots,\varepsilon_n)$, $\varepsilon_j=\pm$. Namely, 
\begin{eqnarray}\label{eq:discreteNoise}
\Big(\sum_{[\omega_n]} B([\omega_n])\ket{\chi}_c\otimes\ket{X_n([\omega_n])}_z\otimes\ket{[\omega_n]}_p\Big)\,\otimes\ket{\phi}_p\otimes\ket{\phi}_p\otimes\cdots, 
 \end{eqnarray}
where the sum is over all random walks $[\omega_n]$ of length $n$, starting at $x_0$ and with position $X_j([\omega_n])$ at their $j^{\rm th}$ step, and $\ket{[\omega_n]}:=\ket{\varepsilon_1}\otimes\cdots\otimes\ket{\varepsilon_n}$. Here $B([\omega_n])$ is the ordered product of transition matrices $B_\pm$ encountered along the walk.

The evolution (\ref{eq:defUbis}) can of course be written not only for pure states but for density matrices. If $\hat\rho_0$ is the system (walker $+$ gyroscope) initial density matrix and if all probes are prepared identically in the state $\ket{\phi}_p$, after $n$ steps the system and the $n$ first probes are entangled, and the density matrix of the total system is of the form $\rho^{\rm tot}_n\otimes \ket{\phi}_p\bra{\phi}\otimes\cdots$. At the next step, the evolution is
\[ \rho^{\rm tot}_{n+1} = U_{0;n+1}\, \big(\rho^{\rm tot}_n\otimes \ket{\phi}_p\bra{\phi}\big)\, U_{0;n+1}^\dag,\]
and all probes after the $(n+1)$-th remain untouched.

\underline{\it Quantum dynamical map.}
The quantum dynamical map \cite{Qmap,Qprocess} for open quantum random walks is obtained by tracing over the outgoing probes. It induces an evolution for the system reduced density matrix 
\begin{eqnarray}\label{eq:Oqrw_map}
 \boxed{ \bar\rho_{n} \to \bar\rho_{n+1}=(B_+\otimes e^{-i\delta P})\, \bar\rho_n\, (B_+^\dag\otimes e^{+i\delta P})+(B_-\otimes e^{+i\delta P})\,\bar\rho_n\, (B_-^\dag\otimes e^{-i\delta P}). }
 \end{eqnarray}
Let us decompose $\bar \rho_n$ as $\bar\rho_n=\sum_{x,y}\, \bar\rho_n(x,y)\otimes \ket{x}_z\bra{y}$ with $\bar\rho_n(x,y)$ a matrix acting on the internal Hilbert space ${\cal H}_c$. Then eq.(\ref{eq:Oqrw_map}) reads
\begin{eqnarray}\label{eq:Oqrw_bis}
 \bar\rho_{n+1}(x,y)= B_+\, \bar\rho_n(x-\delta,y-\delta)\, B_+^\dag + B_-\, \bar\rho_n(x+\delta,y+\delta)\, B_-^\dag.
\end{eqnarray}
This preserves diagonal density matrices, that is, if the system density matrix is initially diagonal and localized in the orbital space,  $\rho_0(x,y)=\bar\rho_0(x)\, \delta(x-y)$, so is it at each step, $\bar\rho_n(x,y)=\bar\rho_n(x)\, \delta(x-y)$.

As suggested by the notation, this map may also be obtained by averaging over quantum trajectories, defined just below, i.e. $\bar\rho_n= \mathbb{E}(\hat \rho_n)$. Of course the two approaches are equivalent because the measure (in the sense of probability theory) on the output probe measurements, and hence on the quantum trajectories, is induced by the probe states. 

\underline{\it Simple quantum trajectory.}
Quantum trajectories \cite{Qtrajec,Qprocess} emerge when measuring the probe after each step. We choose to call {\it simple quantum trajectories} those arising when measuring the probe observable with basis $\ket{\pm}_p$. Let $\hat \rho_n$ be the density matrix of the system (walker + gyroscope) after the $n$-th step of a quantum trajectory. Since the output of the probe measurements are random -- with a probability measure induced by quantum mechanical rules for measurement --,  quantum trajectories are random. In the present case, they are defined by random up-datings 
\begin{equation} \label{eq:qtraj}
\boxed{ \hat \rho_n\to \hat \rho_{n+1}:=\frac{(B_\pm\otimes e^{\mp i\delta P})\, \hat \rho_n\, (B_\pm^\dag\otimes e^{\pm i\delta P})}{p_\pm(n)} ,}
\end{equation}
with probability
\[ p_\pm(n):={\rm Tr}_{{\cal H}_{\rm sys}}\big( (B_\pm\otimes e^{\mp i\delta P}) \hat \rho_n (B_\pm^\dag\otimes e^{\pm i\delta P}) \big). \]
This may also be seen as the evolution of the system density matrix on repeated POVM\footnote{POVM = Positive Operator Valued Measure.} (with two possible outputs $\pm$).

Eq.(\ref{eq:qtraj}) preserves orbital diagonality: if the system density matrix is initially diagonal and localized in the orbital space it remains so at each step, and we may write 
\[ \hat \rho_n = \rho_n\otimes \ket{X_n}_z\bra{X_n},\]
with $\rho_n$ a density matrix on the internal Hilbert space. The random evolution is then
\begin{eqnarray}\label{eq:diag_traj}
\rho_n\to \rho_{n+1}:=\frac{B_\pm\, \rho_n\, B_\pm^\dag}{p_\pm(n)},\quad X_n\to X_{n+1}:=X_n\pm \delta,
\end{eqnarray}
with probability $p_\pm(n):={\rm Tr}_{{\cal H}_c}\big( B_\pm \rho_n B_\pm^\dag\big)$. Notice that in such case, the walker position $X_n$ is slave to the random measurement outputs. We can thus choose to measure either the series of probes or the walker position, an equivalence which is lost for tilted trajectories. 

Simple quantum trajectories are defined on the probability space whose events are the series $(s_1,s_2,\cdots)$ with $s_k=\pm$ depending whether the $k$-th out-going probe is measured in the state $\ket{\pm}_p$. Functions which depend only on the $n$ first data $(s_1,\cdots,s_n)$ -- and leave unspecified the following data $s_j$, $j\geq n+1$, --  are those measurable with respect to the natural filtration ${\cal F}_n$ \cite{probas}, and 
\[ \mathbb{E}[\mathbb{I}_{\{s_{n+1}=\pm\}}|{\cal F}_n]:= p_\pm(n),\]
 the probabilities for $s_{n+1}=\pm$ conditioned on the value of the internal state at the $n$-th step. 

\underline{\it Tilted quantum trajectory.} 
Quantum trajectories depend on which probe observables are measured at each step. The above {\it simple} quantum trajectories correspond to measure $\sigma^z$, but other choices can be done. What we call {\it tilted quantum trajectories} correspond to measure the probe effective spin but in a tilted direction ${\bf u}$, that is, they correspond to measure $\sigma_u:={\bf u}\cdot\sigma$, with ${\bf u}$ a unit vector ${\bf u}\cdot{\bf u}=1$, at each step. 

Let $\ket{\pm^u}$ be the normalized eigen-vectors of $\sigma_u$, and let us parametrize them as above:
\[ \ket{\pm^u} = e^{\pm i\varphi/2} \cos\vartheta/2\, \ket{\pm} \pm e^{\mp i\varphi/2} \sin\vartheta/2\, \ket{\mp} .\]

If $\hat \rho_n$ is the system density matrix after the $n$-th step, and if all probes are prepared in the pure state $\ket{\phi}_p$ as before, the system-probe interaction induces the unitary evolution $\hat\rho_n \otimes\ket{\phi}_p\bra{\phi} \to U\, \hat\rho_n \otimes \ket{\phi}_p\bra{\phi} \, U^\dag$ with $U$ defined in eq.(\ref{eq:defUbis}).
If now a measurement of $\sigma_u$ is performed on the last probe, with output $\pm$, this induces a projection of the system density matrix,
\begin{eqnarray} \label{eq:defTilted}
\hat \rho_n \to \hat\rho_{n+1} = \frac{ U^u_\pm\, \hat\rho_n\, U^{u\, \dag}_\pm}{ p_\pm^{u}(n)}
\end{eqnarray}
with $p_\pm^{u}(n):=\mathrm{Tr}_{{\cal H}_{\rm sys}}(U^u_\pm\, \hat\rho_n\, U^{u\, \dag}_\pm)$ and 
\[ U^u_\pm := (B_+\otimes e^{-i\delta P})\, \langle{\pm^u}\ket{+} + (B_-\otimes e^{+i\delta P})\, \langle{\pm^u}\ket{-}.\]
These operators form a POVM again since
\[ U^{u\, \dag}_+\,U^u_+ + U^{u\, \dag}_-\,U^u_- = \mathbb{I}.\]
Notice that eq.(\ref{eq:defTilted}) does not preserve orbital diagonality, that is, even if the initial system density matrix is diagonal in orbital space  -- so that one can talk about the initial position of the walker -- it becomes non diagonal after a few iterations -- and one may talk about the walker position only once a measurement of the position operator has been performed but this would change the rest of the evolution.

Eq.(\ref{eq:defTilted}) looks simpler in momentum space. Let $\breve{\rho}_n(p,q):= {}_z\bra{p}\hat\rho_n\ket{q}_z$ with $\ket{q}_z$ eigen-states of the momentum operator, $P\ket{q}_z=q\ket{q}_z$. It  acts on ${\cal H}_c$ and is normalized to $\int \hskip -0.1 truecm dp\, \mathrm{Tr}_{{\cal H}_c}( \breve{\rho}_n(p,p))=1$. Then
\begin{eqnarray} \label{eq:rhopq}
\boxed{ \breve{\rho}_{n}(p,q)\to \breve{\rho}_{n+1}(p,q)= \frac{U^u_\pm(p)\, \breve{\rho}_n(p,q)\, U^{u}_\pm(q)^\dag}{ p_\pm^{u}(n)}, }
 \end{eqnarray}
with
\[ U^u_\pm(p):= e^{-i\delta p} \langle{\pm^u}\ket{+}\, B_+ +  e^{+i\delta p}\, \langle{\pm^u}\ket{-}\, B_-,\]
and $p_\pm^{u}(n)=\int dp\, \mathrm{Tr}_{{\cal H}_c}\big( \breve{\rho}_n(p,p)\, U^{u}_\pm(p)^\dag U^u_\pm(p)\big)$ the probability to observe $\pm$ as output for the $\sigma_u$ probe measurement.

Tilted quantum trajectories are  defined in a filtered probability space isomorphic to that for simple quantum trajectories, with the filtration ${\cal F}_n$ naturally associated to the output data up to step $n$ included. We still denote by $s_j=\pm$ the output data on the tilted probe measurements.

It is a simple exercise to check, using $U^{u\, \dag}_+\,U^u_+ + U^{u\, \dag}_-\,U^u_- = \mathbb{I}$, that $n\to \mathrm{Tr}_{{\cal H}_c}\big( \breve{\rho}_n(p,p) \big)$ are martingales, that is
\[ \mathbb{E}\big[  \mathrm{Tr}_{{\cal H}_c}( \breve{\rho}_{n+1}(p,p))\vert {\cal F}_n \big] =\mathrm{Tr}_{{\cal H}_c}\big( \breve{\rho}_n(p,p) \big).\]
Recall that $\mathrm{Tr}_{{\cal H}_c}\big( \breve{\rho}_n(p,p) \big)$ is probability density to get $p$ as output of a measurement of the walker impulsion after $n$ step. This is remarkable because computing the average amounts to trace over the probe degrees of freedom, and hence the momentum probability density $\mathrm{Tr}_{{\cal H}_c}\big( \breve{\rho}_n(p,p) \big)$ is conserved by the evolution if no information is extracted from the probes.

\subsection{Scaling limit}

The scaling limit is the limit in which the number of steps goes to infinity but with both the lattice spacing $\delta$ and the time step duration $\epsilon$ going to zero in such a way that $t=n\epsilon$ is fixed and $\epsilon=\delta^2$. For this limit to exist the transition matrices have to admit the following Taylor expansion in $\sqrt{\epsilon}$:
\begin{eqnarray} \label{eq:expanB}
\boxed{ B_\pm=\frac{1}{\sqrt{2}}\,[\mathbb{I}\pm \sqrt{\epsilon} N - \epsilon(iH\pm M +\frac{1}{2}N^\dag N) + O(\epsilon^{3/2})], }
 \end{eqnarray} 
with $H$ hermitian but not necessarily $N$ and $M$. This is the most general expansion for $B_\pm$ solutions of the unitary constraint $B_+^\dag B_+ + B_-^\dag B_- = \mathbb{I}$ around the symmetric solution $B_\pm=\mathbb{I}/\sqrt{2}$ such that the scaling limit exists. The solutions $B_\pm=\mathbb{I}/\sqrt{2}$ corresponds to standard walks.

Indeed, the general solution to $B_+^\dag B_+ + B_-^\dag B_- = \mathbb{I}$ in the neighborhood of $B_\pm=\mathbb{I}/\sqrt{2}$ is
\[ B_\pm = \frac{1}{\sqrt{2}}\,[\mathbb{I}\pm \sqrt{\epsilon} N_\pm - \epsilon(\pm M_\pm +\frac{1}{2}N_\pm^\dag N_\pm) + O(\epsilon^{3/2})],\]
with $\Re\, (N_+)=\Re\, (N_-)$ and $\Re\, (M_+)=\Re\, (M_-)$ (here $2\,\Re\, (N)=N+N^\dag$). Not all these solutions lead to a consistent scaling limit. Existence of the scaling limit requires $N_+=N_-$ and this limit only depends on the difference $M_+-M_-$. Since $M_\pm$ have identical hermitian part, only the difference of their anti-hermitian components matters in the scaling limit, i.e. $M_\pm=K\pm iH_\pm$ with $K$ and $H_\pm$ hermitian and the scaling limit depends only on $H:=(H_++H_-)/2$. This explains the form we choose in eq.(\ref{eq:expanB}).

Scaling limit of open QRW, that is open QBM, depends on $H$ and $N$, and thus it depends on $3\,\mathfrak{M}^2$ real parameters with $\mathfrak{M}$ the dimension of the internal Hilbert space ${\cal H}_c$, because $H$ is hermitian ($\mathfrak{M}^2$ parameters) and $N$ arbitrary  (2\,$\mathfrak{M}^2$ parameters). The gauge transformations consisting in conjugating the density matrix by a unitary transformations ($\mathfrak{M}^2$ parameters), and that consisting in translating $H$ by a multiple of the identity ($1$ parameter), reduce this number $2\mathfrak{M}^2-1$. This is slightly smaller, by $(\mathfrak{M}-1)$, than the dimension of the space of solutions to the constraint up to gauge transformations. Indeed, writing $B_\pm=\hat V_\pm\, K_\pm$ with $K_\pm$ hermitian and $\hat V_\pm$ unitary (such decomposition is always possible) transforms the constraint $B_+^\dag B_+ + B_-^\dag B_- = \mathbb{I}$ into $K^2_+ + K^2_- =\mathbb{I}$. So, $K_\pm$ can be diagonalised simultaneously and we can write $B_\pm=V_\pm D_\pm W$ with $W,\, V_\pm$ unitary and $D_\pm$ diagonal such that $D^2_+ + D^2_-=\mathbb{I}$. This involves $3\, \mathfrak{M}^2+\mathfrak{M}$ parameters. We again have the gauge freedom consisting in conjugating by a unitary transformation ($\mathfrak{M}^2$ parameters) and multiplying $B_\pm$ by independent uni-modular numbers ($2$ parameters). These yields $2\, \mathfrak{M}^2+\mathfrak{M}-2$ parameters. 
%Up to gauge transformations we can write $B_\pm=V_\pm D_\pm$ with $V_\pm$ unitary and $D_\pm$ diagonal such that $D_+^2+D_-^2=\mathbb{I}$

\section{Open QBM: Quantum trajectories}

We shall derive the stochastic differential equations (SDEs) for the open QBM as the continuous time limit of those for open QRW.  In this Section we first describe those equations for quantum trajectories, that is, the trajectories of the system (walker+gyroscope) density matrix induced by measurements of probe observables at each time step. These are random because output of measurements are random in quantum mechanics.

\subsection{Open QBM simple quantum trajectories}
Simple quantum trajectories are those induced by measuring the probe observable $\sigma^z$ at each step.
For simplicity, we first restrict to a diagonal system density matrix localized in the orbital space (but more general case will be deal with in the case of tilted quantum trajectories), so that $\hat\rho_n=\rho_n\otimes \ket{X_n}_z\bra{X_n}$ in the discrete setting. As was announced in \cite{BBT13}, we claim that in the continuum limit this converges in law to the density matrix $\hat \rho_t$,
\[\hat \rho_t= \rho_t\otimes \ket{X_t}_z\bra{X_t},\]
solution of the coupled stochastic differential equations
\begin{eqnarray}\label{eq:drho}
\boxed{\begin{array}{rcl}
d\rho_t&=&\big(- i[H,\rho_t]+L_N(\rho_t)\big)dt + D_N(\rho_t)\, dB_t, \\ \\
dX_t&=&U_N(\rho_t)\,dt + \, dB_t,
\end{array} }
\end{eqnarray}
with $B_t$ a normalized Brownian motion, and Lindbladian $L_N$ and operator valued diffusion coefficient $D_N$,
\begin{eqnarray}\label{eq:defLDU}
L_N(\rho) &:=&N\rho N^\dag - \half(N^\dag N\rho+ \rho N^\dag N),\nonumber\\
D_N(\rho) &:=&N\rho+\rho N^\dag -\rho\,U_N(\rho),\\
U_N(\rho) &:=&{\rm Tr}_{{\cal H}_c}(N\rho+\rho N^\dag).\nonumber
\end{eqnarray}
These equations are independent of $M$.
Not surprisingly, eqs.(\ref{eq:drho}) are those for continuous time quantum measurement, either named Belavkin's type \cite{belavkin} or master equations \cite{barchielli}. The evolution equation for the internal density matrix is independent of that for the walker position, and the drift term in the walker evolution equation is slave to the internal state. 
%The relative coefficient in the equation for $\rho_t$ and for $X_t$ are important for consistency.

The data one gets by recursively measuring the probe observables $\sigma^z$ is a series of $\pm$. These variables are exchangeable because the observables $\sigma^z$ acting on two different probes commute, and all behaviors are coded into the frequencies of appearance $N_n(\pm)$ of the output $\pm$ after $n$ steps. These are defined by $N_n(\pm) = \sum_{j=1}^n \mathbb{I}_{\{s_j=\pm\}}$ with $\mathbb{I}_{\{s_{j}=\pm\}}$ the characteristic functions on set of events such that $s_{j}=\pm$. Recall that $s_j$ denotes the value of the output measurement at step $j$. In the scaling limit, $N_n(\pm)\simeq n/2+\cdots$, by the law of large number. The sub-leading term in $N_n(+)$ is opposite to that in $N_n(-)$ because $N_n(+)+N_n(-)=n$ and their difference is $\delta^{-1}\,(X_n-X_0)$. Hence,
\[ \epsilon^{1/2}\,\big(2N_n(\pm)- {n}\big)= \pm(X_{t=n\epsilon}-X_0 )+ \cdots,\]
in the scaling limit $n\to\infty$ with $t=n\epsilon$ fixed. This expresses the fact that the walker position is slave to the orbital moves in the case of simple trajectories.

This can simply be generalized with an initial density matrix, still diagonal on the orbital space, but centered at $P$ different positions, that is, $\hat \rho_0=\sum_{j=1}^P\rho_0^{(j)}\otimes \ket{X_0^{(j)}}_z\bra{X_0^{(j)}}$. In that case the system density matrix is going to remain of this form at all time, so that $\hat \rho_t=\sum_{j=1}^P\rho_t^{(j)}\otimes \ket{X_t^{(j)}}_z\bra{X_t^{(j)}}$ describes $P$-coupled walker trajectories. These $P$ trajectories $X_t^{(j)}$, for $j=1,\cdots, P$, are parallel, sample by sample, but their probabilities of occurrence entangle them. This can of course be further generalized to non-diagonal initial density matrices (but we leave to the reader the writing of the quantum trajectory equations in this case).

\medskip

\underline{\it ``Esquisse d'une preuve":}\\
Recall that quantum trajectories are defined on the measurable space formed by all sequences of probe measurement outputs, with probability measure induced by that of quantum mechanics and filtration ${\cal F}_n$ specified by providing the data   of the first $n$-the measurement outputs. Although it is not fully mathematically rigorous\footnote{Because it implicitly assumes the existence of the scaling limit with the property that the continuous process interpolates the discrete one such that $\rho_n=_\mathrm{in\ law} \rho_{(t=n\epsilon)}$ and $X_n=_\mathrm{in\ law}X_{(t=n\epsilon)}$. This method is slightly different from that of ref.\cite{Clement}.},
a simple way to obtain the scaling limit consists in decomposing the process $\rho_n$ as a sum of a martingale $M_n$ plus a predictable process $O_{n}$. This is called a Doob decomposition \cite{probas}. It consists in writing
\[ \rho_n = O_n + M_n,\]
with $M_n$ a ${\cal F}_n$-measurable martingale -- i.e. $\mathbb{E}[M_n|{\cal F}_{n-1}]=M_{n-1}$ --  and $O_n$ a ${\cal F}_{n-1}$-measurable process. This decomposition is always possible because it is enough to define $M_n=\sum_{k=1}^n\pi_k$ with $\pi_k:=\rho_k-\mathbb{E}[\rho_k|{\cal F}_{k-1}]$, which by construction is a martingale, and to set $O_{n}:=\rho_n-M_n$, which by construction is ${\cal F}_{n-1}$-measurable. In the scaling limit the martingale (resp. predictable) contribution is going to converge to the noisy source (resp. the drift) of the SDEs. 

Eqs.(\ref{eq:diag_traj}) coding for the quantum trajectory evolution may be tautologically written as
\begin{eqnarray*}
 \rho_{n+1}&=&\rho_n^{(+)}\, \mathbb{I}_{\{s_{n+1}=+\}} + \rho_n^{(-)}\, \mathbb{I}_{\{s_{n+1}=-\}} ,\\
X_{n+1}&=&X_n+\delta\big(\mathbb{I}_{\{s_{n+1}=+\}} -\mathbb{I}_{\{s_{n+1}=-\}}\big).
\end{eqnarray*}
with 
\[ \rho_n^{(\pm)}:=\frac{B_\pm\, \rho_n\, B_\pm^\dag}{p_\pm(n)},\]
 with $p_\pm(n):={\rm Tr}_{{\cal H}_c}\big( B_\pm \rho_n B_\pm^\dag\big)$, and $\mathbb{I}_{\{s_{n+1}=\pm\}}$ the characteristic functions on set of events such that $s_{n+1}=\pm$.
Using $\mathbb{I}_{\{s_{n+1}=+\}} + \mathbb{I}_{\{s_{n+1}=-\}}=1$ and  $X_{n+1}-X_n=\delta(\mathbb{I}_{\{s_{n+1}=+\}} -\mathbb{I}_{\{s_{n+1}=-\}})$, we may then write
\[ \rho_{n+1}-\rho_{n}= \frac{1}{2}\big( \rho_n^{(+)}+ \rho_n^{(-)}-2\rho_n\big) +\frac{1}{2\delta}\big( \rho_n^{(+)}- \rho_n^{(-)}\big) (X_{n+1}-X_n).\]

Taylor expanding (with the above implicit assumption) the finite difference $\rho_{n+1}-\rho_n$ as $\epsilon\to0$, with $t=n\epsilon$ fixed, yields $\rho_{n}^{(+)}-\rho_{n}^{(-)}=2\sqrt{\epsilon}\,D_N(\rho_n)+\cdots$ and 
\[ \rho_{n+1}-\rho_{n}= \epsilon\, \Big[ -i[H,\rho_n] + L_N(\rho_n) - D_N(\rho_n)\, U_N(\rho_n)\Big] +  D_N(\rho_n)\, (X_{n+1}-X_n),\]
for $\delta=\sqrt{\epsilon}$ and with $L_N\,,D_N\,, U_N$ defined in eq.(\ref{eq:defLDU}). This gives
\[ d\rho_t = \Big[ -i[H,\rho_n] + L_N(\rho_n) - D_N(\rho_n)\, U_N(\rho_n)\Big]\, dt + D_N(\rho_n)\, dX_t,\]
for $d\rho_t:=\rho_{n+1}-\rho_n$ and $dX_t:= X_{n+1}-X_n$ as $\epsilon\to 0$.

On the other hand, since $\mathbb{E}[\mathbb{I}_{\{s_{n+1}=\pm\}}|{\cal F}_{n}]=p_\pm(n)$, the Doob martingale $M_n$, such that $M_{n+1}-M_n =  \rho_{n+1}-\mathbb{E}[\rho_{n+1}|{\cal F}_{n}]$, is 
\[ M_{n+1}-M_n := \frac{1}{2}
\big(\rho_{n}^{(+)}-\rho_{n}^{(-)}\big)\Big(\mathbb{I}_{\{s_{n+1}=+\}} -p_+(n)+p_-(n) -\mathbb{I}_{\{s_{n+1}=-\}}\Big).\]
By construction $\mathbb{E}[ M_{n+1}-M_n|{\cal F}_{n}]=0$ since $M_n$ is a martingale, and $\mathbb{E}[(M_{n+1}-M_n)^2|{\cal F}_{n}]= \epsilon\,D_N(\rho_n)^2+\cdots $, again because $\rho_{n}^{(+)}-\rho_{n}^{(-)}=2\sqrt{\epsilon}\,D_N(\rho_n)+\cdots$. As a consequence, $M_n$  converges to the Brownian martingale,
\[ M_t=\int_0^t D_N(\rho_s)\, dB_s,\]
with $B_t$ a normalized Brownian motion, $dB_t^2=dt$. 

Finally, because $p_+(n)-p_-(n)=\sqrt{\epsilon}\, U_n(\rho_n)+\cdots$,  the scaling limit of the previous formula for the difference $M_{n+1}-M_n$ yields
\[ dM_t = D_N(\rho_t)\,\big( dX_t - U_N(\rho_t)\, dt\big),\]
with $dM_t=M_{n+1}-M_n$ and $dX_t$ as above. Hence, $dX_t - U_N(\rho_t)\, dt =dB_t$, and together with the previous equation relating $d\rho_t$ and $dX_t$, it proves eq.(\ref{eq:drho}) for simple quantum trajectories. 
\cqfd

\subsection{Open QBM tilted quantum trajectories}

We now look at the scaling limit of tilted quantum trajectories, still using the parametrization (\ref{eq:expanB}) for the transition matrices $B_\pm$. At $\epsilon=\delta^2=0$, there is no system-probe interaction, $B_\pm=\mathbb{I}/\sqrt{2}$, and the state of the system is unchanged. However, that of the probe is transformed into $(\ket{+}+\ket{-})/\sqrt{2}$. The probability to get $\pm$ as result of  a measurement of $\sigma_u={\bf u}\cdot\sigma$, with eigen-vectors $\ket{\pm^u}$, is then (at $\epsilon=0$)
\[ p^{0\, u}_\pm:= \frac{1}{2} \Big|\bra{\pm^u}(\ket{+}+\ket{-})\Big|^2= \frac{1}{2}(1 \pm \sin\vartheta\cos\varphi),\]
where we use the parametrization of $\ket{\pm^u}$ introduced before. We assume that the direction ${\bf u}$ is such that these probabilities do not vanish.

Let $\hat \rho_n$ be the system density matrix after $n$ steps, and hence $n$ probe measurements of $\sigma_u$, and as before, let $\breve{\rho}_n(p,q)$ be its representation in momentum space
\[ \breve{\rho}_n(p,q) := {}_z\bra{p}\hat \rho_n \ket{q}_z,\]
with $\ket{p}_z$ and $\ket{q}_z$ eigen-vectors of the momentum operator, say $P\ket{q}_z=q\ket{q}_z$. In the continuous time limit, this reads:
\[ \breve \rho_t(p,q):={}_z\bra{p}\hat \rho_t\ket{q}_z=\int \frac{dxdy}{2\pi}\, e^{i(qx-py)}\, \hat\rho_t(x,y).\]
Its evolution is that given in eq.(\ref{eq:rhopq}).
We are going to argue that in the continuous time limit $\epsilon\to 0$, $n\to\infty$, with $t=n\epsilon$ fixed and $\delta=\sqrt{\epsilon}$, it converges in law to a solution of the stochastic equation 
\begin{eqnarray} \label{eq:Trhopq}
\boxed{ d\breve{\rho}_t(p,q) = L_N^{[p,q]}\big(\breve{\rho}_t\big)\, dt + D_{N,v}^{[p,q]}\big(\breve{\rho}_t\big)\, dB_t, }
\end{eqnarray}
with
\begin{eqnarray}\label{eq:LDUpq}
L_N^{[p,q]}\big(\breve{\rho}_t\big) &:=& -\frac{1}{2}(p-q)^2\,\breve{\rho}_t(p,q) -i(p-q) (N\,\breve{\rho}_t(p,q)+\breve{\rho}_t(p,q)\,N^\dag)\nonumber\\ 
   & & ~~~~~~~~  - i[H,\breve{\rho}_t(p,q)]+ L_N\big(\breve{\rho}_t(p,q)\big),\nonumber\\
D_{N,v}^{[p,q]}\big(\breve{\rho}_t\big) &:=&  \bar v\, (N-ip)\, \breve{\rho}_t(p,q) + v\, \breve{\rho}_t(p,q)\, (N^\dag+iq) - \breve{\rho}_t(p,q)\, U_{N,u}(\breve{\rho}_t),\\
U_{N,v}\big(\breve{\rho}_t\big) &:=& \int dp\, \mathrm{Tr}_{{\cal H}_c}\big( \bar v\, (N-ip) \breve{\rho}_t(p,p) + v\, \breve{\rho}_t(p,p) (N^\dag+ip)\big),\nonumber
\end{eqnarray}
with $B_t$ a normalized Brownian motion and $L_N$ defined in eq.(\ref{eq:defLDU}) above. Eq.(\ref{eq:Trhopq}) are again Belavkin's like equations for quantum trajectories. 

In eq.(\ref{eq:Trhopq}), the parameter $v$, $v\bar v=1$, depends on the direction ${\bf u}$ of the measured probe observable via
\begin{eqnarray} \label{eq:vu}
 v := \frac{\cos\vartheta + i \sin\vartheta\sin\varphi}{(1-\sin^2\vartheta\cos^2\varphi)^{1/2}}.
 \end{eqnarray}
 Eqs.(\ref{eq:Trhopq}) only depend on the unimodular parameter $v$ whereas the measured probes observables depend on the two parameters specifying the direction ${\bf u}$. This reduction of the number of parameters is a consequence of the fact that the probe states in absence of interaction are invariant under rotation along the $x$ axis, because they are $(\ket{+}+\ket{-})/\sqrt{2}$.
Notice that the drift term is independent of this parameter as expected, because averaging over the measurement outputs amounts to trace over the probe degrees of freedom, independently of which probe observable is measured.

It is easy to check that $\mathrm{Tr}_{{\cal H}_c}\big(\breve{\rho}_t(p,p)\big)$ is a local martingale, 
\[ d\, \mathrm{Tr}_{{\cal H}_c}\big(\breve{\rho}_t(p,p)\big) = \mathrm{Tr}_{{\cal H}_c}\big( D_{N,v}^{[p,p]}\big(\breve{\rho}_t\big)\big)\, dB_t,\]
since $\mathrm{Tr}_{{\cal H}_c}\big(L_N^{[p,p]}(\breve{\rho}_t)\big)=0$ as in the discrete settings. In particular, its mean is conserved in time: $\mathbb{E}[ \mathrm{Tr}_{{\cal H}_c}\big(\breve{\rho}_t(p,p)\big)]= \mathrm{Tr}_{{\cal H}_c}\big(\breve{\rho}_0(p,p)\big)$. Recall that $\mathrm{Tr}_{{\cal H}_c}\big(\breve{\rho}_t(p,p)\big)$ is the probability distribution density of observing $p$ in a measurement of the walker momentum. 

\begin{figure}
\centering
\includegraphics[scale=.35]{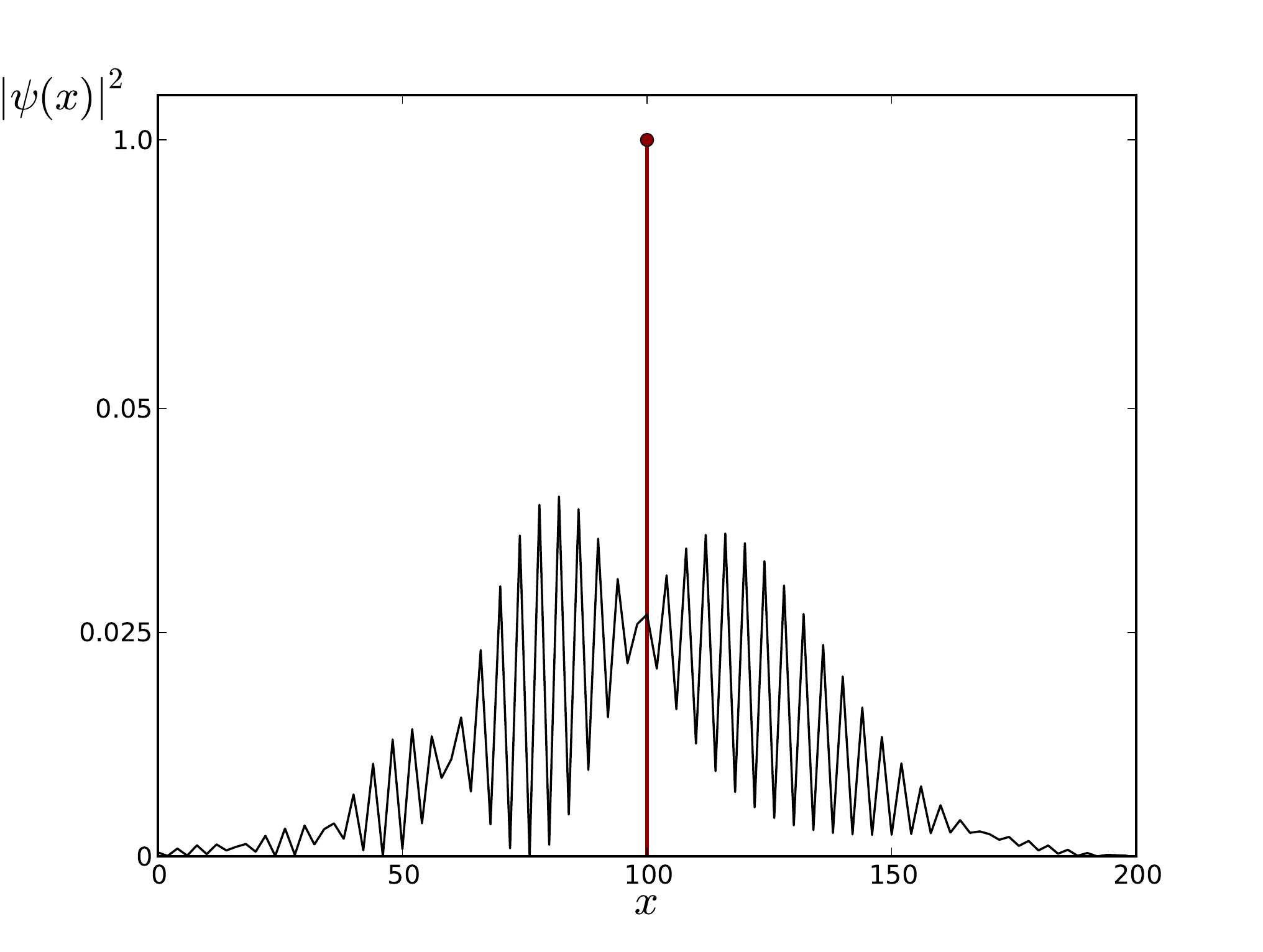}
\includegraphics[scale=.35]{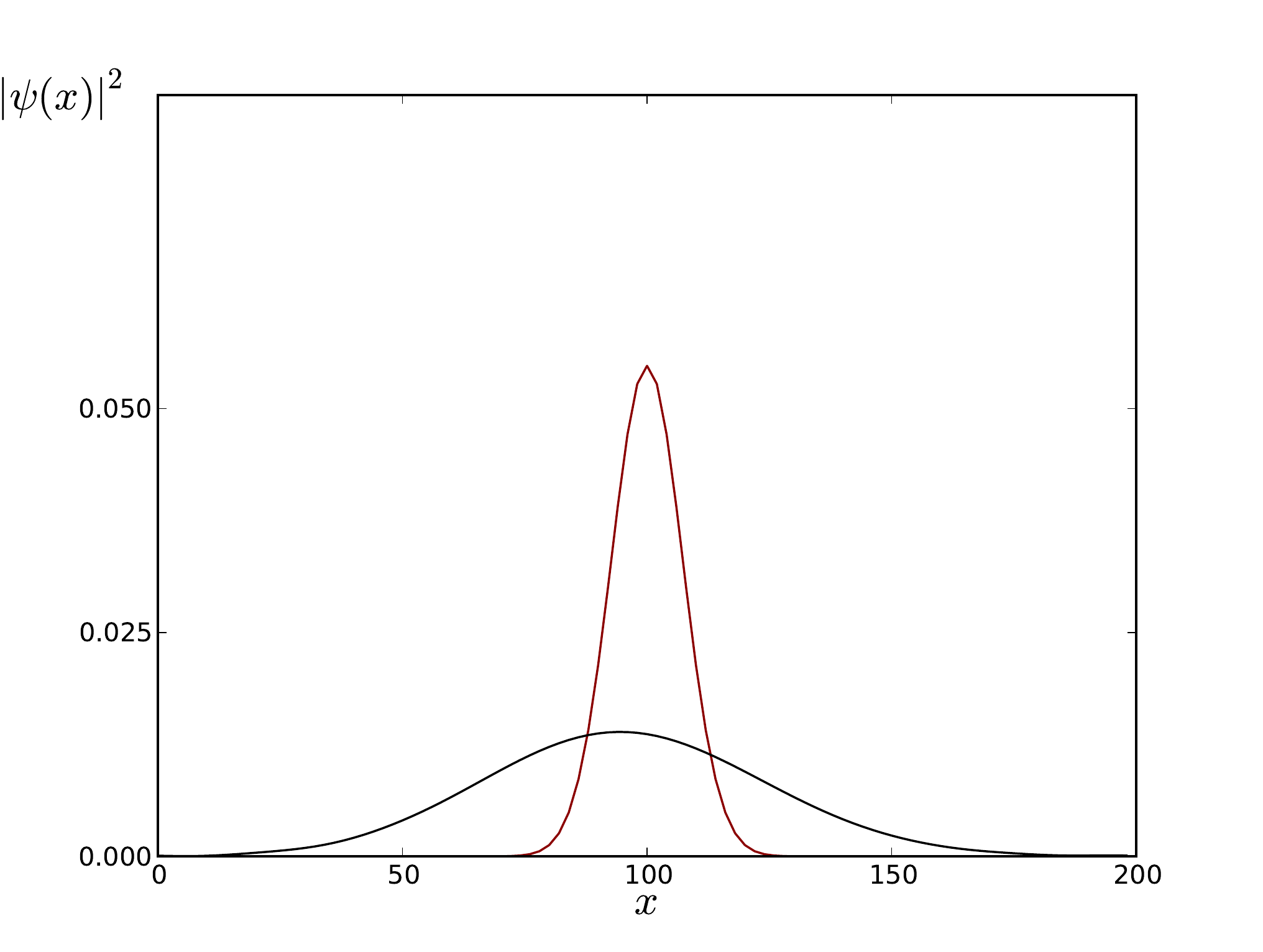}
\caption{\small Evolution of the modulus square of the wave function for {\it discrete} tilted trajectory without internal degree of freedom, as described in Section 1.2. The values of tilting parameters are $\vartheta=0.6=\varphi$. The red curves are the initial conditions, and the black are obtained after $t=1000$ iterations. On the right, a smooth initial condition propagates regularly and no singularity are produced. On the left, a hyper localized initial condition produces oscillations at the lattice scale at any later time, and this spoils the continuous limit.}
 \label{fig:regu}
\end{figure}

Deriving these equations in the scaling limit requires expanding the operators $U^u_\pm(p)$, defined just below eq.(\ref{eq:rhopq}), in power of $\sqrt{\epsilon}$ and $\delta$. Since $\delta$ is coupled to the momenta $p$ via $e^{\pm ip \delta}$, justifying this expansion requires controlling that $p \delta$ remains small. Although clearly not a complete proof, the fact that $\int_\Delta dp\, \mathrm{Tr}_{{\cal H}_c}\big(\breve{\rho}_t(p,p)\big)$ is a positive and bounded martingale for any interval $\Delta$, and thus possesses some regularity property, suggests that  imposing regularity property at initial time is enough to ensure such control.
Indeed, this martingale property and numerical simulations indicate (see Fig.\ref{fig:regu}) that if we start with an initial condition regular in position space (i.e. with a spectral momentum density sufficiently decreasing at large momentum) it will stay so even with tilted measurements and the continuous limit will make sense. If, on the other hand, one starts with a particle localized on one lattice site, the momentum distribution is not going to be bounded in the continuous limit, and the evolution will not smooth anything out (as it would with a classical heat equation, for instance). Hence, we can expect to have a well defined continuous limit only if the initial momentum distribution decreases fast enough at infinity. The tilted (quantum) random walk evolutions do not smooth out wave functions but, and remarkably, they apparently do not self-generate singularity so that it seems enough to impose regularity conditions at initial time only. 

In the case of tilted trajectory, and contrary to the case of simple trajectory, the walker position is not slave to the orbital motion. This is in particular true because the position of the walker is not well defined since the system density matrix is not diagonal in position space. One can nevertheless talk about the frequencies of occurrence $N_n^u(\pm)$  for $\pm$ as output measurements of $\sigma_u$ after $n$ step, $N_n^u(\pm):=\sum_{j=1}^n \mathbb{I}_{\{s_j=\pm\}}$. As before, in the scaling limit their leading contributions are $N_n^u(\pm)=np_\pm^{0\, u}+\cdots$ and we may write
\[ \sqrt{\epsilon}\big( N_n^u(\pm)-n\,p_\pm^{0\, u}\big) = \pm \sqrt{p_+^{0\, u}p_-^{0\, u}}\ Y_n^u,\]
where $Y^u_n$ codes for the deviation from this leading behavior (this equation serves as a definition of $Y^u_n$). In the scaling limit we have
\begin{eqnarray} \label{eq:Ytraj}
 dY_t^u= dB_t + U_{N,v}\big(\breve{\rho}_t\big)\, dt,
 \end{eqnarray}
in a way similar to the untitled case (but $Y^u_t$ is not directly linked to the walker position).

Before giving proofs of the above statement, let us make explicit these equations in case the internal gyroscope is trivial ($N=0$ and $H=0$). Then $L_N^{[p,q]}(\breve\rho_t)=-\frac{1}{2} (p-q)^2\,  \breve \rho_t(p,q)$ and $D_{N,v}^{[p,q]}(\breve\rho_t)= i\big((vq-\bar v p) -(v-\bar v)\vev{p}_t\big)\, \breve \rho_t(p,q)$ with $\vev{p}_t:=\int dp\, p\, \breve \rho_t(p,p)$ and $U_{N,v}(\breve\rho_t)= i(v-\bar v) \vev{p}_t$. Thus
\begin{eqnarray} \label{eq:trajecN=0}
d\breve \rho_t(p,q) &=& -\frac{1}{2} (p-q)^2\,  \breve \rho_t(p,q)\, dt +i\big((vq-\bar v p) -(v-\bar v)\vev{p}_t\big)\, \breve \rho_t(p,q)\, dB_t,\\
dY^v_t&=& i(v-\bar v) \vev{p}_t\, dt +dB_t. \nonumber
\end{eqnarray}
The diagonal components $\breve \rho_t(p,p)$ are local martingales, as it should be, with
\[ d\breve \rho_t(p,p) = i(v-\bar v)\big(p-\vev{p}_t\big)\, \breve \rho_t(p,p)\, dB_t. \]
For any interval\footnote{More precisely, any Borel subset of the real axis.} $\Delta$, the integral $\int_\Delta dp\, \breve \rho_t(p,p)$ is positive and bounded because $\int dp\, \breve \rho_t(p,p)=1$ by normalization. Hence, by the martingale convergence theorem, $\int_\Delta dp\, \breve \rho_t(p,p)$ converges almost surely and in $L^1$ for any interval $\Delta$, i.e. $\breve \rho_t(p,p)$ converges almost surely and in $L^1$ as a distribution on the real axis. The limiting distribution $\breve \rho_\infty(p,p)$ has to be a fixed point of the above evolution equation, that is, it should be such that $(p-\vev{p}_\infty)\, \breve \rho_\infty(p,p)=0$, for $v\not =\bar v$ (i.e. $v\not=\pm1$). Hence, $\breve \rho_\infty(p,p)$ is a Dirac measure peaked at a realization dependent random value $p_\infty$ (for $v\not=\pm1$)
\[ \breve \rho_\infty(p,p):=\lim_{t\to\infty}\breve \rho_t(p,p)= \delta(p-p_\infty).\]
This is in accordance with the analysis of Sect.1, and tilted quantum trajectory provides a model system for non-demolition impulsion measurement. The random impulsion $p_\infty$, which is realization dependent, is the result of this non-demolition measurement. The probability density that $p_\infty$ is equal to a given target $p_*$ (up to $dp_*$) is, by the martingale property, the initial probability density $\breve\rho_0(p,p)$, in accordance with von Neumann rules for quantum measurement.

This simple model (with $H=0=N$) also illustrates the phenomena of decoherence. The series of probes, when not observed or measured, form a reservoir. Tracing over the probes amounts to compute the mean, and eq.(\ref{eq:trajecN=0}) yields
\[ d\mathbb{E}[ \breve \rho_t(p,q) ] = -\frac{1}{2} (p-q)^2\,  \mathbb{E}[\breve \rho_t(p,q)]\, dt .\]
Hence the off-diagonal components vanishes in mean exponentially fast, $\mathbb{E}[ \rho_t(p,q) ] \propto e^{- (p-q)^2t/2}$, with a decoherence time proportional to the inverse distance square between the pointer states (here the momentum eigen-states)\footnote{More generally, one may defined distances between states as the inverse square root of the decoherence time, but this distance is going to dependent on the system-reservoir interaction.}, $\tau_\mathrm{decoherence}= 2/(p-q)^2$, as usual \cite{Qprocess}.
\medskip

\underline{\it ``Esquisse d'une preuve":}\\
The proof relies again in implementing Taylor expansion in the Dood decomposition of $\breve{\rho}_n(p,q)$. Let us decompose
\[ \breve{\rho}_n(p,q)= O_n(p,q) + M_n(p,q),\]
with $M_n(p,q)$ ${\cal F}_n$-measurable martingales and $O_n(p,q)$ ${\cal F}_{n-1}$-measurable predictable processes. By construction,
\begin{eqnarray*}
M_{n+1}(p,q)-M_n(p,q)&=& \breve{\rho}_{n+1}(p,q) - \mathbb{E}[ \breve{\rho}_{n+1}(p,q) |{\cal F}_n],\\
O_{n+1}(p,q)-O_n(p,q)&=& \mathbb{E}[ \breve{\rho}_{n+1}(p,q) |{\cal F}_n]- \breve{\rho}_n(p,q).
\end{eqnarray*}
The differences of the predictable (martingale) terms is going to converge towards the drift (noisy) terms of continuous in time Langevin equations satisfied by the system density matrix, so that
\[ d\breve{\rho}_t(p,q) = dO_t(p,q)+ dM_t(p,q),\]
in the scaling limit. Writing as before
\[ \breve\rho_{n+1}(p,q)=\breve\rho_n^{(+)}(p,q)\, \mathbb{I}_{\{s_{n+1}=+\}} + \breve\rho_n^{(-)}(p,q)\, \mathbb{I}_{\{s_{n+1}=-\}} ,\]
with $\breve{\rho}_{n}^{(\pm)}(p,q):= {U^u_\pm(p)\, \breve{\rho_n}(p,q)\, U^{u}_\pm(q)^\dag}/{ p_\pm^{u}(n)}$, we may present them as
\begin{eqnarray*}
O_{n+1}(p,q)-O_n(p,q)&=& \breve\rho_n^{(+)}(p,q)\, p_+^{u}(n) + \breve\rho_n^{(-)}(p,q)\, p_-^{u}(n)- \breve{\rho}_n(p,q),\\
M_{n+1}(p,q)-M_n(p,q)&=& \frac{1}{2}
\big(\breve\rho_{n}^{(+)}(p,q)-\breve\rho_{n}^{(-)}(p,q)\big)\Big(\mathbb{I}_{\{s_{n+1}=+\}} -p^u_+(n)+p^u_-(n) -\mathbb{I}_{\{s_{n+1}=-\}}\Big).
\end{eqnarray*}

Taylor expanding the predictable term gives
\[ O_{n+1}(p,q)-O_n(p,q) = \epsilon\, L_N^{[p,q]}\big(\breve{\rho}_n\big)+O(\epsilon^{3/2}),\]
with no contribution at order $\sqrt{\epsilon}$ as it should for the existence of the scaling limit. So we get $d\, O_t(p,q)= L_N^{[p,q]}\big(\breve{\rho}_t\big)\, dt$ in the scaling limit.

The martingale differences are going to converge to terms proportional to $dB_t$, so we just have to compute their means, which vanish by construction, and their covariances. Using the fact that $\mathbb{I}_{\{s_{n+1}=\pm\}}$ are characteristic functions on sets with empty intersections and that their means are respectively $p_\pm^u(n)$, we have
\[ \mathbb{E}\big[ \big(M_{n+1}(p,q)-M_n(p,q)\big)^2 |{\cal F}_n\big] =  p_+^u(n)p_-^u(n)\, \big(\breve\rho_n^{(+)}(p,q)-\breve\rho_n^{(-)}(p,q)\big)^2.\]
Then, a lengthy Taylor expansion yields,
\[ \breve\rho_n^{(\pm)}(p,q)= \breve\rho_n(p,q) \pm\sqrt{\epsilon}\, \sqrt{\frac{p_\mp^{0\,u}}{p_\pm^{0\,u}}}\, D_{N,v}^{[p,q]}\big(\breve{\rho}_n\big) +O(\epsilon).\]
Since $\Big({\sqrt{\frac{p_-^{0\,u}}{p_+^{0\,u}}}+\sqrt{\frac{p_+^{0\,u}}{p_-^{0\,u}}}}\Big)^2= ({p_+^{0\,u}p_-^{0\,u}})^{-1}$ and $p^u_\pm(n)=p_\pm^{0\,u}+O(\sqrt{\epsilon})$, we get
\[ \mathbb{E}\big[ \big(M_{n+1}(p,q)-M_n(p,q)\big)^2 |{\cal F}_n\big] = \epsilon\, D_{N,v}^{[p,q]}\big(\breve{\rho}_n\big)^2,\]
and we may write $dM_t(p,q)= D_{n,v}^{[p,q]}\big(\breve{\rho}_t\big)\,dB_t$ in the scaling limit. Gathering $dO_t(p,q)$ and $dM_t(p,q)$ proves eq.(\ref{eq:Trhopq}).

To get the equation for $Y_n^u$, we notice that by definition we have
\begin{eqnarray*} 2\sqrt{\epsilon}\, \big( M_{n+1}-M_n\big)&=&\big(\breve\rho_n^{(+)}(p,q)-\breve\rho_n^{(-)}(p,q)\big)\times\\
&& \times\ \Big(2 \sqrt{p_+^{0\,u}p_-^{0\,u}} \big(Y^u_{n+1}-Y^u_n\big) - \sqrt{\epsilon}(p_+^u(n)-p_+^{0\, u}-p_-^u(n)+p_-^{0\, u})\Big).
\end{eqnarray*}
The formula for $\big(\breve\rho_n^{(+)}(p,q)-\breve\rho_n^{(-)}(p,q)\big)$ and the fact that $p_\pm^u(n)=p_\pm^{0\, u}\pm\sqrt{\epsilon}\, \sqrt{p_+^{0\,u}p_-^{0\,u}}\, U_{N,v}(\breve{\rho}_n)+\cdots$, then yields the formula for $dY^u_t$ in the scaling limit. \cqfd

\section{Open QBM: Quantum dynamical map}

The open QBM maps are quantum dynamical maps acting on the system (walker + gyroscope) states. They apply when no probe measurements are done, or when nobody keeps track of the outputs of the probe measurements, so that the series of probes form a kind of reservoir whose degrees of freedom are traced out. The system is then described by a reduced density matrix. 

\subsection{Open QBM Lindbladian evolution}

We denote by $\bar \rho_t$ the reduced density matrix for the quantum walker and its gyroscope. It is a state on ${\cal H}_{\rm sys}:={\cal H}_c\otimes{\cal H}_z$, with ${\cal H}_z:=L^2(\mathbb{R})$ in the continuum limit.
The open QBM dynamical map is specified by the following Lindblad equation:
\begin{eqnarray} \label{OQBM:lindblad}
\boxed{ \partial_t\bar \rho_t= {\cal L}(\bar \rho_t):= - i[H,\bar \rho_t] -\frac{1}{2} [P, [P,\bar \rho_t]] - i \big(N[P,\bar \rho_t]+[P,\bar \rho_t]N^\dag\big) + L_N(\bar \rho_t),}
\end{eqnarray}
with $P=-i\partial_x$ the momentum operator (which here commutes with $N$) and 
\[L_N(\rho):=N\rho N^\dag - \half(N^\dag N\rho+ \rho N^\dag N).\]
Eq.(\ref{OQBM:lindblad}) is well defined in the sense that the Linbldad operator ${\cal L}$  has the required positivity property to formally\footnote{A more mathematically rigorous statement about complete positiveness would require to deal with operator domains etc.} generate a completely positive map \cite{Qmap}. In this sense eq.(\ref{OQBM:lindblad}) does not suffer from problems with complete positivity as do Markovian approximations of Caldeira-Leggett models \cite{CalLeg,Qprocess}. This can be seen by presenting ${\cal L}$ in the form
\begin{eqnarray} \label{eq:LindbladBis} 
{\cal L}(\bar \rho_t)= -i\big[{\cal H},\bar\rho_t\big] + \big( {\cal P}\,\bar\rho_t\,{\cal P}^\dag - \frac{1}{2}({\cal P}^\dag{\cal P}\, \bar\rho_t + \bar \rho_t\, {\cal P}^\dag{\cal P})\big),
\end{eqnarray}
with ${\cal P}:= P+iN$ and ${\cal H}:= H+\frac{1}{2}(PN+N^\dag P)$.

Eq.(\ref{OQBM:lindblad}) algebraically follows by expanding eq.(\ref{eq:Oqrw_map}) up to order $\epsilon=\delta^2$. Of course justifying mathematically such expansion demands some care since the translation operator $P$ has an unbounded spectrum. Thus this expansion is meaningful only when acting on density matrices whose Fourier transforms have bounded supports, or decreases fast enough at infinity.  It is however instructive to ``phenomenologically" deduce it in the simple case in which the reduced density matrix $\bar \rho_t$ is diagonal in the orbital space, that is, in the continuous limit, 
\[\bar \rho_t= \int dx\, \bar\rho_t(x)\otimes \ket{x}_z\bra{x},\]
 with $\bar\rho_t(x)$ a density matrix on the internal Hilbert space ${\cal H}_c$, and $p(x,t):={\rm Tr}_{{\cal H}_c}\bar\rho_t(x)$ the probability density to find the quantum walker at position $x$ at time $t$. At each time step $\epsilon$, it is updated using the open QRW rules (\ref{eq:Oqrw_bis}), 
\[\bar\rho_{t+\epsilon}(x)= B_-\,\bar\rho_t(x+\delta)\,B_-^\dag+B_+\,\bar\rho_t(x-\delta)\,B_+^\dag,\]
In the continuum limit one imposes the scaling relation $\epsilon=\delta^2$ and $B_\pm=\frac{1}{\sqrt{2}}[\mathbb{I}\pm \sqrt{\epsilon} N - \epsilon(iH\pm M +\frac{1}{2}N^\dag N) + o(\epsilon)$. Taylor expansion then gives :
\begin{eqnarray} \label{eq:FPrho}
\partial_t\bar\rho_t(x)=  - i\big[H,\bar\rho_t(x)\big]+\half \partial_x^2\bar\rho_t(x)- \big(N\partial_x\bar\rho_t(x)+\partial_x\bar\rho_t(x) N^\dag\big)+L_N(\bar\rho_t(x)),
\end{eqnarray}
Doing such Taylor expansion demands the density matrix $\bar\rho_t(x)$ to be regular enough, so that we again recover the condition on the support of the Fourier transform of the density matrix for the expansion to be valid.
Eq.(\ref{eq:FPrho}) coincides with eq.(\ref{OQBM:lindblad}) for diagonal density matrix. It mixes pieces from diffusive Fokker-Planck equation and from Lindbladian quantum evolution for $\bar \rho_t$. It is easily generalized to non-diagonal density matrices.

The probability $p(x,t):=\mathrm{Tr}_{{\cal H}_c}(\bar\rho_t(x))$ is not associated to a Markov process and does not satisfy a linear equation, but $\bar \rho_t$ does. The Markov property emerges when one consider both the orbital and the internal degrees of freedom. Notice that in absence of internal degree of freedom one recovers the standard heat equation $\partial_tp=\half \partial_x^2p$.

%For non-diagonal density matrix, $\bar\rho_t= \int dxdy\, \bar \rho_t(x,y)\otimes \ket{x}_z\bra{y}$, with $\bar \rho_t(x,y):={}_z\bra{x}\bar\rho_t\ket{y}_z$,
%the evolution equation (\ref{OQBM:lindblad}) reads:
%\[ \partial_t\bar \rho_t(x,y)= - i \big[H, \bar\rho_t(x,y)\big]+\frac{1}{2} (\partial_x+\partial_y)^2\bar\rho_t(x,y) - (\partial_x+\partial_y) \big(N\bar\rho_t(x,y) + \bar\rho_t(x,y)N^\dag\big)  + L_N(\bar\rho_t(x,y)),\] 
%This follows from using ${}_z\bra{x}[P,\bar\rho_t]\ket{y}_z=-i(\partial_x+\partial_y)\bar\rho_t(x,y)$, a formula which preserves the diagonality of $\bar\rho_t$ if it happens to be diagonal, that is when $\bar \rho_t(x,y)=\bar\rho_t(x)\, \delta(x-y)$.  
%It can also be deduced using the previous ``phenomenological" approach.

\subsection{From quantum trajectories to Lindbladian evolution}
Since the reduced density matrix $\bar\rho_t$ and its evolution describes situations in which the outputs of the probe measurements are not recorded, eq.(\ref{OQBM:lindblad}) should follow from that of quantum trajectories under averaging over all possible probe measurement outputs. This is what we are doing in this section, that is we show how to derive the Lindblad equation (\ref{OQBM:lindblad}) from that of the quantum trajectories (\ref{eq:drho}). This can be done in two different ways, either using tilted trajectory, say in momentum space, or using simple quantum trajectory, say in position space, and a bit of It\^o calculus. 

Eq.(\ref{OQBM:lindblad}) indeed takes a simple form in Fourier space. Let $\bar \rho_t(p,q):= {}_z\bra{p}\bar\rho_t\ket{q}_z$, where $\ket{q}_z$ are momentum eigen-state $P\ket{q}_z=q\ket{q}_z$. It acts on the internal space ${\cal H}_c$. Eq.(\ref{OQBM:lindblad}) then reads,
\[ \partial_t\bar \rho_t(p,q)= -\frac{1}{2} (p-q)^2\bar \rho_t(p,q) - i (p-q)\big(N \bar \rho_t(p,q)+\bar \rho_t(p,q)N^\dag\big) - i[H,\bar \rho_t(p,q)] + L_N(\bar \rho_t(p,q)).\]
The r.h.s. of the previous equation coincides with the drift term of tilted quantum trajectory (\ref{eq:Trhopq}) so that we can write
\[ \partial_t\bar \rho_t(p,q)= L_N^{[p,q]}(\bar \rho_t),\]
and we have $\bar \rho_t(p,q)=\mathbb{E}[\breve \rho_t(p,q)]$.
The diffusion terms vanish for $p=q$ so that the trace of the diagonal component is time independent $\partial_t\, {\rm Tr}_{{\cal H}_c}\bar\rho_t(q,q)=0$, and this is linked to the martingale property of the analogue component in quantum trajectory.

Next, we look at simple quantum trajectory $(\rho_t,X_t)$ and we restrict ourself to reduced density matrices localized and diagonal in the orbital space. Generalization to non-diagonal density matrix is simple. Averaging over all possible probe measurement outputs corresponds to averaging over realizations of the Brownian motion driving the quantum trajectories. Starting from quantum trajectory realizations we have to reproduce the reduced density matrix by averaging, thus
\begin{eqnarray}\label{eq:defXrho}
\bar\rho_t=  \int {\hskip -0.1 truecm} dx\, \bar\rho_t(x)\otimes \ket{x}_z\bra{x}=\mathbb{E}[\,\rho_t\otimes \ket{X_t}_z\bra{X_t}\,].
 \end{eqnarray}
or alternatively,
\[  \int {\hskip -0.1 truecm} dx\, \bar\rho_t(x) f(x)=\mathbb{E}[\,\rho_t\otimes f(X_t)\,],\]
for any test function $f(x)$. We shall prove that the stochastic equations (\ref{eq:drho}) for simple quantum trajectories imply equations (\ref{eq:FPrho}) for the reduced density matrix reconstructed via eq.(\ref{eq:defXrho}). 
\medskip

\underline{\it ``Esquisse d'une preuve":}\\
Computing the time evolution of $\bar \rho_t(x)$ follows from computing the It\^o derivative of $\rho_t\,f(X_t)$. 
Because $dX_t=U_N(\rho_t) dt + dB_t$ from eq.(\ref{eq:drho}), one has 
\[df(X_t)=\big[U_N(\rho_t)f'(X_t)+\frac{1}{2}f''(X_t)\big]dt + f'(X_t)\,dB_t.\]
 Since $d\rho_t=\big( -i[H,\rho_t]+L_N(\rho_t)\big)dt + D_N(\rho_t)\,dB_t$ from eq.(\ref{eq:drho}), we get using It\^o rules
\begin{eqnarray*}
 d\big[\rho_t\,f(X_t)\big]&=&\Big[\frac{1}{2}\rho_tf''(X_t)+\big(- i[H,\rho_t]+L_N(\rho_t)\big)f(X_t) + (\rho_tU_N(\rho_t) + D_N(\rho_t))f'(X_t)\Big]dt\\
 & & +\Big[ \rho_t f'(X_t)+ D_N(\rho_t)f(X_t)\Big]dB_t,
 \end{eqnarray*}
Since $\rho_tU_N(\rho_t) + D_N(\rho_t)=N\rho_t+ \rho_tN^\dag$ by construction, we may equivalently write
\[  d\big[\rho_t\,f(X_t)\big] = \Big[\frac{1}{2}\rho_tf''(X_t)+\big( -i[H,\rho_t]+L_N(\rho_t)\big)f(X_t) + (N\rho_t+ \rho_tN^\dag)f'(X_t)\Big]dt+[\cdots]dB_t.\]
where we did not write explicitly the noisy term $[\cdots]dB_t$ as we do not need it (its mean vanishes by It\^o convention). Now the above drift term is linear in $\rho_t$ as it should be, and hence we get
\[ \int {\hskip -0.1 truecm} dx\,  f(x)d\bar\rho_t(x) =  \int {\hskip -0.15 truecm} dx\, \Big[\frac{1}{2}\bar\rho_t(x)f''(x)+\big( -i[H,\bar\rho_t(x)]+L_N(\bar\rho_t(x))\big)f(x) + (N\bar\rho_t(x)+ \bar\rho_t(x)N^\dag)f'(x)\Big]dt,\]
By integration by part this gives 
\[d\bar\rho_t(x)=\Big[\frac{1}{2}\bar\rho_t''(x)- i[H,\bar\rho_t(x)]+L_N(\bar\rho_t(x)) - (N\bar\rho_t'(x)+ \bar\rho_t'(x)N^\dag)\Big]dt,\]
which is equivalent to eq.(\ref{eq:FPrho}). A similar computation can be done with multiple trajectories and non-diagonal mixed states.
\cqfd

\section{Open QBM: Quantum stochastic processes}

Quantum stochastic processes describe the system -- with both orbital and internal degrees of freedom -- when no measurement are done on the probes so that they remain entangled with the system after having interacted. Hence, they correspond to dynamical processes on the total system (walker + gyroscope + probes).

Quantum stochastic differential equations (quantum SDE) and quantum noises were introduced in \cite{Para}. They in particular provide a simple framework to construct dilations of quantum dynamical maps. We start by presenting this framework, probably in an oversimplified manner but hopefully useful to amateurs. It is based on canonical operators defined on the line, a concept familiar from field theory. We also explain the relation with the discrete setting we started with, following an approach advocated in \cite{AttalPautrat}. We then present what are the quantum SDE for open QBM. These are obtained from the quantum dynamical map by deciphering some algebraic structures inherent to quantum SDE (which we did not find much developed in the literature). To illustrate the consistency between the different approaches, we provide a derivation of quantum trajectory from quantum SDE. In the literature, this is usually done using the formalism of quantum filtering \cite{Qfilter}. We here provide a (we hope) 
more physical, and hopefully more transparent although less rigorous, derivation based on basic rules of quantum mechanics.

\subsection{Dilation, purification and quantum noise}

Let us go back for a short while to discrete open quantum random walks. Recall that, if all probes are prepared in the same pure state $\ket{\phi}_p$ and if the initial state of the system is $\ket{\chi}_c\otimes\ket{x_0}_z$, the state of the total system in ${\cal H}_{\rm sys}\otimes {\cal H}_p^\infty$ after $n$ steps is highly entangled and may be expanded as in eq.(\ref{eq:discreteNoise}),
\[ \boxed{ \Big(\sum_{[\omega_n]} B([\omega_n])\ket{\chi}_c\otimes\ket{X_n([\omega_n])}_z\otimes\ket{[\omega_n]}_p\Big)\,\otimes\ket{\phi}_p\otimes\ket{\phi}_p\otimes\cdots, }\]
where the sum is over all random walks $[\omega_n] : =(\varepsilon_1,\cdots,\varepsilon_n)$, with $\varepsilon_j=\pm$, of length $n$, starting at $x_0$ and with position $X_j([\omega_n])$ at their $j^{\rm th}$ step, and $\ket{[\omega_n]}_p:=\ket{\varepsilon_1}\otimes\cdots\otimes\ket{\varepsilon_n}$, and $B([\omega_n])$ is defined in eq.(\ref{eq:discreteNoise}). 

Quantum noise theory consists in giving a meaning to these sums in the scaling limit in which the lattice spacing $\sqrt{\epsilon}\to 0$ and $n\to \infty$ but with $t=n\epsilon$ fixed. Naively, since scaling limits of random walks are Brownian motions, one would expect that the scaling limit of the above states could be written in terms of states indexed by samples of Brownian motion:
\[ ``\Big(\sum_{[\omega_t]} \hat B([\omega_t])\ket{\chi}_c\otimes\ket{X_t([\omega_t])}_z\otimes\ket{[\omega_t]}\Big)"\,\otimes\ket{\phi}_p^{\otimes\infty},\]
with $[\omega_t]$ a sample of Brownian motion. Of course such sum makes barely any sense. At best the sum has to be replaced by some kind of a measure, say a Wiener measure or non-singular modification of it, or any other similar path integral measure or else. But it will still be difficult to make sense of the would be infinite dimensional Hilbert space containing states indexed by a Brownian sample\footnote{This is actually the role played by the Fock space ${\cal F}_{0;\infty}$ that we shall introduce in a short while}. One is in a (slightly) better position if one decides to deal with expectations of observables using the above state because then one would sum functions, or matrix elements of the observables, indexed by Brownian sample (of course the sum would have to be replaced by the elusive measure mentioned above). That is, as usual with non commutative geometry, one is in a better position if one considers the non-commutative algebra of observables and the flows that they are subject to, instead of the would be limiting states.

It is this point of view which makes contact with the theory of quantum noise \cite{Qnoise}. Making this scaling limit and its relation with quantum noise theory rigorous has been done by Attal and Pautrat in \cite{AttalPautrat}, at least in the finite dimensional case. Since we are here dealing with quantum walkers on the continuous line, we are implicitly extending their arguments to infinite dimensional cases. Our aim is not to re-derive their results in infinite dimensional settings (assuming that they still apply) but to illustrate -- in maybe more physical terms --  how to make contact with the previous discussions. We first need to introduce the rules of quantum stochastic calculus. We shall present them in a non-rigorous way, hopefully useful for field theorists.

Quantum noise theory and quantum stochastic calculus \cite{Qnoise,Qprocess} are based on the connexion between Brownian motion and Gaussian free field. Let $a(t)$ and $a^\dag(t)$ be canonical free fields with commutation relations 
\[ [a(s),a(t)]=0,\quad [a(s),a^\dag(t)]=\delta(s-t).\]
They act on the bosonic Fock space over $L^2(\mathbb{R})$. We shall denote ${\cal F}_{s;t}$,  $(s<t)$, the Fock space\footnote{Recall that, for any Hilbert space $V$, the associated (bosonic) Fock space is defined as the graded space $\bigoplus_{n\geq 0} {\rm Sym}(V^{\otimes n})=\mathbb{C}\oplus V \oplus {\rm Sym}(V\otimes V) \oplus \cdots$ where ${\rm Sym}(V^{\otimes n})$ is the component of $V^{\otimes n}$ totally symmetric under permutation. In this decomposition, $\mathbb{C}$ refers to the vacuum state, $V$ to the one particle Hilbert space, $Sym(V\otimes V)$ to the two particle Hilbert space, etc. Here we use $V=L^2([s,t))$.} over $L^2([s,t))$, the Hilbert space of square integrable functions on the interval $[s,t)$. We shall only deal with positive times $s,t$. 
The quantum noises are defined by $\xi_t:= \int_0^t ds\, a(s)$ so that
\[ d\xi_t:= \int_t^{t+dt} {\hskip -0.5 truecm} ds\, a(s),\quad [d\xi_t,d\xi_t^\dag]=dt.\]
They also act on the Fock space ${\cal F}_{[0;\infty)}$.

Let $\ket{\Omega_\infty}$ be the vacuum state in ${\cal F}_{0;\infty}$, such that $a(t)\ket{\Omega_\infty}=0$ for all (non negative) time $t$. Vacuum expectation with noise insertions can be computed using Wick's theorem. Since $\bra{\Omega_\infty}d\xi_t\,d\xi_t^\dag\ket{\Omega_\infty}=dt$ and $\bra{\Omega_\infty}d\xi_t^\dag\, d\xi_t\ket{\Omega_\infty}=0$, we deduce the rules,
\begin{eqnarray} \label{eq:0Ito}
 d\xi_t\,d\xi_t^\dag=dt,\quad d\xi_t^\dag\, d\xi_t=0, 
\end{eqnarray}
valid in any vacuum expectation with other operator insertions away from time $t$. These are called quantum It\^o rules (at zero temperature)\footnote{ More generally quantum It\^o rules tell that $d\xi_td\xi_t^\dag=(1+\mathfrak{n}_t)\, dt$ and $d\xi_t^\dag d\xi_t=\mathfrak{n}_t\, dt$ for some $\mathbb{C}$-number $\mathfrak{n}_t$. This actually correspond to look at expectations not in the vacuum state but in thermally activated states. We shall deal with these cases in the following Section.}.

The connection with the scaling limit of iterated interactions goes as follows. The Fock space ${\cal F}_{[0,\infty)}$ describes the Hilbert space of the infinite series of probes. The state corresponding to all probes in the same identical pure reference state $\ket{\phi_\mathrm{ref}}_p$ -- with, in the present case, $\ket{\phi_\mathrm{ref}}=(\ket{+}+\ket{-})/\sqrt{2}$, the probe state in absence of interaction -- corresponds to the Fock vacuum, 
\[ ``\bigotimes_{k=1}^\infty \ket{\phi_\mathrm{ref}}_p \to \ket{\Omega_\infty}".\]
Once the probes have interacted with the system, their states have been deformed and this is described by the action of the canonical operator $a^\dag(t)$. That is: states of the Fock space correspond to deformations of the states of bundle of probes away from their reference state $\otimes_{k=1}^\infty\ket{\phi_\mathrm{ref}}_p$. Because the scaling limit involves the large $n$ small $\epsilon$ limit, with $t=n\epsilon$ fixed, any deformations on the Fock spaces ${\cal F}_{s;t}$ actually involve an infinite number of probes, which we call a bundle of probes, even for intervals $[s,t]$ as small as we want.

One may be interested in a subset of probes, say in all probes between the $m$-th and the $n$-th ones and not consider the probes before the $n$-th or after the $m$-th one. This amounts to look at the embedding of the Hilbert space of states for the $(n-m)$ selected probes in the Hilbert space of all probes which may be described by the factorization
\[ \bigotimes_{k=1}^\infty {\cal H}_p= \big(\bigotimes_{k<m}{\cal H}_p\big)\otimes\big(\bigotimes_{k=m}^n{\cal H}_p\big)\otimes\big(\bigotimes_{k>n}{\cal H}_p\big).\]
In the scaling limit, only considering probes between time $s$ and $t$, with  $m=[s/\epsilon]$ and $n=[t/\epsilon]$, corresponds to only look at states in the reduced Fock space ${\cal F}_{s;t}$ over $L^2([s,t])$. The direct sum decomposition of $L^2(\mathbb{R}_+)$ as $L^2([0,s))\oplus L^2([s,t))\oplus L^2([t,\infty))$ translates into the factorization of Fock spaces
\[ {\cal F}_{0;\infty} = {\cal F}_{0;s}\otimes {\cal F}_{s;t} \otimes {\cal F}_{t;\infty}.\]
Correspondingly the vacuum state factorizes
\[ \ket{\Omega_\infty}= \ket{\Omega_{0;s}}\otimes\ket{\Omega_{s;t}}\otimes\ket{\Omega_{t;\infty}},\]
with $\ket{\Omega_{s;t}}$ the vacuum in ${\cal F}_{s;t}$.
Operators or observables on probes between time $s$ and $t$ are mapped into operators or observables acting on ${\cal F}_{s;t}$. For instance the infinitesimal canonical operators $d\xi_t$ and $d\xi_t^\dag$ acts on ${\cal F}_{t,t+dt}$, that is, they act on probes between time $t$ and $t+dt$.
Notice that the construction of the Fock space, and its vacuum, is relative to the `ground' state $\otimes_{k=1}^\infty\ket{\phi_\mathrm{ref}}_p$.

\subsection{Open QBM quantum stochastic equations}
Quantum stochastic equation describes the evolution on the total Hilbert space of the total system (walker $+$ gyroscope $+$ probes), in such way that it reduces to the quantum dynamical map after tracing over the probes degrees of freedom. In the scaling limit, the evolution is described by a unitary map in ${\cal H}_{\rm sys}\otimes {\cal F}_{0;\infty}$. It is called a dilation of the dynamical map (but dilations are not unique \cite{Qmap,Qprocess}).

The dilation is generated by operators $\mathfrak{U}_t$, (formally) unitary, which represent the (would be) scaling limit of the iteration of the unitary operator $U_{0k}$ coding for the interaction between the system and the $k$-th probes:
\[ ``U_{0n}\cdots U_{02}U_{01} \to \mathfrak{U}_t",\]
in the scaling limit $n\to\infty$, $\epsilon\to0$ at $t=n\epsilon$ fixed. Since by construction, this product acts only on the $n$-th first probes, it acts only on the Fock subspace ${\cal F}_{0;t}$ and leave the forward vacuum state $\ket{\Omega_{t;\infty}}$ invariant.

We claim that the dilation for the open quantum Brownian motion (at zero temperature)\footnote{And assuming that the probes Hilbert spaces is $\mathbb{C}^2$ so that only one quantum noise is involved in the scaling limit.} is generated by the unitary operator $\mathfrak{U}_t$ on  ${\cal H}_{\rm sys}\otimes {\cal F}_{0;\infty}$ defined by the quantum stochastic differential equation (quantum SDE) 
\begin{eqnarray} \label{eq:Udilat}
 \boxed{ d\mathfrak{U}_t\,\mathfrak{U}_t^{-1}:=-i(P-iN^\dag)\, d\xi_t - i(P+iN)\, d\xi_t^\dag -\big(iH+\frac{1}{2}(P^2+2iPN+N^\dag N)\big)dt, }
 \end{eqnarray}
with $d\mathfrak{U}_t:=\mathfrak{U}_{t+dt}-\mathfrak{U}_t$. Equivalently,
\[ d\mathfrak{U}_t\,\mathfrak{U}_t^{-1}=-i{\cal P}^\dag\, d\xi_t - i{\cal P}\, d\xi_t^\dag - \big(i{\cal H} +\frac{1}{2}{\cal P}^\dag{\cal P}\big)\, dt,\]
with ${\cal P}:= P+iN$ and ${\cal H}:= H+\frac{1}{2}(PN+N^\dag P)$ as above. It is formally unitary in the sense that $d(\mathfrak{U}^\dag_t\,\mathfrak{U}_t )=0$. Notice that $\mathfrak{U}_{t+dt}\,\mathfrak{U}_t^{-1}=\mathbb{I}+d\mathfrak{U}_t\,\mathfrak{U}_t^{-1}$ acts non trivially only on the Fock sub-space ${\cal F}_{t;t+dt}$ because $d\xi_t$ and $d\xi_t^\dag$ only acts on ${\cal F}_{t;t+dt}$. This is of course compatible with the fact that the operator $\mathfrak{U}_{t+dt}\,\mathfrak{U}_t^{-1}$ codes for the successive interaction between the system and all probes between time $t$ and $t+dt$ in the scaling limit.

The quantum stochastic process is defined as a flow on operators or observables, i.e. it is defined in the Heisenberg dual picture, via
\[ A\to A_t:=\mathfrak{U}_t^\dag\, A\,\mathfrak{U}_t,\]
for any observable $A$ that we choose to act non trivially on ${\cal H}_{\rm sys}$ only  (but $A_t$ acts on ${\cal H}_{\rm sys}\otimes{\cal F}_{0;\infty}$). That is, we are looking at the flow of system observables, for simplicity. We claim that the quantum SDE for the open quantum Brownian motion is,
\begin{eqnarray}\label{eq:dilatBr}
\boxed{dA_t = i\big[P-iN^\dag, A\big]_t\, d\xi_t + i\big[P+iN,A\big]_t\, d\xi_t^\dag + {\cal L}_*(A)_t\, dt,}
\end{eqnarray}
or alternatively,
\[ dA_t = i\big[{\cal P}^\dag, A\big]_t\, d\xi_t + i\big[{\cal P},A\big]_t\, d\xi_t^\dag + {\cal L}_*(A)_t\, dt, \]
with dual Lindbladian
\begin{eqnarray} \label{eq:Ldual}
{\cal L}_*(A)&:=&  i\big[{\cal H},A\big] +  {\cal P}^\dag A {\cal P} - \frac{1}{2} ({\cal P}^\dag {\cal P} A + A {\cal P}^\dag {\cal P}) \\
&=& i[H,A]-\frac{1}{2}[P,[P,A]]+i\big([P,A]\, N + N^\dag\, [P,A]\big) + N^\dag A N -\frac{1}{2}(A N^\dag N + N^\dag N A).\nonumber
\end{eqnarray}
Recall that ${\cal P}:= P+iN$ and ${\cal H}:= H+\frac{1}{2}(PN+N^\dag P)$. Note that ${\cal L}_*(\mathbb{I})=0$ and this proves the formal unitarity of $\mathfrak{U}_t$. In eq.(\ref{eq:dilatBr}) above the position of the index $t$ matters: the equation involves $\big[{\cal P},A\big]_t=\mathfrak{U}_t^\dag \big[{\cal P},A\big] \mathfrak{U}_t$ or ${\cal L}_*(A)_t=\mathfrak{U}_t^\dag {\cal L}_*(A) \mathfrak{U}_t$ which are different from $\big[{\cal P},A_t\big]$ and ${\cal L}_*(A_t)$.

Using quantum It\^o rules (at zero temperature), we shall show that eq.(\ref{eq:dilatBr}) is the only quantum stochastic equation compatible with the Linbladian (\ref{eq:LindbladBis}) or its dual (\ref{eq:Ldual}). As we shall explain, it is easy to check that it reproduces the quantum dynamical map (\ref{OQBM:lindblad}) once probe degrees of freedom have been traced out, so that eq.(\ref{eq:dilatBr}) is indeed a dilation of the open QBM dynamical map (\ref{OQBM:lindblad}). 

In the scaling limit tracing out the probe degrees of freedom amounts to compute the vacuum expectation values because in this limit those degrees of freedom are identified with the Fock space and because we initially started with the Fock vacuum. Hence we look at the evolution of 
\begin{eqnarray} \label{eq:Evac}
 \mathbb{E}[A_t]:= \bra{\Omega_\infty} A_t \ket{\Omega_\infty},
 \end{eqnarray}
which is an operator acting on ${\cal H}_{\rm sys}$. We have $d\mathbb{E}[A_t]=\bra{\Omega_\infty} dA_t \ket{\Omega_\infty}$ which can be computed using eq.(\ref{eq:dilatBr}). Now, in a way similar to classical It\^o calculus, the noisy term $i\big[{\cal P}^\dag, A\big]_t\, d\xi_t$ and $i\big[{\cal P},A\big]_t\, d\xi_t^\dag$ have zero expectation, because so does $d\xi_t$ and $d\xi_t^\dag$, and because $i\big[{\cal P}^\dag, A\big]_t$ and $d\xi_t$ are independent in the sense that $i\big[{\cal P}^\dag, A\big]_t$ act on the past Fock subspace ${\cal F}_{0;t}$ whereas $d\xi_t$ acts on ${\cal F}_{t;t+dt}$, so that the expectation of their product factorizes. Hence
\[ d \mathbb{E}[A_t] = \mathbb{E}[{\cal L}_*(A)_t]\, dt.\]
By duality this gives
\[ d\bar \rho_t = {\cal L}(\bar \rho_t)\, dt,\]
for the system reduced density matrix $\bar \rho_t:= {\rm Tr}_{ {\cal F}_{0;\infty}} \big( \mathfrak{U}_t\, \rho_0\otimes\ket{\Omega_\infty}\bra{\Omega_\infty}\, \mathfrak{U}_t^\dag \big),$ with $\rho_0$ the initial system density matrix.
This is equivalent to eq.(\ref{OQBM:lindblad}).

Finally, remark that in case where $N$ is trivial, the previous quantum stochastic differential equation reduces to 
\[ dA_t =  i[P, A]_t\, dQ_t  + \big(i[H,A]-\frac{1}{2}[P,[P,A]]\big)_t\, dt,\]
with $dQ_t=d\xi_t+d\xi_t^\dag$. As is well know, $Q_t$ with the vacuum expectation as measure is a representation of a Brownian motion\footnote{We shall elaborate on this fact below when needed.}, so that open QBM reduces to classical stochastic differential equation, driven by a classical Brownian motion, in absence of internal degrees. A quite similar statement applies if $N$ is non trivial but purely imaginary, $N^\dag=-N$.

\medskip

\underline{\it ``Esquisse d'une preuve":}\\
We now argue that eq.(\ref{eq:dilatBr}) is indeed the only possible quantum stochastic equation with only one pair of quantum noises dilating eq.(\ref{OQBM:lindblad}). This computation is purely algebraic -- it applies, directly or with simple generalizations -- to any Lindblad evolution. The inputs are the structural form of the quantum SDE and the choice of quantum dynamical map, that is the choice of the dual Lindbladian. It is based on two facts:\\
-- The first is that general quantum stochastic equations (driven by one pair of quantum noises) are of the form:
\begin{eqnarray} \label{eq:dq-SDE}
dA_t = {\cal Q}^\dag(A)_t\, d\xi_t + {\cal Q}(A)_t\, d\xi^\dag_t + {\cal L}_*(A)_t\, dt, 
\end{eqnarray}
for any system operator $A$, with ${\cal L}_*$ a dual Lindbladian and ${\cal Q}$, ${\cal Q}^\dag$ linear maps on the operator algebra, and $\xi_t$ and $\xi_t^\dag$ quantum noises\\
-- The second are the quantum It\^o rules: $d\xi_td\xi_t^\dag=(1+\mathfrak{n}_t)\, dt$ and $d\xi_t^\dag d\xi_t=\mathfrak{n}_t\, dt$ for some $\mathfrak{n}$.

Consistency conditions for these two properties demand that ${\cal Q}$, ${\cal Q}^\dag$ are (inner) derivatives satisfying non-commutative Leibnitz rules, i.e.
\[ {\cal Q}(AB) = {\cal Q}(A)\, B + A\, {\cal Q}(B),\]
for any operator $A$ and $B$, whereas the dual Lindbladian ${\cal L}_*$ is a non-commutative analogue of second order differential operator satisfying deformed Leibnitz rules which impose that  ${\cal L}_*(AB)-{\cal L}_*(A)\,B-A\,{\cal L}_*(B)$ is a bilinear form of derivatives on $A$ and $B$, i.e.
\[ {\cal L}_*(AB)={\cal L}_*(A)\,B+A\,{\cal L}_*(B) +\sum_{jk} \ell_{jk}\, {\cal Q}_j(A)\,{\cal Q}_k(B),\]
for some derivatives ${\cal Q}_j$ and coefficients $\ell_{jk}$.
These two facts determine the structure of the quantum SDE and of the quantum noise measure, i.e. the quantum It\^o rules.

Indeed, we can compute the derivative of a product $AB$ in two different ways: either using directly the previous equation (\ref{eq:dq-SDE}) but for $AB$, or expanding the derivative $d(AB)$ using quantum It\^o rules. The first way yields (we drop the index $t$ for simplicity)
\[ d(AB) = {\cal Q}^\dag(AB)\, d\xi_t + {\cal Q}(AB)\, d\xi^\dag_t + {\cal L}_*(AB)\, dt, \]
The second yields
\[ d(AB) = (dA)\, B + A\, (dB) +  \big({\cal Q}(A)\,\mathfrak{n}\,{\cal Q}^\dag(B)  + {\cal Q}^\dag(A)(1+\mathfrak{n}) {\cal Q}(B)\big)\, dt .\]
Consistency of these two computations implies that ${\cal Q}(A)$ and ${\cal Q}^\dag(A)$ have to be derivatives (by looking at the terms proportional to the noise), and that ${\cal L}_*(AB)$ has to satisfy (by looking at the drift term)
\begin{eqnarray} \label{eq:Lquadra}
 {\cal L}_*(AB)={\cal L}_*(A)\,B+A\,{\cal L}_*(B)+  {\cal Q}(A)\,\mathfrak{n}\,{\cal Q}^\dag(B)  + {\cal Q}^\dag(A)(1+\mathfrak{n}) {\cal Q}(B) .
 \end{eqnarray}
Now given a quantum dynamical map and its dual Lindbladian, one can compute ${\cal L}_*(AB)-{\cal L}_*(A)\,B-A\,{\cal L}_*(B)$ and identify what the derivatives ${\cal Q}$, ${\cal Q}^\dag$ and the coefficients $\mathfrak{n}$ are. In the simple case\footnote{But extension to general dual Lindbladian is easy, as is the generalization with time dependent coefficient $\mathfrak{n}_t$.} of the dual Lindbladian ${\cal L}_*$ defined in eq.(\ref{eq:Ldual}) we have:
\[ {\cal L}_*(AB)={\cal L}_*(A)\,B+A\,{\cal L}_*(B)- [{\cal P}^\dag,A][{\cal P},B].\]
 From this we learn that $\mathfrak{n}=0$ and (up to an irrelevant phase which may be absorbed in the definition of the quantum noise)
\[ {\cal Q}(A)= i\big[{\cal P},A\big],\quad {\cal Q}^\dag(A)= i\big[{\cal P}^\dag,A\big] \]
Hence, the data of the dual Lindbladian determines the quantum SDE. This is quite the analogue to the fact that a continuous time Markov process is completely determined by the second order differential operator it generates. \cqfd

\subsection{From quantum noise to quantum trajectory}

The aim is here to derive the equations governing (tilted) quantum trajectories from the quantum SDE. This will provides another check of the validity of the quantum SDE as the correct dilation. Recall that the Fock space ${\cal F}_{0;\infty}$ is a modeling of the series of all probes in the scaling limit. So, to go from quantum SDE to quantum trajectory we have to implement measurements of some Fock observables continuously in time. 

We choose to measure continuously in time the observable, $\Theta_t^v:=\bar v\,\xi_t+v\,\xi_t^\dag$ with $v\bar v=1$, linear in the quantum noise. That is, during each time interval $[t,t+dt)$, we assume measuring the observable $d\Theta^v:=\bar v\, d\xi_t+v\, d\xi_t^\dag$, after the interaction between the system and the Fock space has taken place, and we denote by $dY^v_t$ the random output of these measurements\footnote{Of course the spectrum of $\Theta^v_t$ is continuous, so that what we are really talking about it is the measurement output to be in a given interval. Notice also that we can generalized the present analysis with $v$ time dependent.}:
\[ \mathrm{observables}\quad: d\Theta^v_t=\bar v\, d\xi_t+v\, d\xi_t^\dag\ \to\ \mathrm{measurement~ outputs}\quad dY^v_t.\]
The observable $d\Theta^v_t$ acts on ${\cal F}_{t;t+dt}$ and leaves untouched states in the past Fock subspace ${\cal F}_{0;t}$ and in the forward Fock subspace ${\cal F}_{t+dt;\infty}$. We shall later identify the series of discrete probe observables $\Theta^v_t$ corresponds to. Let us assume as before that the Fock state is initially in its vacuum. After having performed all measurements up to time $t$ and got outputs $Y^v_t$, the Fock state is projected on a (random) state $\ket{Y^v_t}\otimes\ket{\Omega_{t;\infty}}$ with $\ket{Y^v_t}\in {\cal F}_{0;t}$ depending on the output measurements. Since interactions after time $t$ leave invariant all past states, once projected the states $\ket{Y^v_t}$ remains unchanged at any later time and we can forget about them. 

Let $\hat \rho_t$ be the density matrix of the system (walker $+$ gyroscope) at time $t$ after having performed all probe  measurements up to time $t$. This is the quantum trajectory system density matrix. Forgetting about the past Fock states which are frozen, the total density matrix of the system plus the forward Fock space ${\cal F}_{t;\infty}$ before interaction with the probes is:
\[ \hat \rho_t \otimes \ket{\Omega_{[t,\infty)}} \bra{\Omega_{[t,\infty)}}.\]
During time $t$ and $t+dt$ interaction takes place between the system and the probes in the Fock sub-space ${\cal F}_{t;t+dt}$. Since $\ket{\Omega_{[t,\infty)}} =\ket{\Omega_{[t,t+dt)}} \otimes \ket{\Omega_{[t+dt,\infty)}}$ and since the interaction between $t$ and $t+dt$ only acts non-trivially on ${\cal F}_{t;t+dt}$, the state $\ket{\Omega_{[t+dt,\infty)}}$ is un-affected by the interaction and only $\ket{\Omega_{[t,t+dt)}}$ gets modified and entangled with the system.
It evolves according to the quantum flow (\ref{eq:Udilat}), so that at time $t+dt$ the total density matrix is
\[ (\mathbb{I}+d\mathfrak{U}_t\,\mathfrak{U}_t^{-1})\,\hat \rho_t \otimes \ket{\Omega_{[t,\infty)}} \bra{\Omega_{[t,\infty)}}\, (\mathbb{I}+d\mathfrak{U}_t\, \mathfrak{U}_t^{-1})^\dag.\]
Because $\ket{\Omega_{[t+dt,\infty)}}$ is unaffected by the interaction this is of the form $\rho^{\rm tot}_{t;t+dt}\otimes \ket{\Omega_{[t+dt,\infty)}} \bra{\Omega_{[t+dt,\infty)}}$ with
\begin{eqnarray}\label{eq:rho-tot}
 \rho^{\rm tot}_{t;t+dt} :=  (\mathbb{I}+d\mathfrak{U}_t\, \mathfrak{U}_t^{-1})\,\hat \rho_t \otimes \ket{\Omega_{[t,t+dt)}} \bra{\Omega_{[t,t+dt)}}\, (\mathbb{I}+d\mathfrak{U}_t\,\mathfrak{U}_t^{-1})^\dag   
 \end{eqnarray}
After such evolution, a measurement of $d\Theta^v_t$ is performed, giving $dY^v_t$ as output and projecting the Fock state on $\ket{dY^v_t}\in{\cal F}_{t;t+dt}$ with $d\Theta^v_t\ket{dY^v_t}=(dY^v_t)\, \ket{dY^v_t}$. As a consequence $\rho^{\rm tot}_{t;t+dt}$ is projected on
\begin{eqnarray} \label{eq:rho-Y}
 \hat \rho_{t+dt} \otimes \ket{dY^v_t}\bra{dY^v_t}.
 \end{eqnarray}
Recall that this is the component of the state in ${\cal H}_{\rm sys}\otimes{\cal F}_{t;t+dt}$, the past Fock state have been projected on states depending on the previous output measurements and the forward states are yet untouched by the interaction.

The map $\hat \rho_t \to \hat \rho_{t+dt}$ defines the flow for the quantum trajectory associated to the continuous measurement of $d\Theta^v_t$. It is random because it depends on the measurement outputs $dY^v_t$. Recall that the probability measure on $dY^v_t$ is that induced by the von Neumann rules for quantum measurements. We have to describe both the random evolution of the outputs and of the system density matrix. Of course these are coupled evolutions. 

We shall first prove that:\\
(i) $dY^v_t$ is Gaussian with mean $i{\rm Tr}_{{\cal H}_{\rm sys}}\big(\hat \rho_t {\cal P}^\dag v-\bar v {\cal P}\,\hat \rho_t\big)\,dt$ and covariance $dt$.\\
(ii) The integrated random variable $Y^v_t:=\int_0^t dY^v_s$ satisfies the classical SDE:
\begin{eqnarray}\label{eq:SDE-dY}
 dY^v_t = {\cal U}_v(\hat\rho_t)\,dt + dB_t,\quad 
 {\cal U}_v(\hat\rho_t):= i\, {\rm Tr}_{{\cal H}_{\rm sys}}\big(\hat \rho_t\, ( {\cal P}^\dag v - \bar v\, {\cal P})\big)
 \end{eqnarray}
with $B_t$ a normalized Brownian motion.
This provides a description of the statistics of the measurement outputs. It is entangled with that for the system density matrix. 

We shall then prove that:\\
(iii) The system density matrix $\hat\rho_t$ for (tilted) quantum trajectory satisfies the classical SDE:
\begin{eqnarray}\label{eq:SDE-drho}
d\hat\rho_t= {\cal L}(\hat \rho_t)\, dt + {\cal D}_v(\hat\rho_t)\, dB_t,
 \end{eqnarray}
with ${\cal L}(\bar \rho_t) :=- i\big[{\cal H},\bar\rho_t\big] + \big( {\cal P}\,\bar\rho_t\,{\cal P}^\dag - \frac{1}{2}({\cal P}^\dag{\cal P}\, \bar\rho_t + \bar \rho_t\, {\cal P}^\dag{\cal P})\big)$, as above, eq.(\ref {eq:LindbladBis}), and 
\[ {\cal D}_v(\hat\rho_t) := i\big(\hat \rho_t {\cal P}^\dag v-\bar v\, {\cal P}\,\hat \rho_t\big)-\hat\rho_t\, {\cal U}_v(\hat\rho_t).\] 
Recall that ${\cal P}= P+iN$.
 
These quantum trajectories coincide with those found by taking the scaling limit of tilted quantum trajectories. Namely, writing them in momentum space gives eq.(\ref{eq:Trhopq}) and eq.(\ref{eq:Ytraj}).
We thus learn that measuring $d\Theta^v_t$ in the continuous theory corresponds to measure $\sigma_u={\bf u}.\sigma$ in the discrete setting with $v$ related to the direction ${\bf u}$ by formula (\ref{eq:vu}). Note also that tilting the quantum trajectories (using $v\not= 1$) just amounts to replace ${\cal P}$ by $\bar v\, {\cal P}$, as expected. In particular, this leaves the Lindbladian ${\cal L}$ invariant. Hence, averaging over all realizations of (tilted) quantum trajectories yields the quantum dynamical map of the open QBM, independently of the tilting parameter $v$. Actually this had to be true as averaging over all trajectory realizations amounts to average over all possible probe measurement outputs and hence and it amounts to trace over all probe degrees of freedom, independently of the choice of the measured observable.
 
\medskip 

\underline{\it ``Esquisse d'une preuve" (I):}\\
The proof of eq.(\ref{eq:SDE-dY}) relies in part on the relation between quantum noises (i.e. bosonic free field) and Brownian motion. For any uni-modular complex number $v$, $v\bar v=1$, all vacuum expectations of $\Theta^v_t:=v\xi_t^\dag+\bar v \xi_t$ coincides with those of the Brownian motion\footnote{The connection is actually deeper but requires going into decomposition of Brownian functional in chaos.}. Indeed, $\Theta^v_t$ is a real Gaussian field as is the Brownian motion, and canonical commutation relations imply $\bra{\Omega_\infty}\Theta^v_{t} \Theta^v_{s}\ket{\Omega_\infty}={\rm min}(t,s)$ so that its covariance coincides with that of the Brownian motion. A direct application of Wick's theorem yields the identification of the multi-point vacuum expectation values of $\Theta^v_t$ with those of the Brownian motion \cite{BrowQfield}. 

The difference between this well known fact and the present setting is that expectations are here computed using the state (\ref{eq:rho-tot}) which is not the vacuum state but a deformation of it induced by the interaction with the system.

Statistics of the measurement outputs $dY^v_t$ is coded into von Neumann rules for quantum measurements. Hence,  
\begin{eqnarray*}
 \mathbb{E}[dY^v_t]&=& {\rm Tr}_{{\cal H}_{\rm sys}\otimes{\cal F}_{t;t+dt}}\big(d\Theta_t^v\rho^{\rm tot}_{t;t+dt}\big),\\
\mathbb{E}[(dY^v_t)^2]&=&{\rm Tr}_{{\cal H}_{\rm sys}\otimes{\cal F}_{t;t+dt}}\big((d\Theta_t^v)^2\rho^{\rm tot}_{t;t+dt}\big).
\end{eqnarray*}
So we compute, keeping only terms of order $dt$ at most\footnote{We present the computation using a language from field theory, e.g. vacuum expectation value, but it can also be formulated using quantum It\^o calculus.}. For the mean we get:
\begin{eqnarray*}
\mathbb{E}[dY^v_t]&=& {\rm Tr}_{{\cal H}_{\rm sys}} \big( \hat\rho_t   \bra{\Omega_{[t,t+dt)}} 
(\mathbb{I}+d\mathfrak{U}_t\,\mathfrak{U}_t^{-1})^\dag \, d\Theta_t^v\,(\mathbb{I}+d\mathfrak{U}_t\,\mathfrak{U}_t^{-1}) \ket{\Omega_{[t,t+dt)}}\big)\\
&=& {\rm Tr}_{{\cal H}_{\rm sys}} \big( \hat\rho_t   \bra{\Omega_{[t,t+dt)}}  \Big(
(i{\cal P}^\dag\, d\xi_t +\cdots)\, d\Theta_t^v + d\Theta_t^v\,(- i{\cal P}\, d\xi_t^\dag +\cdots)\Big)\ket{\Omega_{[t,t+dt)}}\big),
\end{eqnarray*}
where in the last equation we did not keep terms either annihilated by the vacuum state or of order $o(dt)$. Evaluating the last expectation we obtain
\[ \mathbb{E}[dY^v_t]= {\rm Tr}_{{\cal H}_{\rm sys}} \big(\hat\rho_t \big(i v{\cal P}^\dag-i\bar v{\cal P}\big)\big)\, dt = {\cal U}_v(\hat\rho_t)\, dt.\]
For the covariance, we get:
\begin{eqnarray*}
\mathbb{E}[(dY^v_t)^2]&=& {\rm Tr}_{{\cal H}_{\rm sys}} \big( \hat\rho_t   \bra{\Omega_{[t,t+dt)}} 
(\mathbb{I}+d\mathfrak{U}_t\,\mathfrak{U}_t^{-1})^\dag \, (d\Theta_t^v)^2\,(\mathbb{I}+d\mathfrak{U}_t\,\mathfrak{U}_t^{-1}) \ket{\Omega_{[t,t+dt)}}\big)\\
&=& {\rm Tr}_{{\cal H}_{\rm sys}} \big( \hat\rho_t   \bra{\Omega_{[t,t+dt)}} 
\, (d\Theta_t^v)^2\, \ket{\Omega_{[t,t+dt)}}\big) = (v\bar v)\, dt.
\end{eqnarray*}
It is clear from this last equation which reduces the computation to the un-deformed vacuum expectation that the statistics of $dY^v_t$ is Gaussian and coincides with that of a drifted Brownian motion. Hence, proving\footnote{A la physicist...} eq.(\ref{eq:SDE-dY}). Notice that deforming the Fock vacuum by the time dependent evolution operator $\mathfrak{U}_t$ provides an algebraic  analogue of Girsanov's theorem \cite{probas}.
\cqfd
\medskip

\underline{\it ``Esquisse d'une preuve" (II):}\\
We now derive the equation for (tilted) quantum trajectories starting from the quantum SDE plus measurements of $d\Theta^v_t$ on the probes. If, while the system plus the Fock space are in the state $\rho^{\rm tot}_{t;t+dt}$, one measures $d\Theta^v_t$ and get $dY^v_t$ as output then, by quantum mechanics rules, the system state is projected on
\[ \hat \rho_{t+dt}:= \frac{ \bra{dY^u_t} \rho^{\rm tot}_{t;t+dt} \ket{dY^v_t}}{\Pi_{t;t+dt}(dY^v_t)},\]
with
\[ \Pi_{t;t+dt}(dY^v_t) =  {\rm Tr}_{{\cal H}_{\rm sys}} \big(\bra{dY^v_t} \rho^{\rm tot}_{t;t+dt} \ket{dY^v_t}\big),\]
the probability for the output $dY^v_t$ to occur.

Now, since $\rho^{\rm tot}_{t;t+dt}=\hat \rho_t \otimes \ket{\Omega_{[t,t+dt)}} \bra{\Omega_{[t,t+dt)}}$, we may write,
\begin{eqnarray*}
\bra{dY^v_t} \rho^{\rm tot}_{t;t+dt} \ket{dY^v_t} =
\bra{dY^v_t} \big(\mathbb{I}+d\mathfrak{U}_t\,\mathfrak{U}_t^{-1}\big) \ket{\Omega_{[t,t+dt)}}\,\hat \rho_t \, \bra{\Omega_{[t,t+dt)}}\big(\mathbb{I}+d\mathfrak{U}_t\,\mathfrak{U}_t^{-1}\big)^\dag  \ket{dY^v_t}.
\end{eqnarray*}
We have to compute $\bra{dY^v_t} (\mathbb{I}+d\mathfrak{U}_t\,\mathfrak{U}_t^{-1}) \ket{\Omega_{[t,t+dt)}}$, that is
\[ \bra{dY^v_t} \Big(\mathbb{I} -i{\cal P}^\dag\, d\xi_t - i{\cal P}\, d\xi_t^\dag - (i{\cal H} + \frac{1}{2}{\cal P}^\dag{\cal P})dt\Big) \ket{\Omega_{[t,t+dt)}}.\]
Since $\bra{dY^v_t}d\xi_t \ket{\Omega_{[t,t+dt)}}=0$, this only requires computing $\bra{dY^v_t}d\xi_t^\dag \ket{\Omega_{[t,t+dt)}}$ and the overlap $\bra{dY^v_t}\Omega_{[t,t+dt)}\rangle$. This is an exercise with canonical operators. The state $\ket{dY^v_t}$ is by definition an eigen-state of $d\Theta^v_t=\bar v d\xi_t + v d\xi_t^\dag$ while $\ket{\Omega_{[t,t+dt)}}$ is the vacuum for $d\xi_t$. However, going from the canonical pair $(d\xi_t+d\xi_t^\dag)/\sqrt{2}$ and $i(d\xi_t-d\xi_t^\dag)/\sqrt{2}$ to the canonical pair $(\bar v d\xi_t+v d\xi_t^\dag)/\sqrt{2}$ and $i(\bar v d\xi_t-v d\xi_t^\dag)/\sqrt{2}$ is simply a rotation. Hence\footnote{Or alternatively: $(dY^v_t)\, \bra{dY^v_t}\Omega_{[t,t+dt)}\rangle= \bra{dY^v_t}d\Theta^v_t \ket{\Omega_{[t,t+dt)}}=v\, \bra{dY^v_t}d\xi_t^\dag \ket{\Omega_{[t,t+dt)}}$.}, 
\[ \bra{dY^v_t}d\xi_t^\dag \ket{\Omega_{[t,t+dt)}} = \bar v\, (dY^v_t)\, \bra{dY^v_t}\Omega_{[t,t+dt)}\rangle.\]
This implies
\[ \bra{dY^v_t} \big(\mathbb{I}+d\mathfrak{U}_t\,\mathfrak{U}_t^{-1}\big) \ket{\Omega_{[t,t+dt)}}
= \bra{dY^v_t}\Omega_{[t,t+dt)}\rangle\, \Big( \mathbb{I} - i \bar v (dY^v_t)\, {\cal P} - (i{\cal H} + \frac{1}{2}{\cal P}^\dag{\cal P})dt\Big). \]
We now have all ingredients to compute $\hat \rho_{t+dt}$: we just have to gather them. Hopefully, but coherently, the overlap $|\bra{dY^v_t}\Omega_{[t,t+dt)}\rangle|^2$ factorises and simplifies when evaluating both $\bra{dY^v_t} \rho^{\rm tot}_{t;t+dt} \ket{dY^v_t}$ and its trace $\Pi_{t;t+dt}(dY^v_t)$. So we may write
\begin{eqnarray*}
\hat\rho_{t+dt} &=& \frac{\Big( \mathbb{I} - i \bar v (dY^u_t)\, {\cal P} - (i{\cal H} +\frac{1}{2}{\cal P}^\dag{\cal P})dt\Big)\,\hat\rho_t\, \Big( \mathbb{I} + i  v (dY^v_t)\, {\cal P}^\dag + (i{\cal H} - \frac{1}{2}{\cal P}^\dag{\cal P})dt\Big)}{{\rm Tr}_{{\cal H}_{\rm sys}} \big(\cdots\big)}\\  &~& \\
&=& \hat \rho_t + {\cal L}(\hat\rho_t)\, dt + {\cal D}_v(\hat\rho_t)\, (dY^v_t) - {\cal D}_v(\hat\rho_t)\, {\cal U}_v(\hat\rho_t)\,dt,
\end{eqnarray*}
where we used $(dY^v_t)^2=dt$, and where ${\rm Tr}_{{\cal H}_{\rm sys}} (\cdots)$ refers to the trace of the numerator, and ${\cal L}$, ${\cal D}_v$ and ${\cal U}_v$ have been defined above in eqs.(\ref{eq:SDE-drho}). Finally, using $dY^u_t= {\cal U}_v(\hat\rho_t)\, dt +dB_t$ yields eq.(\ref{eq:SDE-drho}) for $\hat\rho_t$. 
\cqfd

\section{Generalizations}

We generalize open QBM with in-homogeneous transition matrices, in higher dimensions, and with probes prepared not in pure states but in mixed states (say at finite temperature). Of course these three generalizations can be mixed together, but we leave this to the dedicated readers. 

As for the homogeneous case, we can develop the theory along (at least) three interconnected lines: quantum trajectories (assuming that some probe observable is measured after each iteration), quantum dynamical maps for the reduced system states (assuming that we do not measure the probes or do not keep track of their measurements so that we trace over the probe degrees of freedom), and quantum stochastic equations (assuming that we neither trace over the probe degrees of freedom nor we measure probe observables). And all these can be done either in the discrete or continuous formulation.

\subsection{Inhomogeneous landscape}

Let us first start with the in-homogeneous 1D case. The framework is the same as for the homogeneous case except that the transition matrices $B_\pm(x)$ are position dependent. The system Hilbert space is ${\cal H}_{\rm sys}:={\cal H}_c\otimes{\cal H}_z$ and the probe Hilbert space ${\cal H}_p$. We assume -- for a while -- that the probes are all prepared in the pure state $\ket{\phi}_p$. In the discrete setting of open QRW, the system-probe interaction is assumed to be such that its action on states $ \ket{\chi}_c\otimes\ket{x}_z\otimes\ket{\phi}_p$ gives
\[\boxed{(B_+(x)\ket{\chi}_c)\otimes\ket{x+\delta}_z\otimes\ket{+}_p +(B_-(x)\ket{\chi}_c)\otimes\ket{x-\delta}_z\otimes\ket{-}_p,}\]
for any $\ket{\chi}_c\in{\cal H}_c$. As in the homogeneous case, $B_\pm(x)$ depends linearly on $\ket{\phi}_p$. Notice that we use the transition matrices evaluated at the starting position. Unitarity constraint then reads
\[B_+^\dag(x) B_+(x)+ B_-^\dag(x) B_-(x)=\mathbb{I},\]
all matrices $B_\pm$ being evaluated at the same point. In the scaling limit, we assume the usual expansion for the transition matrices but with $H$ and $N$ position dependent, i.e.
\[B_\pm(x)=\frac{1}{\sqrt{2}}\big[\mathbb{I}\pm\sqrt{\epsilon} N(x) - \epsilon\big(iH(x)\pm M(x) +\frac{1}{2} N^\dag(x)N(x)\big)+O(\epsilon^{3/2})\big].\]
We shall describe quantum trajectories, quantum dynamical maps and quantum stochastic equations. As above, all these are going to be independent of $M$.

\underline{\it Quantum trajectories.} 
Let us first describe quantum trajectories. For simplicity we assume measuring the probe observables diagonal in the basis $\ket{\pm}_p$ so that we are actually describing simple quantum trajectories (but generalization to tilted quantum trajectories is clear), and we assume that the density matrix is initially diagonal and localized in space (and it remains so). 
If at the $n$-th step, the density matrix is $\rho_n\otimes \ket{X_n}_z\bra{X_n}$, then at the $n+1$-th step it is 
\[ \rho_{n+1}^{(\pm)}\otimes \ket{X_n\pm \delta}_z\bra{X_n\pm\delta },\quad 
\rho_{n+1}^{(\pm)}=\frac{B_\pm(X_n)\rho_nB^\dag_\pm(X_n)}{p_\pm(n)},\]
with probability 
\[ p_\pm(n)= {\rm Tr}_{ {\cal H}_{\rm sys}}\big( B_\pm(X_n)\rho_nB^\dag_\pm(X_n)\big).\]
These probabilities sum to one, $p_+(n)+p_-(n)=1$, thanks to the relation $B_+^\dag(x) B_+(x)+ B_-^\dag(x) B_-(x)=\mathbb{I}$, and for this to be true it was important to evaluate the transition matrices at the starting position.

The scaling limit can be taken as before, and we get the coupled SDE's:
\begin{eqnarray}\label{eq:drhoX}
\boxed{\begin{array}{rcl}
d\rho_t&=&\Big( -i\big[H(X_t),\rho_t\big]+L_{N(X_t)}(\rho_t)\Big)dt + D_{N(X_t)}(\rho_t)\, dB_t, \\ \\
dX_t&=&U_{N(X_t)}(\rho_t)\,dt + \, dB_t.
\end{array} }
\end{eqnarray}

For $\rho_t$ trivial, i.e. ${\cal H}_c=\mathbb{C}$ and $N$ is a $\mathbb{C}$-number, we recover classical SDE's with a drift: \[ dX_t=U(X_t)dt + dB_t,\quad U(x)=2\,\Re( N(x) ).\]
Hence, eq.(\ref{eq:drhoX}) provides a natural quantum generalization of a noisy particle moving in a non-uniform landscape.It\^o

\underline{\it Quantum dynamical map.} 
The associated discrete quantum dynamical map, valid for the mean density matrix $\bar\rho$, is again
\[ \bar\rho_n \to \bar\rho_{n+1}=(\mathbb{I}\otimes e^{-i\delta P})\,B_+\, \bar\rho_n\, B_+^\dag(\mathbb{I}\otimes e^{+i\delta P}) +
(\mathbb{I}\otimes e^{+i\delta P})\, B_-\,\bar\rho_n\, B_-^\dag(\mathbb{I}\otimes e^{-i\delta P}),\]
but with a careful ordering of the operators: one first acts with $B_\pm$ and then with the translation operators $e^{\mp i\delta P}$. Here the operators $B_\pm$ act on both component of ${\cal H}_c\otimes {\cal H}_z$ according to $B_+(\ket{\chi}_c\otimes\ket{x}_z):= (B_\pm(x) \ket{\chi}_c)\otimes\ket{x}_z$. They do not commute with $P$.
For density matrix diagonal in position space,  $\bar \rho_n=\sum_x \bar\rho_n(x) \otimes \ket{x}_z\bra{x}$, this map reads:
\[ \bar \rho_n(x)\to \bar\rho_{n+1}(x)= B_-(x+\delta)\bar\rho_n(x+\delta) B^\dag_-(x+\delta)+ B_+(x-\delta)\bar\rho_n(x-\delta) B^\dag_+(x-\delta),\]
and this can be extended to more general density matrix by linearity. Again, notice that the rule is that we act with the transition matrices $B_\pm(x)$ evaluated at the starting position (not at the final position), and this is compatible with the relation $B_+^\dag(x)B_+(x)+B_-^\dag(x)B_-(x)=\mathbb{I}$.

The scaling limit, $\epsilon=\delta^2\to 0$, $n\to \infty$ at $t=n\epsilon$ fixed, can be taken as before. For diagonal in space reduced density matrix, $\bar\rho_t:=\int dx\, \bar\rho_t(x)\otimes \ket{x}_z\bra{x}$, it yields
\[\boxed{ \partial_t \bar \rho_t(x)= - i[H(x),\bar\rho_t(x)]+ \frac{1}{2} \partial_x^2\bar\rho_t(x) - \partial_x\big( N(x)\bar\rho_t(x)+\bar\rho_t(x) N^\dag(x)\big) + L_{N(x)}(\bar\rho_t(x)),}\]
with $L_{N(x)}$ as above.
This can equivalently be written as a Lindblad equation in ${\cal H}_c\otimes L^2(\mathbb{R})$,
\begin{eqnarray} \label{eq:Lind1D}
\boxed{\partial_t \bar\rho_t=-i\big[H,\bar\rho_t\big] -\frac{1}{2} \big[P,[P,\bar\rho_t]\big] - i\big[P, N\bar\rho_t+\bar\rho_t N^\dag\big] + L_{N}\big(\bar\rho_t\big),}
\end{eqnarray}
with $N$ the operator acting on ${\cal H}_c\otimes L^2(\mathbb{R})$ as $N\, (\ket{\chi}_c\otimes \ket{x})= (N(x)\ket{\chi}_c)\otimes \ket{x}$, and similarly for $H$. These equations preserve the normalization condition $\int dx\, {\rm Tr}_{{\cal H}_c}(\bar\rho_t(x))=1$, or ${\rm Tr}_{{\cal H}_{\rm sys}}(\bar \rho_t)= 1$, as they should. They also code for the evolution of non-diagonal in space density matrix.
\medskip

\underline{\it Quantum noise.}
Equation (\ref{eq:Lind1D}) can explicitly be written in a Lindblad form, namely $\partial_t \bar\rho_t={\cal L}\big(\bar\rho_t\big)$ with
\begin{eqnarray*}
{\cal L}(\bar \rho_t)= -i\big[{\cal H},\bar\rho_t\big] + \big( {\cal P}\,\bar\rho_t\,{\cal P}^\dag - \frac{1}{2}({\cal P}^\dag{\cal P}\, \bar\rho_t + \bar \rho_t\, {\cal P}^\dag{\cal P})\big),
\end{eqnarray*}
with ${\cal P}:= P+iN$ and ${\cal H}:= H+\frac{1}{2}(PN+N^\dag P)$. The only difference with eq.(\ref{eq:LindbladBis}) is that now the operator $P$ and $N$ do not commute. As a consequence one may repeat the algebraic construction justifying the quantum stochastic equation. For any system observable $A$, it again reads
\begin{eqnarray} \label{eq: SDE_1D}
 dA_t = i\big[{\cal P}^\dag, A\big]_t\, d\xi_t + i\big[{\cal P},A\big]_t\, d\xi_t^\dag + {\cal L}_*(A)_t\, dt, 
 \end{eqnarray}
with ${\cal P}=P+iN$ and the same dual Lindbladian as in eq.(\ref{eq:Ldual}),
\[{\cal L}_*(A):=  i[{\cal H},A]+  {\cal P}^\dag A {\cal P} - \frac{1}{2} ({\cal P}^\dag {\cal P} A + A {\cal P}^\dag {\cal P}),\]
 but with careful ordering of the non-commutative operators. Here the quantum It\^o rules are again $d\xi_td\xi_t^\dag=dt$ and $d\xi_t^\dag d\xi_t=0$. The associated unitary flow is generated by $\mathfrak{U}_t$ with
\[d\mathfrak{U}_t\,\mathfrak{U}_t^{-1}=-i{\cal P}^\dag\, d\xi_t - i{\cal P}\, d\xi_t^\dag - \big(i{\cal H} +\frac{1}{2}{\cal P}^\dag{\cal P}\big)\, dt,\]
with $A_t=\mathfrak{U}_t^\dag\, A\, \mathfrak{U}_t$.

Let us look at this quantum flow in the simplest, but interesting, case in which the internal space is trivial ${\cal H}_c=\mathbb{C}$ but with $N$ and $H$ position dependent. This means that ${\cal H}_{\rm sys}=L^2(\mathbb{R})$ and $N$ and $H$ are operators acting on state $\ket{x}_z$ by multiplication by $N(x)$ (complex) and $H(x)$ (real). Let us choose $A$ to be an operator acting diagonally on $\ket{x}_z$ by multiplication by some function $A(x)$, i.e. $A\ket{x}_z:=A(x) \ket{x}_z$. By the stochastic equation (\ref{eq: SDE_1D}), it evolves into an operator $A_t$ acting on $L^2(\mathbb{R})\otimes{\cal F}_{0;\infty}$. Because they are all diagonal in the position basis, $A$ commutes with $N$ and $H$, and eq.(\ref{eq: SDE_1D}) becomes
\[ dA_t = (\partial A)_t\, dQ_t + \big( \frac{1}{2} (\partial^2 A)_t + (U(\partial A))_t\big) dt,\]
with $\partial A$ the operator acting by multiplication by $A'(x)$, the derivative of $A(x)$, and  $U=N+N^\dag$ the operator acting by multiplication by $U(x)=2\Re\, N(x)$ and $dQ_t=d\xi_t+d\xi^\dag_t$. Since $Q_t$ is a representation of the Brownian motion in the Fock space, $dQ^2_t=dt$, the above equation is solved by
\[ A_t = A(X_t),\quad dX_t= U(X_t)\, dt + dQ_t,\quad X_{t=0}=x.\]
Of course in this case this SDE coincides with that for quantum trajectories. This shows that quantum stochastic equations but with trivial internal Hilbert space reproduce all classical stochastic differential equations. However, even in this simple case, one gets more general flows if one considers operators on $L^2(\mathbb{R})$ not commuting with $N$ and $H$, that is, operators not diagonal in the position basis such as those involving the momentum operator.

\subsection{Higher dimensions}

We now look at the generalization to higher dimensions. The construction is quite similar, more indices and transition matrices $B_\pm^\mu(x)$ coding for a move starting at $x$ and up/down in the $\mu$-direction.

In the scaling limit the (simple) quantum trajectories are
\begin{eqnarray}\label{eq:drhoXmu}
\boxed{\begin{array}{rcl}
d\rho_t&=&\big(- i[H,\rho_t]+L_{N}(\rho_t)\big)dt + D_\mu(\rho_t)\, dB_t^\mu, \\ \\
dX_t^\mu&=&U^\mu(\rho_t)\,dt + \, dB_t^\mu,
\end{array} }
\end{eqnarray}
with $dB^\mu_t\, dB^\nu_t=G^{\mu\nu}\, dt$ (we take $G^{\mu\nu}$ constant and $G_{\mu\nu}G^{\nu\sigma}=\delta^\sigma_\mu$) and
\begin{eqnarray*}
 G^{\mu\nu}D_\nu(\rho) &=& (N^\mu\rho+\rho\; N^{\mu\dag}) -\rho\, U^\mu(\rho),\\
U^\mu(\rho)&=& {\rm Tr}_{{\cal H}_c}(N^\mu\rho+\rho\; N^{\mu\dag}),\\
L_{N}(\rho)&=&G_{\mu\nu}\big(N^\mu\rho\; N^{\nu\dag}- \frac{1}{2}(N^{\nu\dag}N^\mu\rho+\rho\; N^{\nu\dag}N^\mu)\big).
\end{eqnarray*}

The Lindblad equation for the reduced density matrix $\bar\rho_t=\int dx\, \bar\rho_t(x)\otimes \ket{x}_z\bra{x}$ is:
\[ \partial_t \bar\rho_t(x)=- i[H(x),\bar\rho_t(x)] + \frac{1}{2} G^{\mu\nu}\partial_\mu\partial_\nu\bar\rho_t(x) - \partial_\mu\big( (N^\mu\bar\rho_t)(x)+(\bar\rho_t N^{\mu\dag})(x)\big) + L_{N(x)}(\bar\rho_t(x)),\]
or equivalently
\begin{eqnarray}\label{eq:lind_DD}
\boxed{\partial_t \bar\rho_t=-i\big[H,\bar\rho_t\big] -\frac{1}{2} G^{\mu\nu}\big[P_\mu,[P_\nu,\bar\rho_t]\big] - i\big[P_\mu, N^\mu\bar\rho_t+\bar\rho_t N^{\mu\dag}\big] + L_{N}(\bar\rho_t).}
\end{eqnarray}
with $P_\mu:=-i{\partial}/{\partial x^\mu}$, the translation operator in the $\mu$-direction, and $N^\mu$ operator acting diagonally on position state with $N^\mu(\ket{\chi}_c\otimes\ket{x}_z)=(N^\mu(x)\ket{\chi}_c)\otimes\ket{x}_z$. Alternatively the Lindblad operator in eq.(\ref{eq:lind_DD}) can be written as:
\begin{eqnarray*}
{\cal L}_D(\bar\rho_t): = -i\big[{\cal H}_D,\bar\rho_t\big] + G^{\mu\nu}\Big( {\cal P}_\mu\,\bar\rho_t\,{\cal P}_\nu^\dag - \frac{1}{2}({\cal P}^\dag_\nu{\cal P}_\mu\, \bar\rho_t + \bar \rho_t\, {\cal P}^\dag_\nu{\cal P}_\mu)\big),
\end{eqnarray*}
with ${\cal P}_\mu:= P_\mu+iG_{\mu\nu}N^\nu$ and ${\cal H}_D:= H+\frac{1}{2}(P_\mu N^\mu+N^{\mu\dag} P_\mu)$.

The associated stochastic differential equation is, for any system operator $A$,
\begin{eqnarray*} 
 dA_t = i\big[{\cal P}_\mu^\dag, A\big]_t\, d\xi^\mu_t + i\big[{\cal P}_\mu,A\big]_t\, d\xi_t^{\mu\dag} + {\cal L}_D^*(A)_t\, dt, 
 \end{eqnarray*}
with ${\cal L}_D^*$ dual to ${\cal L}_D$, and quantum noises $d\xi^\mu_t d\xi_t^{\nu\dag}\, =G^{\mu\nu}\, dt$ and $d\xi_t^{\mu\dag}\, d\xi^\nu_t =0$.

\subsection{Finite temperature}

Let us now briefly extend the previous study to the case in which the probes are prepared not in pure states but in mixed states. Semi classicality arises as the high temperature limit. To change gears, we choose to start from quantum SDE and work directly in the continuum. 

The fact that the probes are in mixed states modifies the measure on the quantum noises $\xi_t$ and $\xi_t^\dag$. That is, it changes the way to compute expectations of observables acting on the Fock space ${\cal F}_{0;\infty}$ by replacing the vacuum state by a thermal state\footnote{We shall distinguish objects (measure, Lindbladian,...) related to finite temperature with a prime (i.e. $\mathbb{E'}$, ${\cal L'}$,...).}. On Fock spaces such as ${\cal F}_{0;\infty}$, the thermal state is characterized by occupation numbers $\mathfrak{n}_s$ such that $\mathbb{E'}[a^\dag(s)a(s')]=\mathfrak{n}_s\, \delta(s-s')$ for $a^\dag(s)$ and $a(s)$ canonical operators. Instead of using vacuum expectation values as in eq.(\ref{eq:Evac}), thermal expectations  are defined using the thermal state
\[ \mathbb{E'}[\cdots] = \mathrm{Tr}_{{\cal F}_{0;\infty}}\big( \Lambda\,\cdots\big),\]
with $\Lambda$, hermitian and normalized, such that $\mathfrak{n}_s\, a^\dag(s)\Lambda= (1+\mathfrak{n}_s)\, \Lambda a^\dag(s)$. As a consequence, $\mathbb{E'}[d\xi^\dag_td\xi_t]= \mathfrak{n}_t\, dt$ and $\mathbb{E'}[d\xi_t d\xi_t^\dag]= (1+\mathfrak{n}_t)\, dt$, from which we deduce the quantum It\^o rules,
\begin{eqnarray} \label{eq:Etherm}
 d\xi^\dag_t\, d\xi_t = \mathfrak{n}_t\, dt,\quad d\xi_t d\xi_t^\dag= (1+\mathfrak{n}_t)\, dt,
\end{eqnarray}
valid at finite temperature. Zero temperature limit corresponds to $\mathfrak{n}_t=0$. Although we can deal with time dependent occupation numbers, we shall restrict ourselves to time independent occupation numbers $\mathfrak{n}$, for simplicity.

To illustrate the inter-relations between the structures we have developed, we here start from an {\it a priori} on the quantum SDE and derive the rest (the complete quantum SDE, the dynamical map and the quantum trajectory) from this hypothesis.  So, we choose the quantum SDE to be of the form similar to eq.(\ref{eq:dq-SDE}), that is
\begin{eqnarray}\label{eq:dAtemp}
\boxed{ dA_t = {\cal Q}^\dag(A)_t\, d\xi_t + {\cal Q}(A)_t\, d\xi^\dag_t + {\cal L'}_*(A)_t\, dt, }
 \end{eqnarray}
for any system operator $A$, with by hypothesis,
\[ {\cal Q}(A)= i\big[{\cal P},A\big],\quad {\cal Q}^\dag(A)= i\big[{\cal P}^\dag,A\big], \]
where ${\cal P}=P+iN$ as above but with a dual Lindbladian $ {\cal L'}_*$ different from that of zero temperature. This dual Lindbladian is however completely determined, up to a derivative, by the deformed Leibnitz rule (\ref{eq:Lquadra}),
\[ {\cal L'}_*(AB)={\cal L'}_*(A)\,B+A\,{\cal L'}_*(B)+  {\cal Q}(A)\,\mathfrak{n}\,{\cal Q}^\dag(B) + {\cal Q}^\dag(A)(1+\mathfrak{n}) {\cal Q}(B).\]
Indeed, it is clear that, given the derivatives ${\cal Q}(\cdot)$ and ${\cal Q}^\dag(\cdot)$, any two solutions of this equation differ by a derivative which, assuming it is inner, can be written as a commutator. We know that the zero temperature Lindbladian ${\cal L}_*$ is a particular solution for $\mathfrak{n}=0$. Since the term proportional to $\mathfrak{n}$ is obtained from that proportional to $(1+\mathfrak{n})$ by exchanging the role played by ${\cal P}$ and ${\cal P}^\dag$, a particular solution of the above equation can be written as a linear combination of the zero temperature Lindbladian and that with the role ${\cal P}$ and ${\cal P}^\dag$ exchanged. As a consequence, the general solution of this equation is
\begin{eqnarray}\label{eq:Ldtemp}
{\cal L'}_*(A) := i\big[{\cal H'}, A\big] + \mathfrak{n}\big( {\cal P}A {\cal P}^\dag  - \frac{1}{2} ({\cal P} {\cal P}^\dag  A + A {\cal P} {\cal P}^\dag ) \big)
+(1+ \mathfrak{n}) \big( {\cal P}^\dag A {\cal P} - \frac{1}{2} ({\cal P}^\dag {\cal P} A + A {\cal P}^\dag {\cal P}) \big) \nonumber
\end{eqnarray}
Eqs.(\ref{eq:dAtemp}) and (\ref{eq:Ldtemp}) define the quantum SDE for the open QBM at finite temperature.

By duality eq.(\ref{eq:dAtemp}) tells us what is the associated quantum dynamical map (obtained by tracing out the probe Fock space). Namely,
\begin{eqnarray}\label{eq:Ltemp} 
\partial \bar \rho_t = {\cal L'}\big(\bar \rho_t\big),
\end{eqnarray}
with
\[
\boxed{ {\cal L'}(\bar \rho_t) := -i\big[{\cal H'}, \bar \rho_t\big] + \mathfrak{n} \big( {\cal P}^\dag \bar \rho_t {\cal P} -  \frac{1}{2} ({\cal P} {\cal P}^\dag  \bar \rho_t + \bar \rho_t {\cal P} {\cal P}^\dag ) \big) 
+ (1+ \mathfrak{n}) \big( {\cal P}\bar \rho_t {\cal P}^\dag  - \frac{1}{2} ({\cal P}^\dag {\cal P} \bar \rho_t + \bar \rho_t {\cal P}^\dag {\cal P}) \big). } 
\]
Recall that ${\cal P}=P+iN$. We can then expand the above Lindbladian as in eq.(\ref{OQBM:lindblad}). Actually not much computation has to be done because the term proportional to $(1+ \mathfrak{n})$ has already been computed and that proportional to $\mathfrak{n}$ is obtained from the latter by exchanging $N$ and $-N^\dag$. Hence, we get:
\[ {\cal L'}(\bar \rho_t) = -i[H',\rho_t] -\frac{\kappa}{2}[P,[P,\bar \rho_t]] - i([P,N'\,\bar \rho_t+\bar \rho_t\,N'^\dag]) + L'_N(\bar \rho_t),\]
with $\kappa=1+2\mathfrak{n}$, and $N'= (1+ \mathfrak{n}) N - \mathfrak{n}\, N^\dag$ and $L'_N=(1+ \mathfrak{n}) L_N +\mathfrak{n}\, L_{N^\dag}$.

The stochastic equations for simple quantum trajectory can then be found by consistency.  Indeed, for simple trajectory the system density matrix is of the form $\rho_t\otimes \ket{X_t}_z\bra{X_t}$ and the stochastic differential equation it satisfies should be of the form,
\begin{eqnarray} \label{eq:TrajTemp}
\boxed{
\begin{array}{rcl} d\rho_t &=& \big(-i[H',\rho_t]+L'_N(\rho_t)\big)\, dt + D'_N(\rho_t)\, \sqrt{\kappa}\, dB_t,\\
  ~& ~& \\
dX_t &=& U'_N(\rho_t)\, dt + \sqrt{\kappa}\, dB_t ,
\end{array} }
\end{eqnarray}
for some diffusion coefficient $\kappa$ and $D'_N$, and potential $U'_N$, such that its mean $\mathbb{E'}[\rho_t\otimes \ket{X_t}_z\bra{X_t}]$ time evolves linearly according to the Lindbladian ${\cal L'}$. To impose this consistency condition one only has to reproduce (using classical It\^o calculus) the computation validating eq.(\ref{eq:defXrho}) but for the Lindbladian ${\cal L'}$. Again, by linearity, almost no computation has to be done. The term with double $P$-commutator imposes that $\kappa=1+2\mathfrak{n}$. Consistency of the terms linear in $P$ demands that $\rho\, U'_N(\rho)+\kappa\, D'_N(\rho)= N'\, \rho+ \rho\, N'^\dag$ with $N'= (1+\mathfrak{n})N- \mathfrak{n}N^\dag$. Since $D'_N$ is traceless this yields
\begin{eqnarray}\label{eq:U'D'}
U'_N(\rho)&=&\mathrm{Tr}_{{\cal H}_c}(N'\rho+\rho N'^\dag),\\
\kappa\, D'_N(\rho)&=& N'\, \rho+ \rho\, N'^\dag - \rho\,U'_N(\rho),\nonumber
\end{eqnarray}
and $L'_N$ is the thermal Lindbladian associated to $N$ and $N^\dag$ previously defined.

The classical limit corresponds to large occupation numbers\footnote{The fact that the random classical processes require large temperature is clear from the fact that if the probes were prepared in some state at zero temperature and were interacting classically with the system, then no randomness in the output would be present.}: $\mathfrak{n}\gg 1$. Then, quantum noises commute since,
\[ d\xi_td\xi_t^\dag\simeq \mathfrak{n}\, dt,\quad d\xi_t^\dag d\xi_t\simeq \mathfrak{n}\, dt.\]
Furthermore, the potential $U'_N$ vanishes in this limit because $N'$ is then purely imaginary since $N'\simeq \mathfrak{n}(N-N^\dag)$, and the walker trajectories are purely Brownian,
\[ dX_t\simeq \sqrt{2\mathfrak{n}}\, dB_t .\]
The Lindbladians also become quasi-classical in the sense that they can written in terms of double commutators only and $D'_N(\rho)\simeq\frac{1}{2}[N-N^\dag,\rho]$. We leave to the dedicated readers the pleasure to study the approach to classicality.

\section{The case of a spin half gyroscope}

We now consider the case -- the simplest -- of open QBM with a two level system as gyroscope, i.e. ${\cal H}_c=\mathbb{C}^2$. Let $\ket{\uparrow\downarrow}_c$ be a specified basis of state in ${\cal H}_c$, and $\sigma^{1,2,3}$ the standard Pauli matrices in that basis (not to be confused with the Pauli matrices acting on the probe Hilbert space which we introduced earlier but which we denoted by $\sigma^{x,y,z}$).

We do not claim presenting an exhaustive exploration of all properties of open QBM even in this simplest case. Although one certainly should explore it in some details, according to different facets (quantum trajectories, quantum dynamical map, quantum SDE), we shall restrict ourselves to analyze simple quantum trajectory at zero temperature because their behaviors are already rich enough. Some of results presented below were announced in \cite{BBT13}, and we provide more detailed proofs for them.

Let us recall their equations:
\begin{eqnarray}
d\rho_t&=&\big(- i[H,\rho_t]+L_N(\rho_t)\big)dt + D_N(\rho_t)\, dB_t, \label{eq:bisA}\\ 
dX_t&=&U_N(\rho_t)\,dt + \, dB_t, \label{eq:bisB}
\end{eqnarray}
with $L_N(\rho) = N\rho N^\dag - \half(N^\dag N\rho+ \rho N^\dag N)$, and $D_N(\rho) = N\rho+\rho N^\dag -\rho\,U_N(\rho)$ with $U_N(\rho) = {\rm Tr}_{{\cal H}_c}(N\rho+\rho N^\dag)$. Here $X_t$ is the walker position and $\rho_t$ is a $2\times2$ matrix representing the internal density matrix on ${\cal H}_c$. It can be parametrized as $\rho_t=\frac{1}{2}\big(\mathbb{I}+\vec{Q}\cdot\vec{\sigma})$ with $\vec{Q}^2\leq1$, so that it is parametrized by a point in a ball called the Bloch sphere. 

\subsection{Moduli space and regimes}

The moduli of open QBM are the matrices $H$ and $N$, with $H$ hermitian but not necessarily $N$, up to unitary conjugation and translation of $H$ by the identity. Furthermore, a translation of $N$ by a complex multiple of the identity can be absorbed in a redefinition of $H$ up to a translation of the potential $U_N$ by a constant, which simply adds a trivial drift to the walker motion. Hence we can choose both $H$ and $N$ traceless. For ${\cal H}_c=\mathbb{C}^2$, this leaves a moduli space of dimension $6$. 

Even in this simplest case, the moduli space is already quite large and we are not yet able to present a complete picture of its landscape. However, as mentioned earlier, eq.(\ref{eq:bisA}) is a known equation in quantum optics \cite{belavkin,barchielli}, encoding continuous time monitoring \cite{Qprocess}. We use this to extract, as much as we could, pieces of information on their behaviors. The behavior of the walker position is then slave to that of the internal density matrix via eq.(\ref{eq:bisB}).

One may try to organize patterns in the moduli space depending whether $N$ is hermitian or not. This corresponds to the cases $A$ or $B$, and their sub-cases, below. 

{\bf Cases  A}: If $N$ is hermitian, then both $H$ and $N$ are diagonalizable and what matters is whether they are diagonalizable in the same basis or not.

--- Suppose that $H$ and $N$ are both diagonalizable in the same basis. Let $\ket{\alpha}_c$ be the collection of states forming this basis. Eq.(\ref{eq:bisA}) then preserves diagonal density matrices, and each diagonal element $\rho_{\alpha\alpha}$ is a bounded martingale (it is easy to check that the drift term vanishes for diagonal density matrices), and hence they converge almost surely and in $L^1$ at large time. Provided that all real parts of the diagonal elements of $N$ are non zero, the limiting internal density matrix is a projector on one of the basis states, i.e. $\lim_{t\to\infty}\rho_t=\ket{\alpha_\infty}_c\bra{\alpha_\infty}$ for some random target state $\ket{\alpha_\infty}_c$. This corresponds to non-demolition measurements analyzed in \cite{BB11,BBB12}. The collapse to the target state $\ket{\alpha_\infty}_c$ is exponentially fast, so that the drift term in eq.(\ref{eq:bisB}) becomes constant exponentially fast but with a random asymptotic value, depending on the target state $\ket{\alpha_\infty}$, and the walker behavior is simple enough.

\begin{figure}
\centering
\includegraphics[scale=0.37]{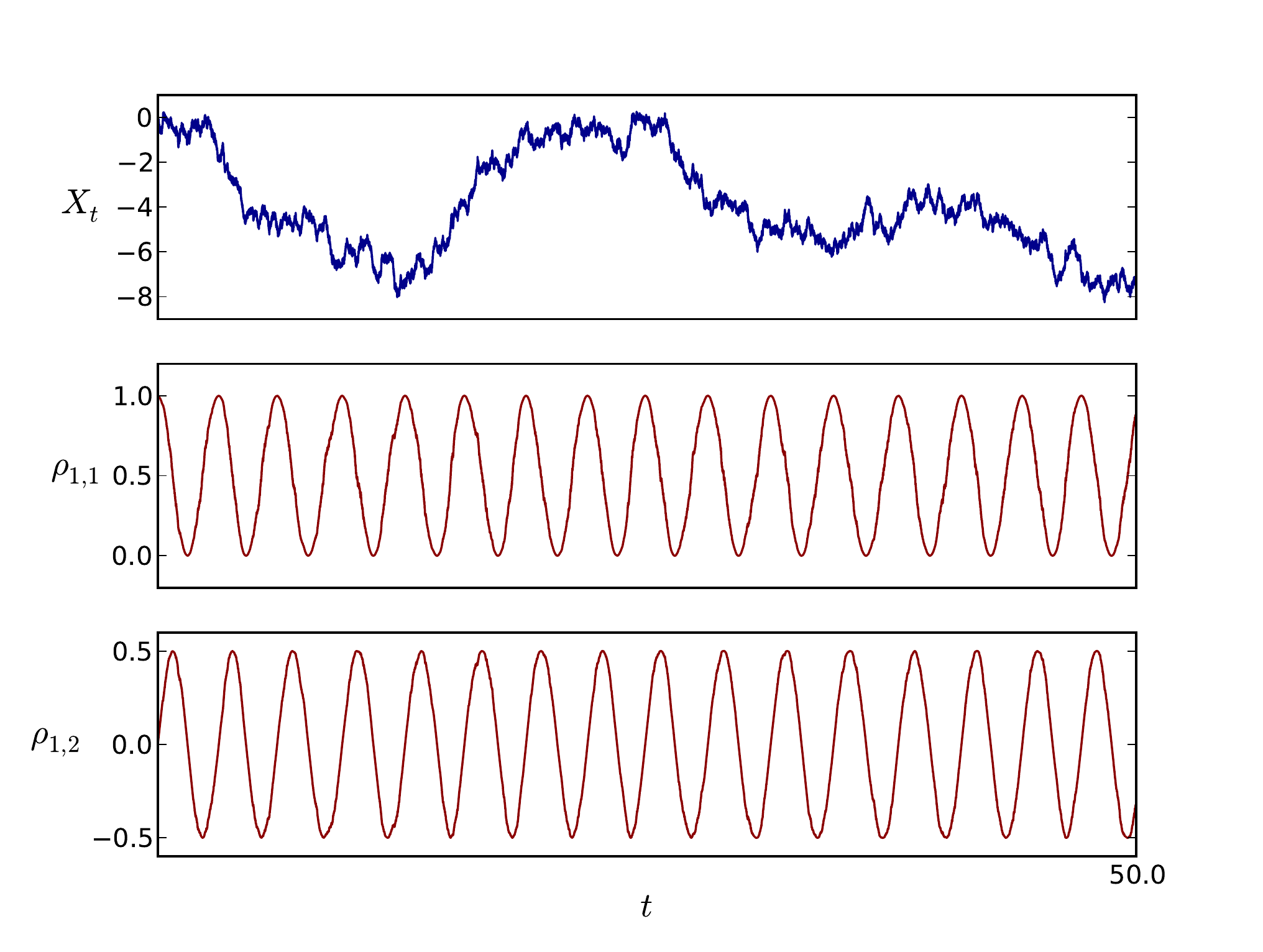}
\includegraphics[scale=0.37]{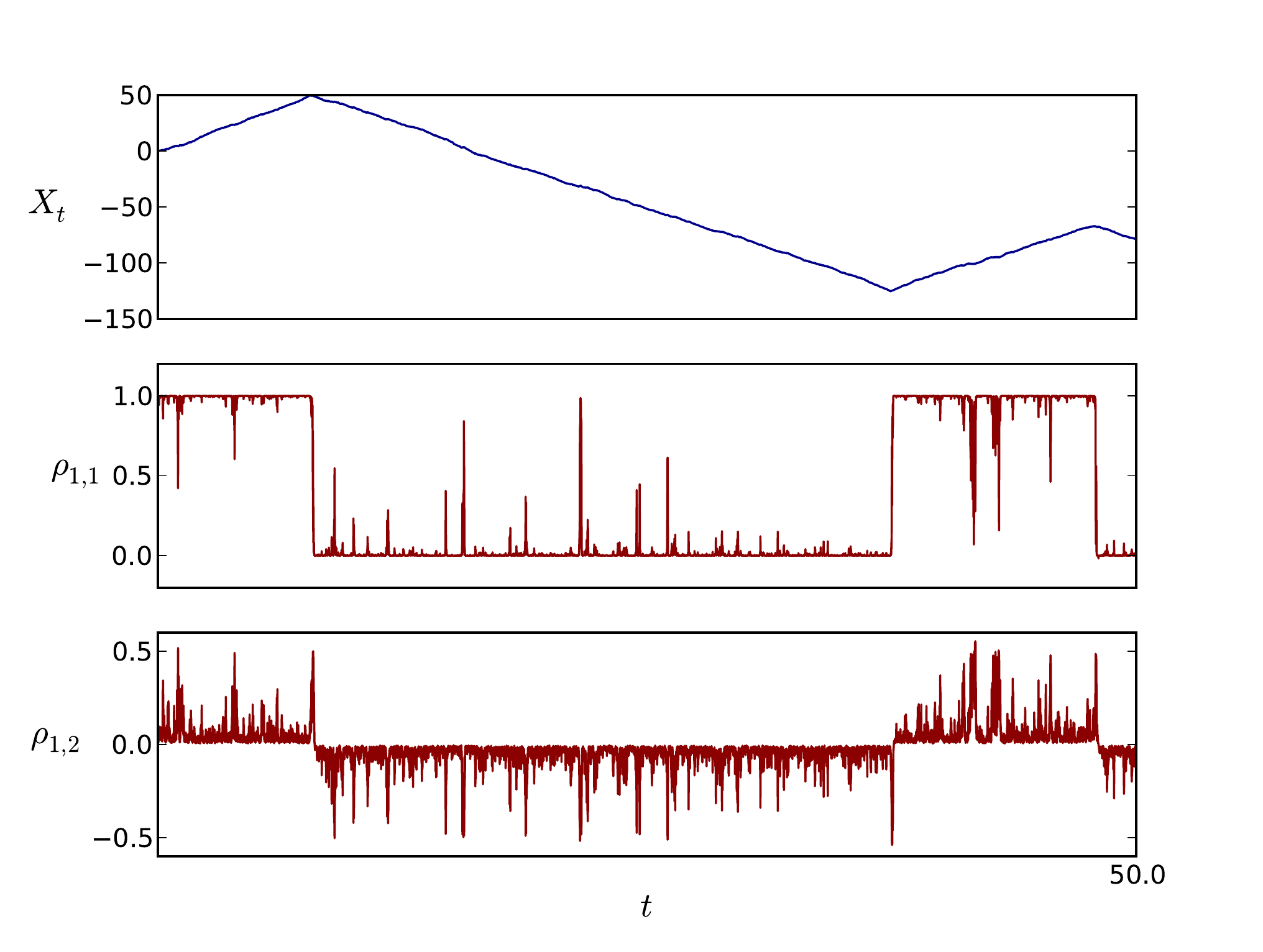}
\caption{\small {\bf Cases A:} Open quantum walk trajectories for $H=-\sigma^2$ and $N=a \sigma^3$. On the left, $a=0.1$, the Rabi oscillations dominate and the walker trajectory is decoupled from the internal evolution. On the right $a=3$ and the measure dominates, we are in the ballistic diffusion regime and the walker trajectory is tightly coupled to the gyroscope evolution.}
\label{fig:ballistic_rabi}
\end{figure}

-- Suppose that $H$ and $N$ are both hermitian but do not commute, and hence cannot be diagonalized in the same basis. This is the case presented in \cite{BBT13}, but in a quite compact form without much details. Two mechanisms are then in competition. If only $H$ were present then the evolution would consist in oscillations between the Hamiltonian eigen-states, called Rabi oscillations. If only $N$ was present, the internal density matrix would behave as in the previous case with exponentially fast collapses on eigen-projectors of $N$, i.e. the dynamics generated by $N$ is that of a non-demolition measurement of the observable $N$. If both $H$ and $N$ are present these two evolutions are in competition and the resulting behavior depends on the relative values of the characteristic time scales of each of these dynamical processes. Let $\tau_\mathrm{oscill}$ be the time scale associated to the Rabi oscillations, and let $\tau_\mathrm{collapse}$ be that of the progressive collapses associated to the non-demolition 
measurements of $N$.\\
 If $\tau_\mathrm{collapse}\gg\tau_\mathrm{oscill}$, the measurement has not enough time to take place during any tiny bit of Rabi oscillation, and as a result, the internal density matrix still oscillates, with small noisy contributions due to the influence of the $N$-dynamics, and correspondingly the walker position has a quasi-Brownian behavior. See Fig. \ref{fig:ballistic_rabi}. \\
 If $\tau_\mathrm{collapse}\ll\tau_\mathrm{oscill}$, the measurement is rapidly effective so that the internal density matrix collapses rapidly on one of the two eigen-projectors of $N$ but the $H$-dynamics then induces jumps from one eigen-projectors to the other. These jumps occur at random time intervals. They are manifestations of Bohr quantum jumps \cite{Bohr} observed in quantum optics \cite{Qjump,Qjump2}. As a consequence the walker position follows a random see-saw trajectory, with a ballistic behavior at intermediate scale quite different from the previous quasi-Brownian behavior. See Fig. \ref{fig:ballistic_rabi}.

 This is the case that we are going to describe more precisely in the following.

{\bf Cases B}: $H$ is hermitian but not $N$. Then it matters whether $N$ is diagonalizable or if it possesses a non-trivial Jordan cell (that is, if it can be reduced to a triangular but not diagonalizable matrix). 

\begin{figure}
\centering
\includegraphics[scale=0.37]{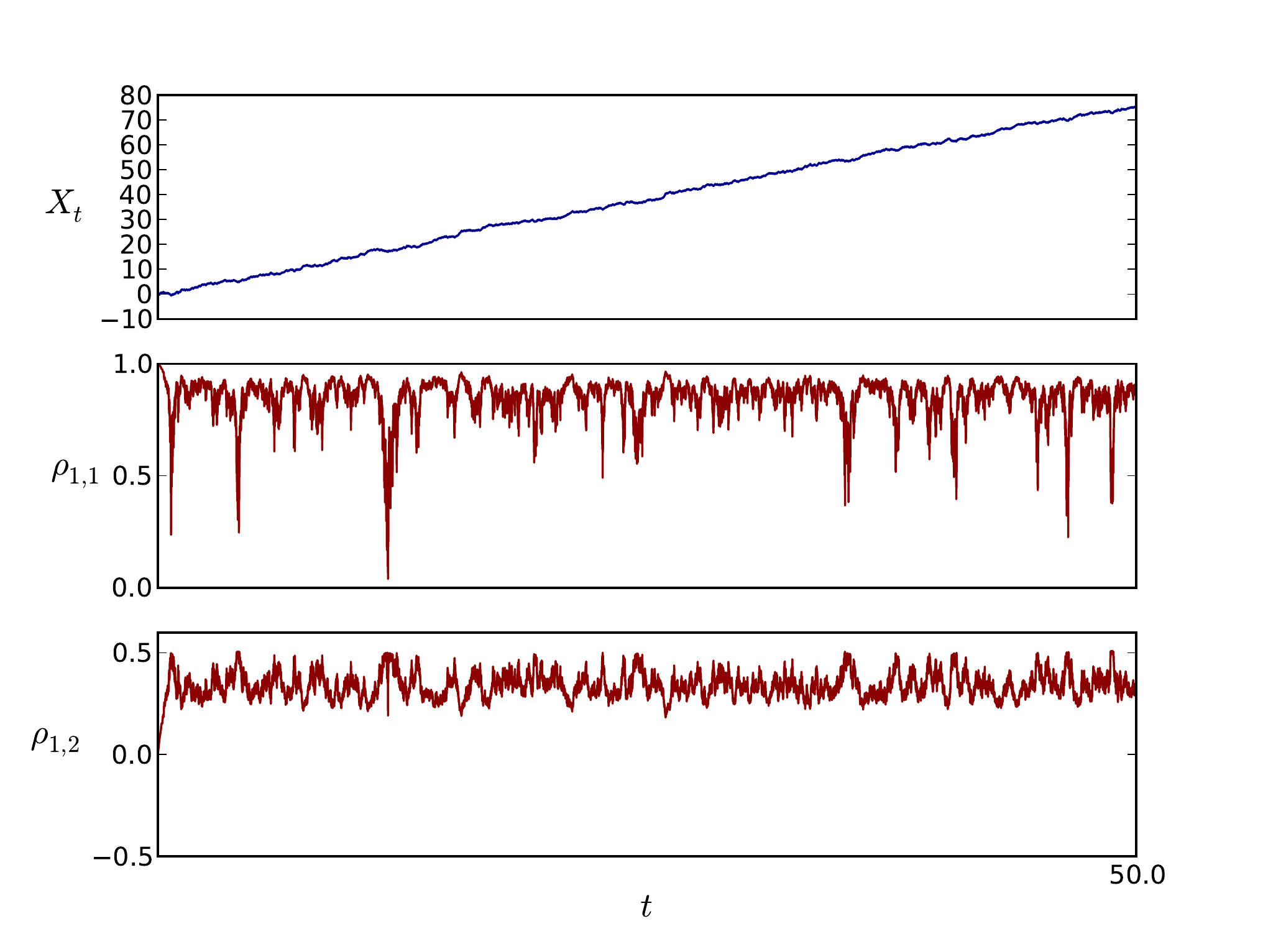}
\includegraphics[scale=0.37]{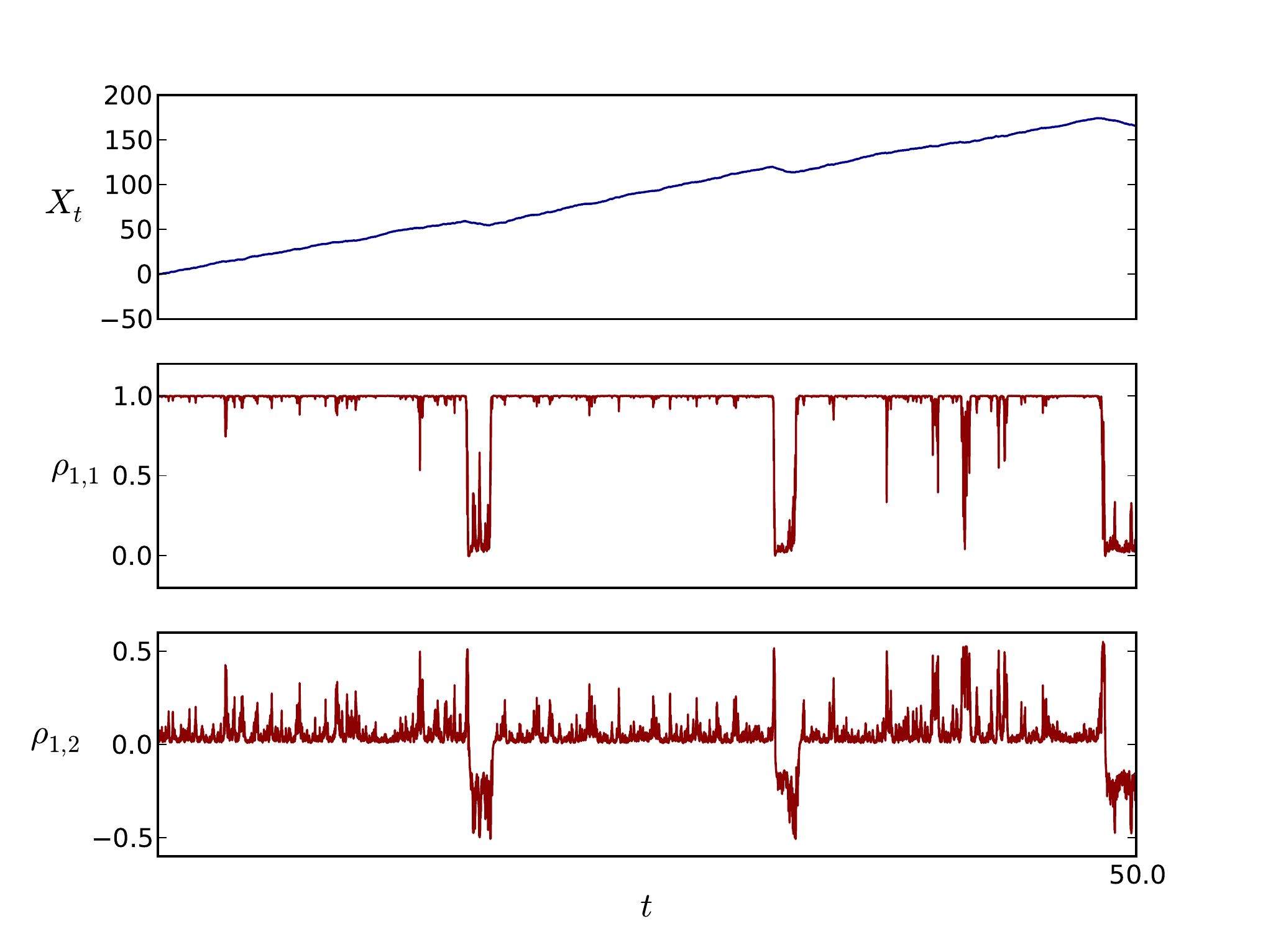}
\caption{\small {\bf Cases B:} Open quantum walk trajectories for $H=-\sigma^2$ and $N=a \sigma^3+b\sigma^+$. On the left, $a=0$ and $b=2.0$, spontaneous emission dominates and one state is so much favored that the Rabi oscillations are completely erased. On the right $a=3$ and $b=0.5$, we are thus in the ballistic diffusion regime with the addition of a preference for one state via spontaneous emission of the other which effectively asymetrizes the walk.}
\label{fig:nilpotent}
\end{figure}

--- If $N$ has a non-trivial Jordan cell, we may identify it with $\sigma^+=\left(\begin{smallmatrix} 0 &1 \\ 0 & 0 \end{smallmatrix}\right)$ in some basis. Assuming that the Hamiltonian $H$ is still hermitian in this new basis, we generically have to examine two cases: either $H\propto \sigma^3$ or $H\propto \sigma^2$. For $N\propto \sigma^+$, the $N$-dynamics amounts to populate one of the two basis states (physically representing emission or absorption). So, if $H$ is diagonal in this basis, not much happens and the internal density matrix rapidly converges to the projector on this selected basis state. If $H$ is not diagonal in this basis, two dynamical processes are in competition, Rabi oscillations induced by $H$ and emissions induced by $N$. So, there is a progressive change of regime from pure Rabi oscillations to pure emission when increasing the strength of emissions (i.e. of the $N$-dynamics). But this does not induce dynamical patterns as interesting as the previous one. See Fig \ref{fig:nilpotent}.

--- If $N$ is non-hermitian but diagonalizable the situation is more complex and, we must admit, not completely understood. One gets, in most cases, a mix of the two previous effects. We have seen that adding a multiple of the identity to $N$ was equivalent to a redefinition of the Hamiltonian so we can just take it to be traceless. An example of a subclass of such matrices is the upper trigonal matrices of the form  $N=a\sigma^3+b\sigma^+=\left(\begin{smallmatrix} a &b \\ 0 & -a \end{smallmatrix}\right)$, i.e. matrices that are just a sum of the two matrices discussed above. We do not claim to describe exhaustively all the parameter space with this simpler subset but rather hope that it will give the reader a feeling of what happens more generally. The evolution is then slightly more subtle and gives rise to a competition between Rabi evolution, measurement along $\sigma^3$ and spontaneous emission. This complex superposition of effects can be understood in limiting cases, for example in the case of a small perturbation of the ballistic diffusion regime. For large $a$ and $b=0$, one is in this see-saw regime previously discussed. When $b$ is slowly increased, one state is favored compared to the other by spontaneous emission. This has the effect of asymetrizing the previous situation where both states had the same probability. See Fig.\ref{fig:nilpotent}. Let us emphasize the fact that this discussion is heuristic and not flawless as it misses interesting special values of the parameters. Indeed, for the line $ab=2$ (in a basis where $H=\sigma_2$), the Lindbladian takes a special form and the evolution is dramatically different from what our naive ``perturbative'' discussion would suggest. This is yet an other example showing that even with a 2-dimensional internal space, open QBM have an extremely rich structure which we hope will be unraveled soon.

\medskip

In the following we provide a detailed presentation of the behaviors of simple open QBM trajectories when $N$ and $H$ are both hermitian but not commuting. Up to conjugation we can choose $N\propto \sigma^3$, diagonal in the basis $\ket{\uparrow\downarrow}_c$, and $H\propto \sigma^2$, and this choice ensures that we will be able to restrict ourselves to real density matrices. We set:
\[ H=-\omega_0\, \sigma^2,\quad N=a\, \sigma^3.\]
We take $\omega_0>0$.
The time scale of the Rabi oscillation is $\tau_\mathrm{oscill}:=\omega_0^{-1}$, and that of the collapse induced by the $N$-dynamics is $\tau_\mathrm{collapse}:=a^{-2}$. As we shall show below, there are two regimes: for $a^2<\omega_0$, the internal density matrix oscillates almost regularly and the walker position is quasi-Brownian, for $a^2>\omega_0$, the internal density matrix is subject to random jumps between two values, asymptotically close to the eigen-projectors of $N$, and the walker trajectories have see-saw profiles.

\subsection{Ballistic regime at mesoscopic time scale}

Eqs.(\ref{eq:bisA},\ref{eq:bisB}), with $H=-\omega_0\, \sigma^2$ and $N=a\, \sigma^3$, are compatible with reality of the internal density matrix, and we restrict our analysis to such class of density matrices for simplicity. We parameterize them as 
\[ \rho_t=\half(\mathbb{I}+ Q_1\sigma^1+ Q_3\sigma^3),\]
with $Q_1^2+Q_3^2\leq 1$. Eqs.(\ref{eq:bisA}) then reads:
\begin{eqnarray*}
dQ_3&=& 2\omega_0\, Q_1\,dt + 2a(1-Q_3^2)\, dB_t\\
dQ_1&=& -2(\omega_0\, Q_3+a^2\, Q_1)\,dt - 2a\, Q_1Q_3\, dB_t
\end{eqnarray*}

The first property of these trajectories is that they rapidly converge to pure states (which correspond to density matrices of rank one). Of course these states are random and still evolve in time. To prove convergence to pure states, we consider the determinant $\Delta_t:=\det\rho_t$. We have:
\[ d\Delta_t^{1/2} = \Delta_t^{1/2}\big[ -2a^2\, dt - 2a Q_3^2\, dB_t\big],\]
so that $\Delta_t^{1/2}$ is a sub-martingale (because the drift in the above equation is always negative). It is of course bounded, and therefore it converges almost surely and in $L^1$ \cite{probas}. Its mean decreases exponentially fast $\mathbb{E}[\Delta_t^{1/2}]=\Delta_0^{1/2}\, e^{-2a^2t}$. Since, $\Delta_t^{1/2}$ is non-negative, its limit is zero, $\lim_{t\to\infty}\Delta_t^{1/2}=0$ almost surely, and the limiting internal density matrix is of rank one. This convergence to pure states is a particular example of more general results obtained in \cite{Maas}.

Informed by this property, we describe $\rho_t$ as a pure state, so that $Q_1^2+Q_3^2=1$ and we use the parametrization $Q_1=\sin\theta$, $Q_3=\cos\theta$. The angle $\theta$ satisfies
\begin{equation} \label{eq:dtheta}
d\theta_t=-2(\omega_0+a^2\sin\theta_t\cos\theta_t)\,dt - 2a\sin\theta_t\, dB_t
\end{equation}
This is a classical SDE for a variable on the unit circle, and we may use all standard results for such equations. The following discussion relies freely classical formul\ae\ for $1$-dimensional diffusions (see
e.g. the third reference in \cite{probas}).

\begin{figure}
\centering
\includegraphics[scale=0.04]{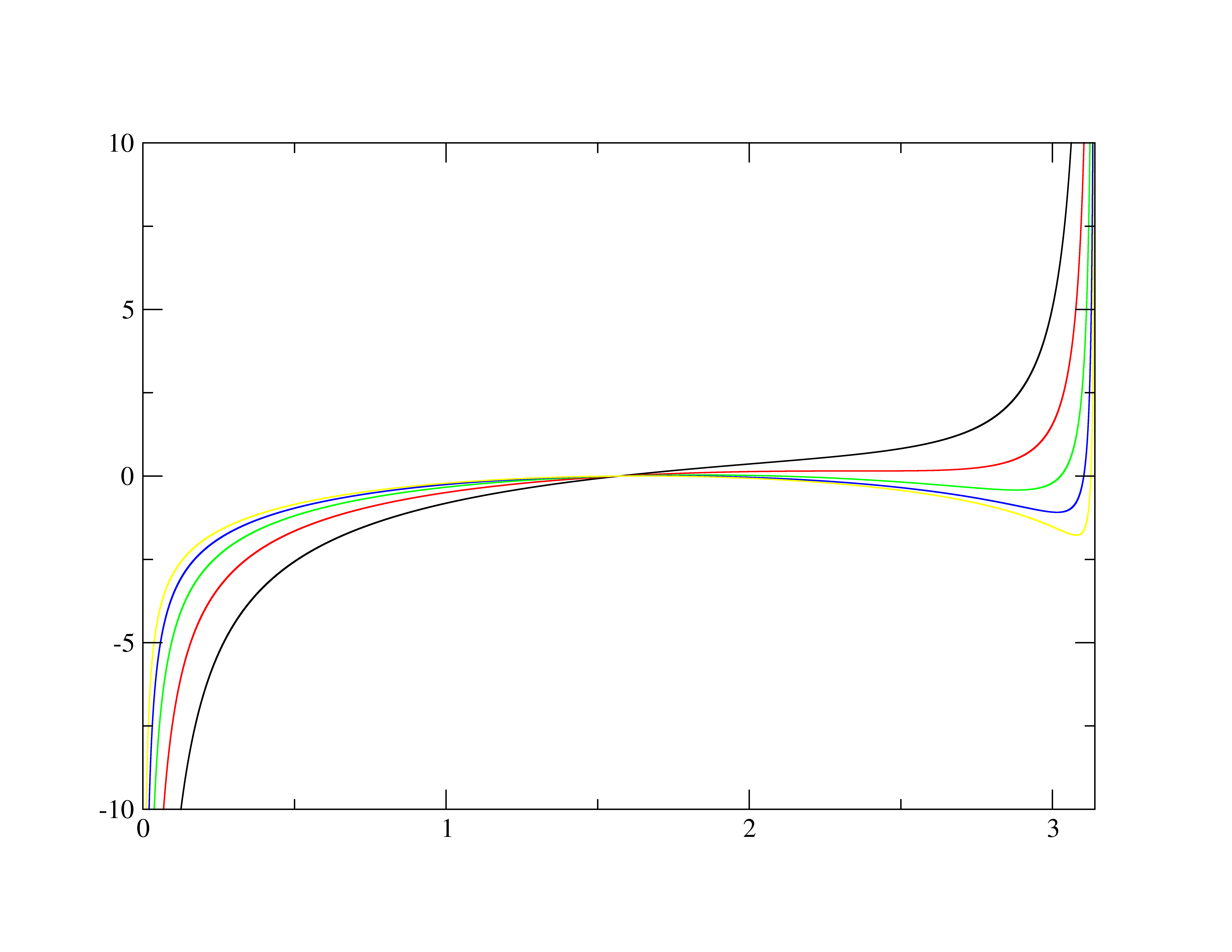}

\caption{\small The potential $W(\theta)$, represented for $a^2=\omega_0$
  (black curve), $a^2=2\omega_0$ (red curve, critical case), $a^2=4\omega_0$,
  $a^2=8\omega_0$ and $a^2=16\omega_0$ (yellow curve).}
\label{fig:potentiel}
\end{figure}

The existence of two different regimes --- one, for $a^2< 2\omega_0$, in which
the internal matrix oscillates almost regularly, and the other, for $a^2> 2\omega_0$, in which it jumps randomly from one state to another --- can be grasped as follows. Presenting eq.(\ref{eq:dtheta}) in the form $d\theta_t= f(\theta_t)\, dt + g(\theta_t)\, dB_t$ with $f(\theta):= -2(\omega_0+a^2\sin\theta\cos\theta)$ and $g(\theta)=- 2a\sin\theta$, we deduce that it possesses an invariant measure proportional to
\[ e^{-2W(\theta)}\, g^{-2}(\theta)\, d\theta := e^{2\int^\theta du\, f(u)/g^2(u)}\, g^{-2}(\theta)\, d\theta.\]
Here $W$ is given by
\[ 2 W(\theta)=\log | \sin \theta| - (\frac{ \omega_0}{a^2}) \cot \theta. \] The
shape of $W$ is different depending on whether $a^2<2\omega_0$ or $a^2>
2\omega_0$, and this is an echo of the two different regimes\footnote{We are
  talking about different regimes and not about phase transition as the
  bifurcation is not sharp.} for the behaviors of the quantum trajectories. The
function $W$ (which is $\pi$-periodic), has no extrema in $]0,\pi[$ for
$a^2\leq 2\omega_0$, whereas for $a^2>2\omega_0$ it possesses a minimum and a
maximum. See Fig \ref{fig:potentiel}.

Hence by ergodicity, for $a^2\ll\omega_0$, the angle $\theta_t$ winds around the
unit circle almost regularly, whereas for $a^2\gg\omega_0$, this angle spends
most of its time in the neighborhood of the minima and jumps \`a la Kramers
between them. The maxima  on the circle are always close to $\pi/2$ and
$3\pi/2$. The minima on the circle, which we denote by $\theta^*$
(asymptotically close to $0^-$ for $a^2\gg\omega_0$, namely $\theta^*\simeq
-\omega_0/a^2$) and $\theta^*+ \pi$, are such that when $\theta_t$ is close to
them the internal density matrix is close to the pure eigen-projectors of $N$,
i.e.  $\rho_t\simeq \ket{\uparrow}_c\bra{\uparrow}$ for $\theta\simeq 0^-$ and
$\rho_t\simeq \ket{\downarrow}_c\bra{\downarrow}$ for $\theta\simeq
\pi^-$. As a consequence, for $a^2\gg \omega_0$, the internal system spends most
of this time in one of the two states $\ket{\uparrow\downarrow}_c$ with sharp
transitions between them.

The walker position evolution is governed by eq.(\ref{eq:bisB}) which here reads: $d X_t= 2a\cos\theta_t\, dt + dB_t$. When $\theta_t\simeq 0^-$ ($\theta_t\simeq\pi^-$), it is subject to a constant positive (negative) drift $\pm 2a$ dominating over the Brownian noise for $a$ large enough. This yields the see-saw profiles observed in Fig. \ref{fig:ballistic_rabi}.

We provide a more precise description of these behaviors below. From this we
learn that the mean time $\bar \tau_\mathrm{wait}$ the internal system spends in
either of these states before jumping to the other one is
\begin{eqnarray} \label{Twait}
 \bar \tau_\mathrm{wait} = a^2/\omega_0^2 \propto \tau^2_\mathrm{oscill}/\tau_\mathrm{collapse},
 \end{eqnarray}
for $a^2\gg \omega_0$, and that the times in between jumps are distributed exponentially.
\medskip 

\underline{\it ``Esquisse d'une preuve" (I):}\\
A feature of (\ref{eq:dtheta}) that we want to note is that beside the
obvious symmetry $\theta \rightarrow \theta +2 \pi$ their is a symmetry
$\theta \rightarrow \theta + \pi$. To be more precise, if $B_t \rightarrow
\tilde{B}_t=-B_t$ is the standard, distribution preserving, involution of
Brownian motion and if $\theta_t$ is a solution of (\ref{eq:dtheta}) with noise
$B_t$ and value $\theta_0$ at time $0$ then $\theta_t +\pi$ is a solution of
(\ref{eq:dtheta}) with noise $\tilde{B}_t$ and value $\theta_0 +\pi$ at time $0$.
This allows us to concentrate on initial conditions  $\theta_0 \in [0,\pi]$. 

The next key observation is that there is an arrow for the evolution of
$\theta_t$: when $\theta =k\pi$ for some integer $k$, the noise term vanishes,
while the drift term is negative (assuming $\omega_0 >0$) so that there is an
irreversible gate: $\theta_t$ will easily transit from $(k\pi)^+$ to
$(k\pi)^-$ but the other direction is forbidden. 

We turn this hand-waving argument to a more rigorous one as follows. 
For $\theta_- < \theta < \theta_+$, let $P_{[\theta_-, \theta_+]}(\theta)$ be the probability that a
trajectory started at $\theta$ exits the interval $[\theta_-, \theta_+]$ at
$\theta_-$. By standard probabilistic arguments, 
$P_{[\theta_-,\theta_+]}(\theta_t)$ is a martingale, so that 
\[
2a^2\sin^2\theta\, \frac{d^2  P_{[\theta_-, \theta_+]}(\theta)}{d \theta^2}
-2(\omega_0+a^2\sin\theta\cos\theta)
\frac{d P_{[\theta_-, \theta_+]}(\theta)}{d \theta}=0.
\]
It follows that $ P_{[\theta_-, \theta_+]}(\theta) \propto
\int^\theta d\vartheta\, e^{2W(\vartheta)}$ whenever this integral makes sense. 
The precise value is fixed by boundary conditions. As long as the integrals
converge, one gets 
\[ 
P_{[\theta_-,\theta_+]}(\theta)=\frac{\int_\theta ^{\theta_+} d\vartheta\, e^{2W(\vartheta)}}{\int^{\theta_+}_{\theta_-} d\vartheta\, e^{2W(\vartheta)}}.
\]

Since $2W(\theta)=\log | \sin \theta| -\frac{ \omega_0}{a^2} \cot
\theta$, this formula holds as long as there is no $k\pi \in\,
]\theta_-, \theta_+]$. Indeed, $W(\theta)$ is $\pi$-periodic and $\lim_{\theta\to\pi^-}
W(\theta) = +\infty$ while $\lim_{\theta\to0^+} W(\theta) =-\infty$. The rate of growth
at $\theta \rightarrow \pi^-$ is such that the integral $\int^{\pi^-} d\vartheta
e^{2W(\vartheta)}$ is divergent. Hence the above formula for $P_{[\theta_-,
  \theta_+]}(\theta)$ is valid for $k\pi \leq \theta_- < \theta \leq \theta_+ <
(k+1)\pi$ for each integer $k$. The other cases are obtained by limiting and
consistency arguments. They entail that if $\theta_t \leq k\pi$ then $\theta_s
\leq k \pi$ for every $s \geq t$ (no return to higher $\theta$'s is possible), and
that whatever $\theta_0$ is, the event $\lim_{t\rightarrow +\infty} \theta_t
=-\infty$ has probability $1$.

Due to the divergence of $W(\theta)$ at $\pi^-$, the average time $ \bar
\tau_\mathrm{flat}$ it takes to the angle $\theta_t$ to travel from $\pi$ to $0$ is
finite and given by 
\[
\bar \tau_\mathrm{wait} = 2\int_0 ^\pi d\vartheta\,
e^{2W(\vartheta)} \int_\vartheta^\pi e^{-2W(\theta)}\, g^{-2}(\theta)\, d\theta.
\]
Note that the second integrand involves the invariant measure. 
Although a (known) martingale argument can be used to get this formula, we do not spell it out here because
it is a special case of the martingale argument we shall give below when computing
the probability distribution function of the time it takes to the angle $\theta_t$ to travel from
$\theta \in [0,\pi]$ to $0$. To estimate $ \bar \tau_\mathrm{wait}$ for
${ \omega_0}\ll{a^2}$ we set $\cot \theta := \cot
\vartheta - (\frac{a^2}{\omega_0})\,u$ in the second integral and get
\[ \bar \tau_\mathrm{wait} = \frac{a^2}{2\omega_0^2}
\int_{0}^{+\infty} du e^{-u}\int_{0}^{\pi} d\vartheta
\sqrt{u^2\sin ^2 \vartheta -\frac{\omega_0}{a^2}\, u \sin 2\vartheta
  +\frac{\omega_0^2}{a^4}\ }.\]
For large ${a^2}/{\omega_0}$ we thus have
\[ \bar \tau_\mathrm{wait} \sim \frac{a^2}{2\omega_0^2}
\int_{0}^{+\infty} du\, e^{-u} u \int_{0}^{\pi} d\vartheta\, \sin \vartheta =  \frac{a^2}{\omega_0^2},\]
as announced above.

\textit{Remark}: It is not difficult to see that the subleading term in $\tau_\mathrm{flat}$ involves
$\log({a^2}/{\omega_0})$ but a systematic expansion is quite
cumbersome. However \cite{luck:private}, borrowing a trick from disordered
systems and random matrices \cite{luck-et-al}, one can rewrite $\bar \tau_\mathrm{wait}$
as the ratio of two simple integrals (of the Bessel function type), leading to a
 routine systematic expansion. The probabilistic aspect of this trick is not
 transparent to the authors, but an important tool on the random matrix side is
 to recognize $\bar \tau_\mathrm{wait}$ as the real part of an analytic function whose
 imaginary part could have some interesting probabilistic meaning in our case. 

\underline{\it ``Esquisse d'une preuve" (II):}\\
 We now turn to the full distribution of jump times for arbitrary starting and
 ending points when $ \frac{a^2}{\omega_0} \rightarrow +\infty$. We take $0 \leq
 \theta_f \leq \theta_i \leq \pi$ and write $\bar T(\theta_i \rightarrow
 \theta_f):=\inf \{t, \theta_t < \theta_f | \theta_0=\theta_i\}$ for the (random)
 time it takes for $\theta_t$ to travel from $\theta_i$ to $\theta_f$. If
 $\theta \in [\theta_f, \theta_i]$ one can split a path contributing to $\bar
 T(\theta_i \rightarrow \theta_f)$ as a path from $\theta_i$ to $\theta$
 and then a path (independent of the previous one by the strong Markov property)
 to go from $\theta$ to $\theta_f$. So one can write
\[\mathbb{E}[e^{-u\,\bar T(\theta_i \rightarrow \theta_f)/ \bar
  \tau_\mathrm{flat}}]=e^{-\int_{\theta_f}^{\theta_i} \varphi(\theta,u)\, d\theta }\] 
for some (nonnegative) function $\varphi(\theta,u)$. By the strong Markov
property, one gets that
\[t\to e^{-u\,t/ \bar \tau_\mathrm{wait}} e^{-\int_{\theta_f}^{\theta_t} \varphi(\theta,u)\,
  d\theta }\] 
is a martingale on $t \leq \bar T(\theta_i \rightarrow \theta_f)$, and an
application of It\^o's formula yields
\[ -\frac{u}{\bar \tau_\mathrm{wait}}-f(\theta)\,\varphi(\theta,u)
-\frac{1}{2}g(\theta)^2\Big(\frac{\partial \varphi}{\partial
    \theta}(\theta,u)-\varphi(\theta,u)^2\Big) =0. \]
For large $\bar \tau_\mathrm{wait}$ the first term is negligible, and keeping only the
dominant contributions in the second and third terms one is lead to the limiting
equation 
\[\cos \theta\, \varphi(\theta,u) +\sin \theta \Big(\varphi(\theta,u)^2- 
  \frac{\partial \varphi}{\partial \theta}(\theta,u)\Big)=0. \] 
The general solution is $\varphi(\theta,u)=\frac{\sin \theta}{C(u)+\cos \theta}$ where
$C(u)$ is an arbitrary function still to be determined. A detailed computation
of $C(u)$ is quite cumbersome. We content with the following heuristic argument:
if $\theta_i$ and $\theta_f$ are both very close to $0$ or to $\pi$, so close
that $a^2 \sin \theta \ll \omega_0$ for $\theta \in [\theta_f,\theta_i]$, the
time to travel from $\theta_i$ to $\theta_f$ is deterministic and $T(\theta_i
\rightarrow \theta_f)\simeq \frac{\theta_i-\theta_f}{2\omega_0}$. As the region
close to $\pi$ is responsible for the scaling of $\bar \tau_\mathrm{wait}$, we choose
that region to match\footnote{Matching with the region close to $0$ would lead to an inconsistent
formula.} with the formula for $\varphi(\theta,u)$, i.e. we write
$\int_{\theta_f}^{\theta_i} \varphi(\theta,u)\, d\theta \simeq
u\,\frac{\theta_i-\theta_f}{2\omega_0}$. This fixes $C(u)=1+\frac{2\omega_0}{u}$.
Note that this ensures that $\varphi(\theta,u)$ is positive and finite on
$[0,\pi]$. Finally, for $a^2 \gg \omega_0$ we obtain
\[ \mathbb{E}[e^{-u\,\bar T(\theta_i \rightarrow \theta_f)/ \bar
  \tau_\mathrm{flat}}] \sim \frac{\omega_0+ u \cos^2 \theta_i/2}{\omega_0+ u \cos^2
  \theta_f/2}.\] 
The Laplace transform of the above formula is easily done. 
In the limit $a^2 \gg \omega_0$, it yields that the distribution of $\bar T(\theta_i
\rightarrow \theta_f)/ \bar \tau_\mathrm{wait}$ is the mixture of a Dirac $\delta$-peak
at $0$ (weight $\frac{\cos^2 \theta_i/2}{\cos^2 \theta_f/2}$) and an exponential
distribution with parameter $\frac{\omega_0}{\cos^2 \theta_f/2}$ (weight
$1-\frac{\cos^2 \theta_i/2}{\cos^2 \theta_f/2}$),
\[ 
\mathbb{P}[\bar T(\theta_i \rightarrow \theta_f) \in B ]
= (\frac{\cos^2 \theta_i/2}{\cos^2 \theta_f/2}) \, \mathbb{I}_{\, 0 \in B} + (1-\frac{\cos^2 \theta_i/2}{\cos^2 \theta_f/2})
(\frac{\omega_0}{\cos^2 \theta_f/2}) \int_B  e^{-\frac{\omega_0}{\cos^2
    \theta_f/2}\, s} \, ds
\]
for any Borel set $B$.
Physically, the meaning of the $\delta$-peak that appears in the limit $a^2 \gg \omega_0$ is that as soon as
$\theta_i < \pi$, there is a finite probability to go to $\theta_f$ without being
trapped in the well. However, there is also a probability to get trapped, and
then the time spend is exponential. The fact that the parameter of the
exponential distribution depends on $\theta_f$ means that the well is only
logarithmically deep and not really localized even in the limit $a^2 \gg
\omega_0$. The most interesting special case is when $\theta_i=\pi$ and
$\theta_f=0$, then
\begin{equation}\label{Pexp}
 \mathbb{P}[\bar T(\pi\rightarrow 0) > s] = e^{-\omega_0\, s}.
 \end{equation}
As a consequence, the Markov property implies that if $a^2 \gg \omega_0$ and
$\theta_0=0$ the successive passages times at $\theta=-k\pi$, $k=0,1,\cdots$,
converge to a standard Poisson process with parameter $\omega_0$. \cqfd

\subsection{Diffusive regime at macroscopic time scale}

The ballistic behavior of simple quantum trajectories (with $H=-\omega_0\, \sigma^2$ and $N=a\, \sigma^3$) we described above occurs at intermediate mesoscopic time scale. At asymptotically large time the walker position is Gaussian in accordance with the central limit theorem proved in \cite{attal2} for open quantum random walks. 

We claim, and prove below, that $X_t/\sqrt{t}$ is Gaussian with zero mean and covariance $(1 + 4a^4/\omega_0^2)$:
\begin{eqnarray} \label{eq:Xpdf}
\lim_{t\to\infty} \mathbb{E}[\,e^{ikX_t/\sqrt{t}}\,]\propto e^{-D_{\rm eff}\, k^2/2},
\end{eqnarray}
for any real $k$, with effective diffusion constant $D_{\rm eff}=1 + 4a^4/\omega_0^2$, for $a^2>2\omega_0$. In the limit of large $a$ we are interested in, the effective diffusion constant is asymptotically large and much bigger than the `bare' diffusion constant (here normalized to $1$). 
\medskip

\underline{\it ``Esquisse d'une preuve":}\\
We aim at finding the large behavior of $X_t/\sqrt{t}$, and this is encoded into the large time behavior of its generating function $\mathbb{E}[e^{ikX_t/\sqrt{t}}]$. Since $X_t$ is coupled to $Q_{1,3}(t)$, we introduce:
\[ E_t(k):= \mathbb{E}[e^{ikX_t}],\ R_t(k):=\mathbb{E}[Q_3(t)e^{ikX_t}],\ S_t(k):=\mathbb{E}[Q_1(t)e^{ikX_t}].\]
These satisfy linear equations, for the same reason as the mean density matrix satisfies a Lindblad linear equation. Namely,
\begin{eqnarray*}
\partial_t E_t &=& 2ika\, R_t - \frac{k^2}{2}\, E_t,\\
\partial_t R_t  &=& 2\omega_0\, U_t - \frac{k^2}{2}\, R_t + 2ika\, E_t,\\
\partial_t S_t &=& -(2a^2+\frac{k^2}{2})\, S_t - 2\omega_0\, R_t.
\end{eqnarray*}
We change variables replacing $F_t(k)$ by $F_t(k)e^{-\frac{k^2}{2}t}$ for all three functions $E,\ R$ and $S$. We then get three new equations for the new functions $E,\ R$ and $S$ (we use the same name/labeling for these new functions). These equations are the same as above but without the terms containing ${k^2}/{2}$. They are again linear equations for the three variables $E,\ R$ and $S$ that we commonly denote by $\vec F$ (as a three dimensional vector). The characteristic equation for the eigenvalues $\gamma$ of this linear equation reads:
\[ \gamma^3+ 2a^2\gamma^2 + (4k^2a^2+4\omega_0^2)\gamma+ 8k^2a^4=0.\]
We denote by $\gamma_0(k)$ and $\gamma_\pm(k)$ the three eigen-values, and by $\vec V_0(k)$ and $\vec V_\pm(k)$ the corresponding eigen-vectors.  At $k=0$ the eigen-values are $\gamma_0(0)=0$ and $\gamma_\pm(0)=-a^2\pm\sqrt{a^4-4\omega^2_0}$ (which is negative for $a^2>2\omega_0$).

Since we are interested in the distribution of $X_t/\sqrt{t}$, we look at the large time behavior of $\vec F_t$ but evaluated at $k/\sqrt{t}$. Because $\vec F_t$ is solution of the above linear system, we may write
\[ \vec F_t\big(\frac{k}{\sqrt{t}}\big)= \vec V_0\big(\frac{k}{\sqrt{t}}\big)\, e^{\gamma_0(\frac{k}{\sqrt{t}})t}+
\vec V_+\big(\frac{k}{\sqrt{t}}\big)\, e^{\gamma_+(\frac{k}{\sqrt{t}})t}+
\vec V_-\big(\frac{k}{\sqrt{t}}\big)\, e^{\gamma_-(\frac{k}{\sqrt{t}})t},\]

For $k$ small, we have $\gamma_*(k)=\gamma_*(0)+\delta_* k^2+\cdots$ perturbatively in $k^2$. In particular $\gamma_0(k)= -\frac{2a^4}{\omega_0^2} k^2+\cdots$ and $\gamma_\pm(k)= \gamma_\pm(0)+\cdots$. As a consequence only the first term proportional to $\vec V_0$ survives the large time limit, and $\lim_{t\to\infty} \vec F_t(\frac{k}{\sqrt{t}})= \vec V_0(0)\, e^{\delta_0\, k^2}$. One verifies that $\vec V_0(0)=(1,0,0)$ and therefore $\lim_{t\to\infty} \mathbb{E}[e^{ikX_t}]\propto e^{-D_{\rm eff}\, k^2/2}$, with effective diffusion constant $D_{\rm eff}= 1 + 2\delta_0=1 + 4a^4/\omega_0^2$, as claimed in eq.(\ref{eq:Xpdf}) above.
\cqfd

%\section{Conclusion}
%If needed.................

%\vskip 1.5 truecm

%{\bf Acknowledgements}:  This work was in part supported by ANR contract ANR-2010-BLANC-0414. 

\end{document}